\documentclass[final,5p,times,twocolumn,authoryear]{elsarticle}

\usepackage{amssymb}
\usepackage{amsmath}
\usepackage{url}

\newcommand{\DMunit}{$\rm pc\,cm^{-3}$}

\newcommand{\RMunit}{$\rm rad\,m^{2}$}

\journal{Astronomy and computing}

\begin{document}

\begin{frontmatter}



\title{Unlocking the hidden potential of pulsar astronomy} 


\author{
{D. Kaur},$^{1}$
{G. Hobbs},$^{2}$
{A. Zic},$^{2}$
{J. R. Dawson},$^{2,3}$
{J. Morgan},$^{1}$
{W. Ling},$^{2}$
{S. Camtepe},$^{4}$
{J. Pieprzyk},$^{4,5}$ \\
{M. C. M. Cheung}$^{2}$
\\
$^{1}$CSIRO Space and Astronomy, Australia Telescope National Facility, 26 Dick Perry Ave, Kensington WA 6151, Australia\\
$^{2}$CSIRO Space and Astronomy, Australia Telescope National Facility, PO Box 76, Epping NSW 1710, Australia\\
$^{3}$School of Mathematical and Physical Sciences and Astrophysics and Space Technologies Research Centre, Macquarie University, 2109, NSW, Australia\\
$^{4}$CSIRO Data61, PO Box 76, Epping, NSW 1710, Australia\\
$^{5}$Institute of Computer Science, Polish Academy of Sciences, Poland
}
\begin{abstract}
Pulsars have traditionally been used for research into fundamental physics and astronomy. In this paper, we investigate the expanding applications of radio pulsars in societal and industrial domains beyond their conventional scientific roles. We describe emerging applications in positioning, navigation, timing and synchronization, random number generation, space weather monitoring, public engagement, antenna calibration techniques, and leveraging extensive pulsar data sets generated by large-scale observatories. Such pulsar data sets have already been used to demonstrate quantum-computing algorithms. 

We evaluate the potential for compact radio receiver systems for pulsar detection by describing optimal observing bands. We show that relatively simple and compact receiver systems can detect the brightest pulsar, Vela. The equivalent of an $\sim 4$\,m-diameter dish with a small bandwidth operating around 700\,MHz would be able to detect many more pulsars. Such a detector would be able to localise itself to around 10\,km using pulsar navigation techniques.
 
The space weather community requires direct measurements of the integrated electron density at a range of solar elongations. The only method to get model-independent values is through pulsar observations and we explore the possibility of measuring dispersion measures (DMs) (and rotation measures) with a range of telescopes (observing from low to mid-frequencies) as well as using a typical model to predict the variation of the DM as a function of solar radii. We review how pulsars can be used to produce random sequences and demonstrate that such sequences can be produced using the scintillation properties of pulsars as well as from pulse.

\end{abstract}




\end{frontmatter}



\section{Introduction}

Pulsars are neutron stars that emit beams of electromagnetic radiation, which can be observed as a regular sequence of pulses. The majority of known pulsars were discovered and studied using radio telescopes, but the emission from some pulsars can be detected across the electromagnetic spectrum.

Pulsars are point radio sources at well-known locations in the sky. They emit across the radio spectrum (although typically their emission is weaker at higher frequencies). The pulse emission is usually polarized (with both linear and circular components) and the integrated pulse profiles are stable in time. Around $3000$ pulsars are now cataloged in the ATNF pulsar catalog \citep{mhth05} meaning that multiple pulsars will be above the horizon at any location at any time, and can be observed with a sufficiently large telescope.

Pulsars are intrinsically interesting as astrophysical objects. They represent an end-state of stellar evolution and they allow studies of extremely high densities and magnetic fields. The regular sequence of pulses can also be used to study the interstellar medium, to test theories of gravity \citep{Kramer2021}, and to search for nanohertz-frequency gravitational waves \citep{Xu2023, Reardon2023, EPTA2023, Agazie2023}. These astronomical uses of pulsars have been well described in numerous papers e.g., \cite{man2017} and in books such as \cite{lk12}.

\begin{figure}
\centering
\includegraphics[width=9cm]{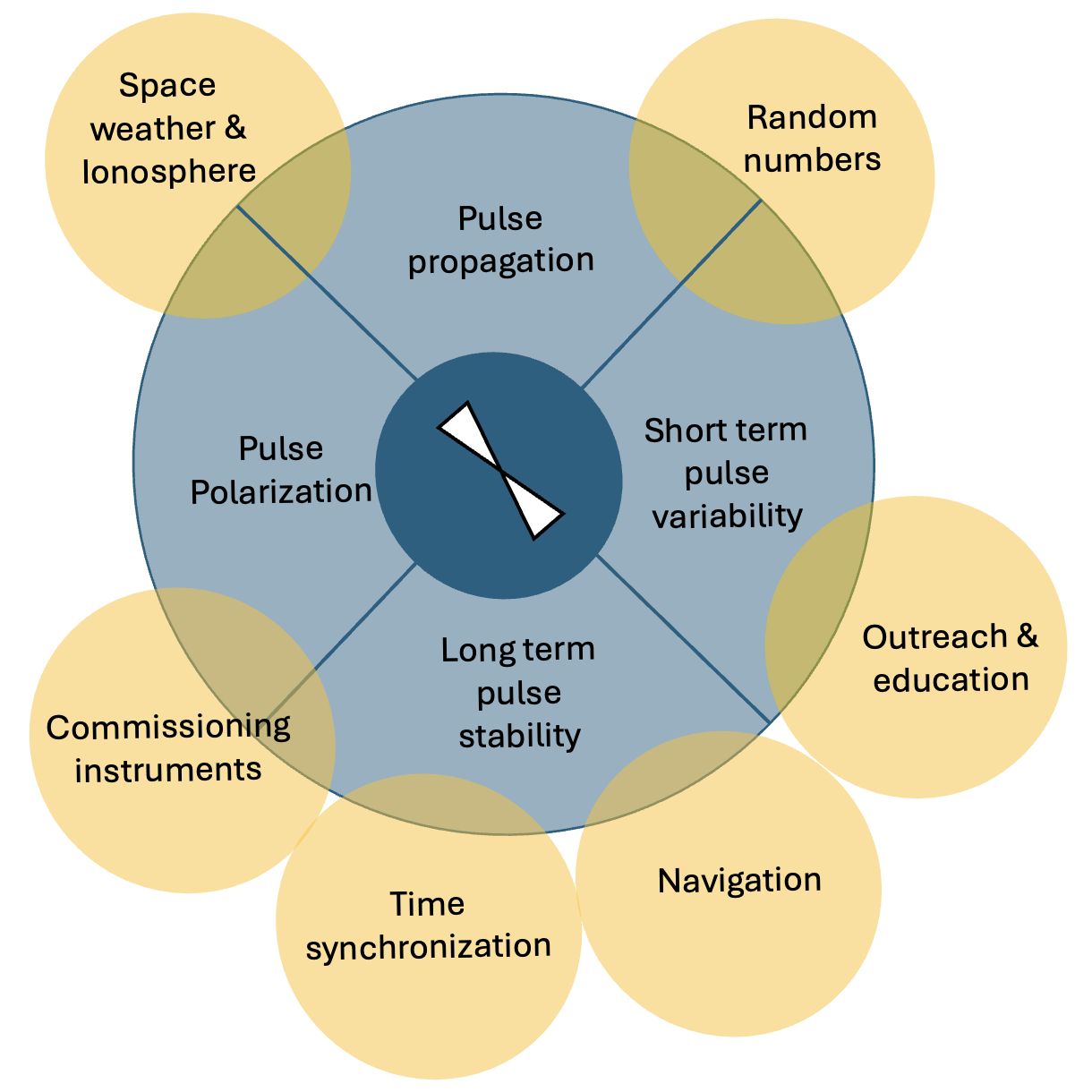}
\caption{A diagram illustrating an overview of pulsar applications as described in this paper where, different pulsar observing properties can be utilized for various applications. Each overlapping circle represents the interconnection across various fields.}
\label{fig:overview}
\end{figure}

In this paper we review the variety of uses of pulsars that have been considered outside of traditional astrophysical research. We identify which pulsars are most suitable for these applications, the requirements for observing systems, and explore the various pulsar properties that can be used. Numerous applications of pulsars have been proposed. The majority of these are based either on the stability (extremely stable rotation) or on the instability (because of the intrinsic emission process itself, and also signal propagation effects) of the pulsar signals. In Figure \ref{fig:overview} we summarize the links between these properties and the corresponding applications.

Applications that make use of a pulsar's stability are usually based on the pulsar timing methodology (e.g., \citealp{hem06}). Although individual pulses show variations in their amplitudes and shapes, the average pulse proﬁle (i.e., the integrated pulse proﬁle), which is obtained by adding several thousands of individual pulses, is typically very stable. They can therefore be used to measure the pulse time of arrival (ToA). These ToAs are essentially the building blocks of pulsar timing. The difference between the observed and predicted arrival times is known as the timing ``residual". For the fastest spinning pulsars (millisecond pulsars) observed with a large telescope, the measurement can have $\sim\,\mu$s precision, and these ToAs can be predicted accurately over many years of observations. Studying how accurately the ToAs can be predicted forms the basis of studies of the solar system and Earth orientation, pulsar-based navigation (Section \ref{positioning and navigation}), time determination, and synchronization (Section \ref{time synchronization}).

The observed pulsar signal is usually polarized and contains information on the properties of the inter-stellar and inter-planetary medium that it has propagated through. The primary propagation effects are pulse dispersion, scintillation, and scattering. Such observations are relevant for studies of space weather and the Earth's ionosphere (Section~\ref{space weather} and \ref{ionosphere}). The polarized nature of the pulsar emission is used for calibrating and commissioning new instrumentation (Section \ref{calibrating instrumentation}). Despite the extreme stability of ensembles of pulses when averaged together, the emission from individual pulses is not perfectly stable due to intrinsic emission processes and signal propagation effects. This is used in applications relating to unpredictability and randomness (Section \ref{randomness}).

Data sets obtained in order to search for pulsars (and other transient astronomical objects) are necessarily stored with high time resolution. Such data sets are often archived (and over 3\,PB of such data spanning over 30\,years of observations are stored in the Parkes data archives; \citealp{hmm+11}). These data sets can be used to study the changing radio frequency interference environment around telescopes and to support the development of methods to mitigate such signals. These data sets have unique properties. They are mostly ``noise-like" (and hence incompressible) and are usually stored with only 1 or 2 bits per sample. They contain rare ``events'' (the properties and times of astronomical discoveries have been published) and also hold the possibility for new discoveries. 

There are, however, challenges when using pulsars for practical applications. Distant radio sources such as pulsars are weak, which means they cannot easily be detected with a receiver system with small collecting area and/or small observing bandwidth. Only a handful of sources are likely detectable with small-scale, compact receiver systems. With existing radio recording systems it is relatively easy to mask the pulsar signal under radio frequency interference. The propagation of the radio signal through the interstellar medium leads to effects such as dispersion, scintillation and scattering that also need to be accounted for when detecting the pulsars. We discuss these challenges along with different applications in respective sections.

\section{Detecting pulsars}
\label{detecting pulsars}

Pulsars are non-trivial to detect. They generally have relatively low flux densities (95\% have pulsed flux densities\footnote{Pulsed flux density refers to the average flux density contained within the pulsed radio emission.} lower than $\sim 5$\,mJy at 1400\,MHz), their signals are dispersed, and the detection system needs to account for their periodic nature. Research observatories now have standard observing modes for pulsar observations and pulsars can be detected routinely from $\sim 50$\,MHz with the Murchison Widefield Array (MWA; \citealp{Tingay2013,Wayth2018}) and Low Frequency Array (LOFAR; \citealp{2013Van}) to around 4\,GHz with telescopes such as Murriyang, CSIRO's Parkes Radio telescope, and for magnetars, up to 8\,GHz with the Deep Space Network (DSN) antennas \citep{pam+19}. While most pulsars exhibit steep spectral energy distributions, leading to lower flux densities at higher frequencies, pulsars have also been detected at millimeter wavelengths $\sim100$--$200$\,GHz with the Atacama Large Millimeter/submillimeter Array (ALMA), and Institut de Radioastronomie Millim\'etrique (IRAM) 30\,m radio telescopes \citep{Liu+2019,Torne+2017}.

For many of the applications described here, it is useful to determine the most compact observing system that could be used to detect a pulsar. Relatively small dishes for e.g., 12\,m-diameter antennas at the Parkes \citep{Sarkissian2017} and Jodrell Bank observatories routinely monitor the brightest pulsars around 1\,GHz. Amateur astronomers have also used software defined radios and small dishes to successfully detect bright pulsars such as the Vela pulsar\footnote{https://www.rtl-sdr.com/detecting-pulsars-rotating-neutron-stars-with-an-rtl-sdr/}.

\begin{figure*}[hbt!]
    \centering
    \includegraphics[width=\textwidth]{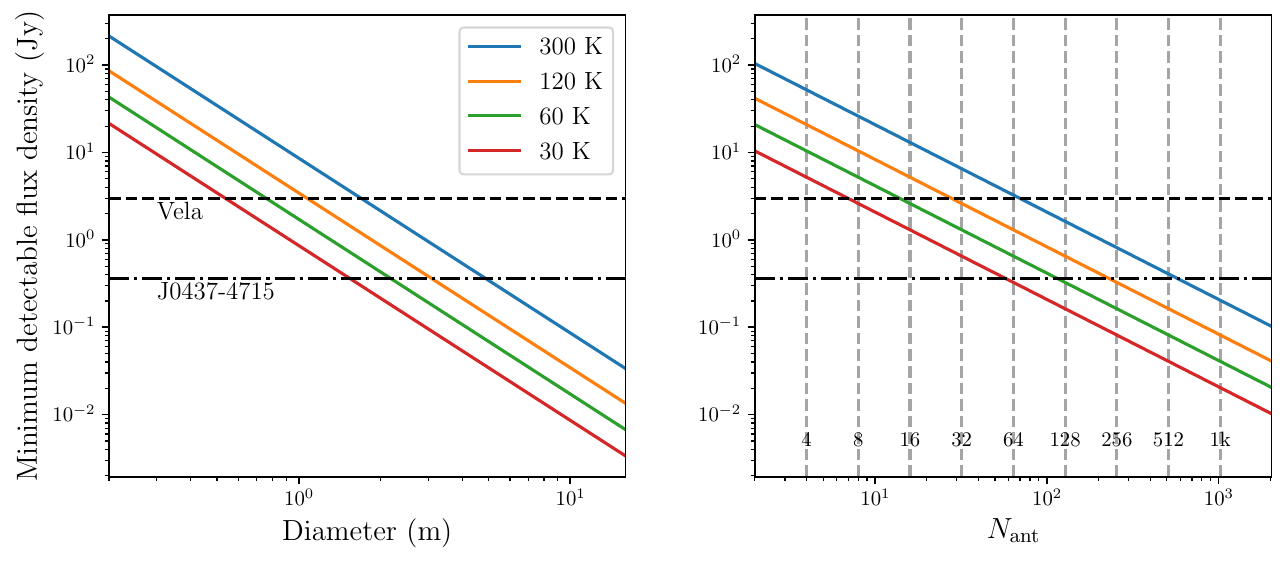}
    \caption{\small{(left) The minimum detectable (i.e. S/N $>8$) flux density as a function of antenna diameter for a system with system temperatures of 300\,K (blue) 120\,K (orange), 60\,K (green), and 30\,K (red), assuming 10\,minutes of integration, a bandwidth of 10\,MHz, a 1\% duty cycle, and two polarizations. The horizontal lines correspond to the 700\,MHz flux densities of PSR~J0437$-$4715 (dash-dotted) and the Vela pulsar (PSR~J0835$-$4510; dashed). (Right) The minimum flux density as a function of the number of elements for an aperture-array telescope at 700\,MHz. For guidance, we have indicated power-of-two numbers of receiving elements with dashed vertical lines.}}
    \label{fig:flux}
\end{figure*}

\begin{figure*}[hbt!]
    \centering
    \includegraphics[width=\textwidth]{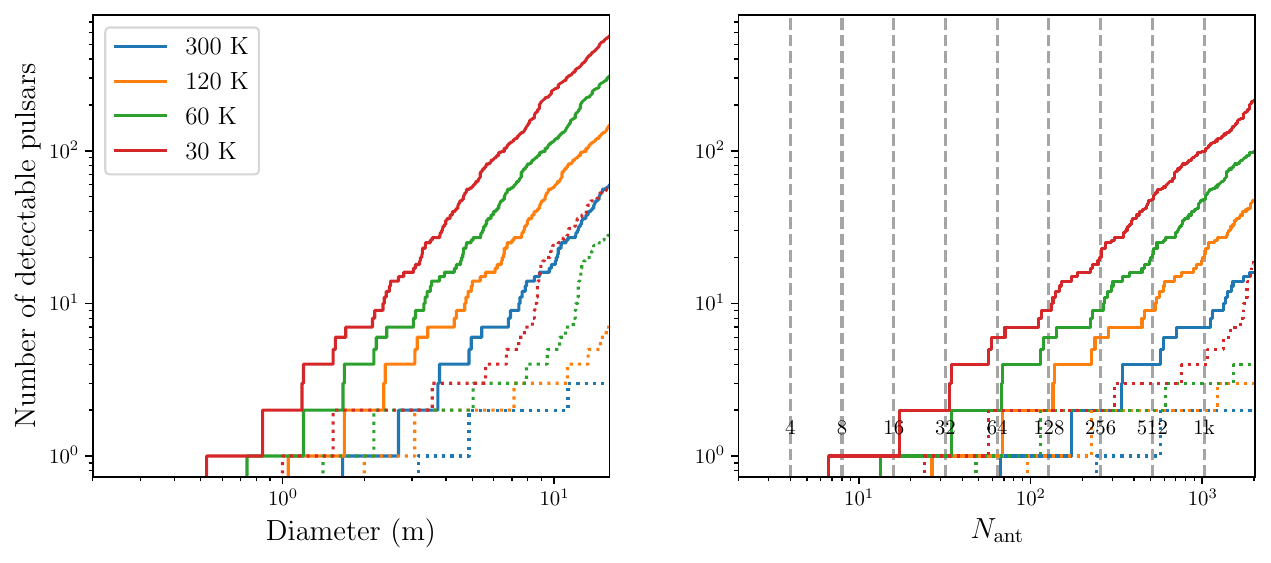}
    \caption{\small{(left) The number of detectable pulsars for a small parabolic antenna (left) and an aperture array system (right) at 700\,MHz, for a range of system temperatures, under the same assumptions as for Figure \ref{fig:flux}. The colours are as in Figure \ref{fig:flux}. The dotted curves indicate the millisecond pulsar population, while the solid curves indicate the overall pulsar population.}}
    \label{fig:ndet_psr}
\end{figure*}

The sensitivity of a radio telescope can be determined through the radiometer equation, which can be used to determine the minimum pulsed flux density $S_{\rm min}$ above a certain $S/N$:
\begin{equation}
S_{\rm min} = \frac{2k_{B}T_{\rm sys}}{A_{\rm eff} \sqrt{n_p\Delta\nu \tau}}\sqrt{\frac{\delta}{1-\delta}}(S/N),
\end{equation}
where $2k_{B}$ is the Boltzmann's constant, $T_{\rm sys}$ is the system temperature, $A_{\rm eff}$ is the effective collecting area, $n_p$ is the number of independent polarisations recorded, $\Delta \nu$ is the observing bandwidth, $\tau$ is the observing duration, $\delta$ is the pulse duty cycle, which is W/P, where W is the pulse width and P is the pulse period, and $S/N$ is the minimum desired signal-to-noise ratio. For example, to detect the four brightest pulsars in the sky at 700\,MHz (setting a limiting flux density of 540\,mJy for an S/N of 8) with a bandwidth of 10\,MHz and a 300\,K system temperature, a 3.8\,m antenna would be required.
 
As pulsars typically have a steep radio spectral energy distribution \citep{Bates2013}, receivers operating at lower frequencies may outperform those operating at higher frequencies. \cite{bwk+07} used an early prototype system of the MWA to detect giant pulses from the Crab pulsar with 6\,MHz of bandwidth operating at 200\,MHz with three tiles each of 16 dipoles. Additionally, Woodchester Observatory, located near Adelaide, Australia, carried out observations of the Vela pulsar (PSR~J0835$-$4510) with a dedicated 2.3-meter dish and confirmed the detection of the glitch in PSR~J0835$-$4510 \citep{ATel2024}. The low-frequency brightness of radio pulsars has been exploited by a small-scale pulsar detector composed of 16 dual-polarisation antennas installed in Gdansk, Poland, which serves as a `Pulsar Clock\footnote{The pulsar clock honours Johann Hevelius, whose grave is just below the clock.}'. The system operates at a frequency of 300\,MHz, beamform and processes 64\,MHz of bandwidth to detect six of the brightest pulsars in visible to the Northern hemisphere. 

The expected sensitivity of the Gdansk `Pulsar Clock' can be estimated using the radiometer equation\footnote{We have been unable to find any literature describing the S/N of the pulse profiles achieved by the system.}. For an aperture-array system composed of coherently-added signals from $N_c$ antennas and incoherently-added signals from $N_i$ antennas, the effective area is:
\begin{equation}
A_{\rm eff} = \left( N_c \sqrt{N_i}\right)\frac{1}{2\pi}\left(\frac{c}{f}\right)^2,
\end{equation}
where $c$ is the speed of light, and $f$ the observing frequency. For 16 coherently combined antennas operating at 300\,MHz we obtain $A_{\rm eff} \approx 2.5$\,m$^2$. Assuming T$_{\rm sys} \sim 100$\,K (T$_{\rm sky}$ must be a significant contributing factor) then we obtain a limiting flux density at 300\,MHz of $\sim 1$\,Jy. This leads to seven pulsars in the catalogue that could potentially be detectable.

In Figure \ref{fig:flux}, we present the minimum detectable flux density (assuming a detection threshold S/N of 8) of a similar system operating at 700\,MHz for a range of system temperatures $T_{\rm sys}$, assuming a modest 10\,minutes of integration and 10\,MHz of bandwidth. In Figure \ref{fig:ndet_psr}, we present the number of detectable pulsars with such a system based on flux densities given in the ATNF pulsar catalog \citep{mhth05}.

\section{Positioning and navigation}
\label{positioning and navigation}

Positioning (the determination of location, velocity, and orientation), and navigation (determining the means to reach a desired position from the current position) are essential for numerous applications. For solar system exploration, spacecraft orbits are modelled using ephemerides that contain the relative positions and motions of the major solar system bodies. The spacecraft position is measured using direct communications to and from the spacecraft from Earth or from other spacecraft, inertial navigational systems and star trackers. Citations to the relevant literature are given in \cite{hpr21}, which also notes the challenges of fully autonomous space navigation. On the Earth, and in low-Earth orbit, the Global Navigation Satellite Systems (GNSS) are used for positioning and navigation. GNSS can be augmented for higher precision in specific regions. However, our reliance on GNSS is problematic: GNSS is vulnerable to jamming, spoofing, hacking, and space weather effects, with limited coverage in certain regions \citep[see][for further discussion]{czly18}. Jamming and spoofing of GNSS signals and likely mitigation strategies are detailed in \cite{srcfr2016}, the impact on maritime traffic is described in \cite{sjs+22}, and \cite{sre16} described the impact of space weather events.

Could observations of pulsars overcome the vulnerabilities of GNSS and provide a Galactic-scale navigation and positioning system? This concept has been explored for several decades \citep[e.g.][]{ssd+72}. Advantages of using pulsars for positioning, navigation, and timing include that they are Galactic objects and hence cannot be directly affected by any human interference (although their signals could potentially be spoofed) as well as providing navigational capability on Galactic-scales. Pulsars also are broadband radio sources and therefore harder to jam than narrow band navigation systems such as GNSS satellites.

A spacecraft system with a pulsar detector can be fully autonomous with knowledge of 3-D spatial coordinates and time. There are many publications and patents relating to the determination of position and local time for a spacecraft using observations of pulsars for example, \citep{dhy+13, poting2018, Mengfan2019, hww+20}. 

For 3-dimensional positioning, if the time is known precisely, observations of only 3 pulsars are required. If not, then four pulsars are required. The concepts have been demonstrated in practice using the Insight-HXMT Satellite \citep{zzl+2019} and the Station Explorer for X-ray Timing and Navigation Technology (SEXTANT), which is a technology demonstration enhancement to the Neutron Star Interior Composition Explorer (NICER) mission \citep{rww2017}. These are based on X-ray detections of pulsars. Beckett (2006\footnote{\url{https://digitalcommons.usu.edu/smallsat/2006/All2006/47}}) argued that X-ray observations were preferable over radio observations on the basis that detection of radio pulsars requires large telescopes that are impractical for navigation; and frequency-dependent delays introduced by the interstellar medium limit the positional accuracy.

These arguments must be revisited, as pulsars can be detected with compact and small-scale radio receivers (with integration times comparable with those used for the X-ray detections described by \citealt{rww2017}). Frequency-dependent propagation effects are readily dealt with in all modern radio pulsar observing systems. For example, dispersion smearing can be accounted for using coherent dedispersion methods; scintillation can be mitigated using wide-bandwidth observations; pulse smearing due to scattering is more challenging, but some methods are available to de-scatter a pulse profile (see, e.g., \citealt{wdv2013}). 

Radio-based observations have some advantages over X-ray observations including: 1) the ability to demonstrate an end-to-end system on Earth, 2) the ability to use such a system for navigation on a planet where the atmosphere will block X-rays (such as the Earth), 3) the large number of pulsars that can potentially be used and 4) the low cost and longevity of radio-frequency infrastructure.

The positional accuracy, in the direction of a given pulsar, depends on the uncertainty on the ToA measurements, $\sigma_{\rm ToA}$, as $c \sigma_{\rm ToA}$ where, $c$ is the speed of light. $\sigma_{\rm ToA}$ depends on the S/N of the observation and the structure of the pulse profile, but an estimate can be obtained assuming a Gaussian pulse shape with full-width at half-maximum $W_{50}$ (with units of time). In this case, 
\begin{equation}
    \sigma_{\rm ToA} \approx \frac{1}{2}\frac{W_{50}}{{\rm S/N}} 
\end{equation} 
Taking a typical width $W_{50} = P/10$ (where $P$ is the pulse period) and a detection threshold S/N of 8, we have $\sigma_{\rm ToA} \approx P / 160$. For a typical pulse period $P = 1$\,s, we have $\sigma_{\rm ToA} \approx 6$\,ms and therefore $\sigma \sim 2000$\,km. Taking instead a ${\rm S/N}=8$ detection of the brightest millisecond pulsar, PSR~J0437$-$4715, with a period of $5.576$\,ms, we have $\sigma \sim 11$\,km. With multiple observations or higher sensitivity it may be possible to achieve positional accuracies better than 1\,km. Space-based X-ray experiments are now achieving a precision of $\sim$10\,km (the goals of the SEXTANS experiment, \citealt{rww2017}, and predicted in papers such as \citealt{dhy+13}).

\begin{figure*}[hbt!]
    \centering
    \includegraphics[width=\textwidth]{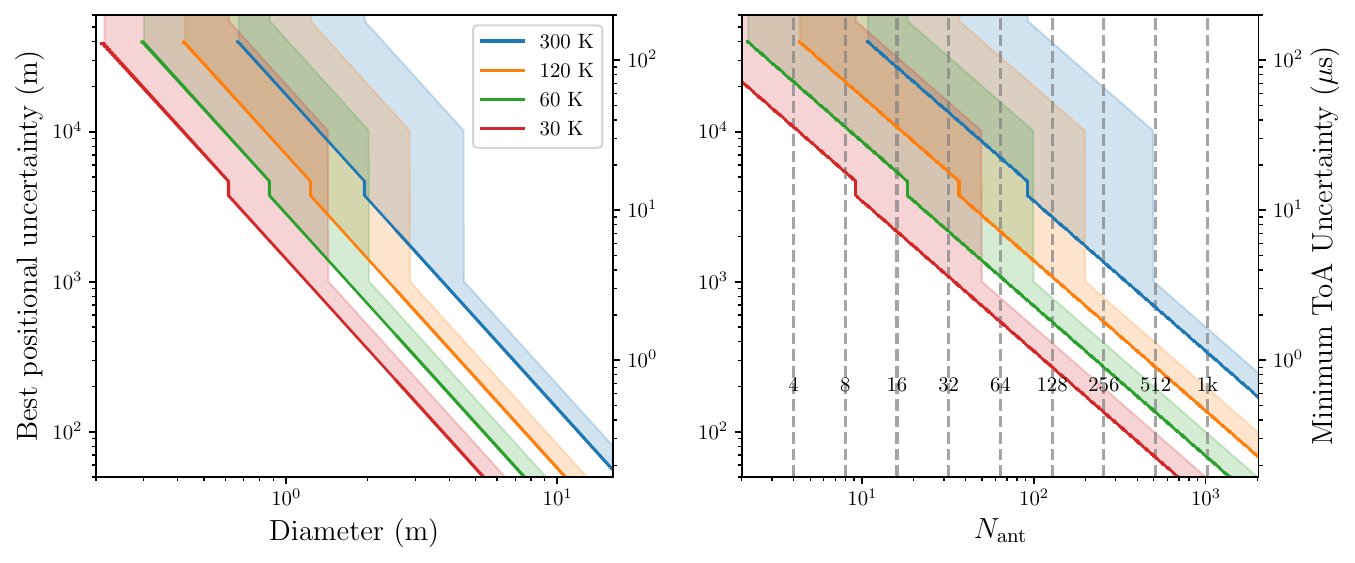}
    \caption{\small{Best-case positional uncertainties achievable with a small-scale radio detector system as a function of detector size, for a range of system temperatures (colours are as in Figure \ref{fig:flux}), using 64\,MHz of bandwidth and 1\,hour of integration, and assuming a 1\% pulse duty cycle. The thick, solid lines show the best 1-dimensional accuracy achievable with any pulsar, while the shaded regions indicate the range of accuracy achievable with the three-best detectable pulsars. Large jumps in the width of this band represent antenna diameters or number of elements where higher-quality pulsars for navigation become detectable.}}
    \label{fig:dx_psr}
\end{figure*}

Even though the majority of the literature is around space-based navigation using X-ray detectors, there have been studies of radio observations of pulsars for terrestrial navigation. Publicly available observations from the Parkes telescope have been used to demonstrate methods to localise Parkes (and, as the position of Parkes is known, to demonstrate the effectiveness of the methods). 
\cite{poting2018} showcased their ability to localize the Parkes telescope with an impressive precision of $\sim$1\,km. They highlighted that the observation duration required to attain a similar level of uncertainty using the full 56-element NICER telescope is approximately equivalent to that of a 4\,m radio antenna. 

For ground-based, terrestrial navigation it is possible to make use of a digital elevation model of the landscape and then search for all possible positions where the antenna may be located. Again using publicly available data from the Parkes telescope and cartographic information, \cite{hww+20} showed that Parkes can be located with a precision of $\sim 100$\,m, but note that this relies on high quality observations with a large-diameter antenna. We note that pulsar parameters are measured relative to the solar system barycentre and hence the positions are determined in the barycentric reference frame. Through knowledge of the Earth's orbit and orientation, such positions can be converted to terrestrial reference frames.

\begin{table}
\centering
\caption{\label{tab:psrnav}Properties of pulsars likely of use in navigation and positioning experiments. The following numbers assume a 256-element array with a system temperature of 60\,K, bandwidth of 64\,MHz, center frequency of 700\,MHz, and integration time of 1\,hour. The final column $\Delta x$ gives the expected one-dimensional positional accuracy achievable with this setup.}

\begin{tabular}{cccccc}
\hline
PSR J & $P0$ & $S_{1400}$ & $S_{700}$ & $\Delta x$\\
      & (s)  & (mJy)      & (mJy)     & (km) \\
\hline
J0437$-$4715 & 0.0057 & 364 & 550 & 0.27 \\
J0835$-$4510 & 0.0893 & 3100 & 5000 & 0.33 \\
J1939$+$2134 & 0.0015 & 68 & 240 & 0.39 \\
J1644$-$4559 & 0.4551 & 300 & 1200 & 3.94 \\
J0332$+$5434 & 0.7145 & 203 & 615* & 4.57 \\
J2145$-$0750 & 0.0161 & 5.5 & 27 & 7.95 \\
J1752$-$2806 & 0.5625 & 48 & 350 & 9.07 \\
\hline
\end{tabular}
\end{table}

The exploration of smaller or innovative radio detectors has garnered attention in recent research. \cite{bhg+14} delved into the feasibility of employing pulsars for aircraft navigation, proposing the integration of radio receivers into aircraft wings. They emphasized the system's potential resilience to jamming, spoofing, extreme space weather, failures, and political interference. Initial analysis suggested a promising positional accuracy ranging between 200\,m and 2\,km, highlighting its potential applicability, particularly for ocean crossings.

Above we have primarily considered positioning, navigation, and timing using a portable system. However, large-aperture radio telescopes commonly observe pulsars as part of long-term monitoring programs.  Over decades of observations it is possible to predict the ToAs for some pulsars to $<$1\,$\mu$s. This precision, over such long durations, allow comparison of different solar system ephemerides and to place constraints on the masses of solar-system objects (e.g., \citealt{gllc19, cgl+18, chm+10}). These measurements require knowledge of the Earth orientation parameters and hence can also be used to constrain any errors in those parameters. This has not been described in depth in the literature. With long-term observations with large antennas it is possible to determine pulsar positions using timing methodology, very long baseline interferometry (VLBI) and to compare those with results from Gaia \citep{lzl+22} and to constrain the rotation that relates the celestial and ecliptic frames (\citealp{mcc13, wch+2017}).

\section{Time and frequency determination and synchronization}
\label{time synchronization}

Pulsars are typically observed at observatories equipped with controlled and stable timing systems. At the Parkes observatory, the timing system is based on a hydrogen maser, synchronized with GPS satellites (for details see \cite{Hobbs+2020}). The data packets are then time-stamped using this observatory time, and these time-stamps propagate through the processing path. Correction files convert the observatory time-stamps into Terrestrial Time (TT), which is based on International Atomic Time (TAI), referred to as TT(TAI). These can then be further converted to the more precise TT standard realized by the International Bureau of Weights and Measures (BIPM), TT(BIPM).

In most pulsar timing experiments, the recorded times (see \cite{wsv+22}) are corrected for the Einstein delay, which accounts for variations between the observatory clock and the Solar System barycenter. The timing residuals formed using a given timing model for a pulsar include variations caused by any errors in the transfer of the observatory clock to TT(BIPM), and any errors in that terrestrial time standard. Unless there is prior knowledge about the properties of the pulsar, then the timing residuals also contain information on the rotation of the pulsar. \cite{avramenko2007}, \cite{hcm12}, and \cite{hgc+20} have used observations of millisecond pulsars spanning decades to disentangle errors in terrestrial time standards from irregularities in the pulsar spin down. However, \cite{hgc+20} concluded that `pulsar-based timescales are unlikely to significantly improve the stability of the best atomic timescales in the next decade but will remain a valuable independent check on them'. \cite{msb+23} demonstrated how the observatory time standard for the MeerKAT telescope could be extracted directly from the pulsar observations. They were able to recover the signal to within 50\,ns for the majority of their 2.5\,yr data span.

The `{\it PULCHRON}' system\footnote{\url{https://www.unoosa.org/documents/pdf/icg/2019/icg14/09.pdf}} generates a real-time timescale. Observations of millisecond pulsars from the European Pulsar Timing Array \citep{dcl+16} are used to provide steering (on a monthly timescale) of a hydrogen maser. This concept is based on the assumption that short-to-medium-term stability is provided by a hydrogen maser, whereas the pulsars provide the long-term stability. They note that years to decades will be needed before the value of the `{\it PULCHRON}' timing system will be realized.

The above studies assume that the pulsar properties need to be determined simultaneously with the search for any errors in the terrestrial time standards. As the pulse period of a pulsar (and its spin-down rate) is intrinsically unknown then any absolute timing offset, or a linear or quadratic drift in the terrestrial time standard, will not be distinguishable from an error in the spin properties of the pulsar. However, it is possible to compare the measured pulse arrival times at multiple observatories in order to transfer time information from one observatory to another or to synchronize the timing systems.

In order to check the absolute timing system at the FAST telescope, simultaneous observations of PSR~J0953$+$0755 were carried out with the FAST and Parkes radio telescopes. Both telescopes were sensitive enough to detect individual pulses from the pulsar. Since these pulse intensities vary significantly, it was possible to cross-correlate the data sets with $\mu$s precision in order to determine the relative timing offset between the observatories. (As described in \cite{dhg+22} these data sets were also used to demonstrate the extraction of random number sequences from pulsars; see Section~\ref{randomness} of this paper.)

If two observatories are able to measure a precise and accurate absolute time, then events at two geographically separated sites can be synchronized within the measurement uncertainties of the pulse timing. \cite{qiu2021pulsar} explored applications related to the power grid, in which existing synchronization systems are `susceptible to temporary or permanent failures due to \ldots factors such as cyber-attack and electromagnetic interferences \ldots'. They note that accurate time is a requirement for grid control and note that signals from the GNSS satellites are currently used. The community is concerned by the possibility of GPS failure through system failures, space weather, jamming and spoofing. They explore the use of pulsars as an independent method and note that large scale observatories could be used to receive the pulsar signal and then transmit the timing signal through Precision Time Protocol technology. 

\cite{hurd1974} and more recently \cite{refId0} describe how time synchronization (at the then 100\,ns level, and now significantly better) can be achieved using VLBI. It is likely that for relative time transfer VLBI-style methods will be superior to using pulsar observations. However, VLBI requires large data volumes to be transferred from one site to the other before the synchronization can occur. In contrast, the variability of pulsar emission (described further below) implies that well separated telescopes could be synchronized to start a process without the requirement for communication between the sites (for instance, sites on Earth and Mars could independently start a process when a specific pulsar emits a giant pulse).

\section{Space weather}
\label{space weather}
Solar activity drives dynamical changes in the space environment around Earth and other bodies in the solar system. Solar drivers of space weather include (1) enhanced emission of extreme ultraviolet radiation, particularly during solar flares, (2) changes in the magnetized solar wind structure (including abrupt changes associated with coronal mass ejections), and (3) the emission of solar energetic particles traveling near the speed of light. These drivers cause space weather effects that manifest in a number of ways to impact and threaten technological infrastructure \citep{Schrijver2015}. For instance, Earthbound coronal mass ejections (CMEs) can cause severe geomagnetic storms. Ionizing radiation from solar flares and geomagnetic storms can dynamically change the ionosphere and thermosphere, impacting satellite drag and GNSS accuracy. 

In-situ measurements of solar wind and CME properties come from probes at large distances from the Sun ($\gtrsim$\,0.3\,AU). However, each spacecraft only provides a single point measurement of the heliosphere at a given time. Therefore, further observational constraints on the solar wind and CMEs are required. In addition to in-situ measurements, existing methods to measure solar wind and CME magnetic fields and densities include radio observations of gyrosynchrotron emission \citep{Carley+2017}, band splitting of solar type II radio bursts \citep{Mahrous2018}, white-light coronagraph imaging \citep{Billings1966}, interplanetary scintillation (IPS; \citealt{Hewish/Dennison1967}), and remote sensing via monitoring of Faraday rotation and dispersion (e.g. \citealt{Kooi2021, Howard2016, You2012}). Faraday rotation measurements are potentially of very high significance, since the magnetic field orientation of an incoming CME has a critical impact on its geoeffectiveness as discussed by \citep{Kooi2022}.

\begin{figure}[hbt!]
    \centering
    \includegraphics[height=0.23\textheight, width=0.49\textwidth]{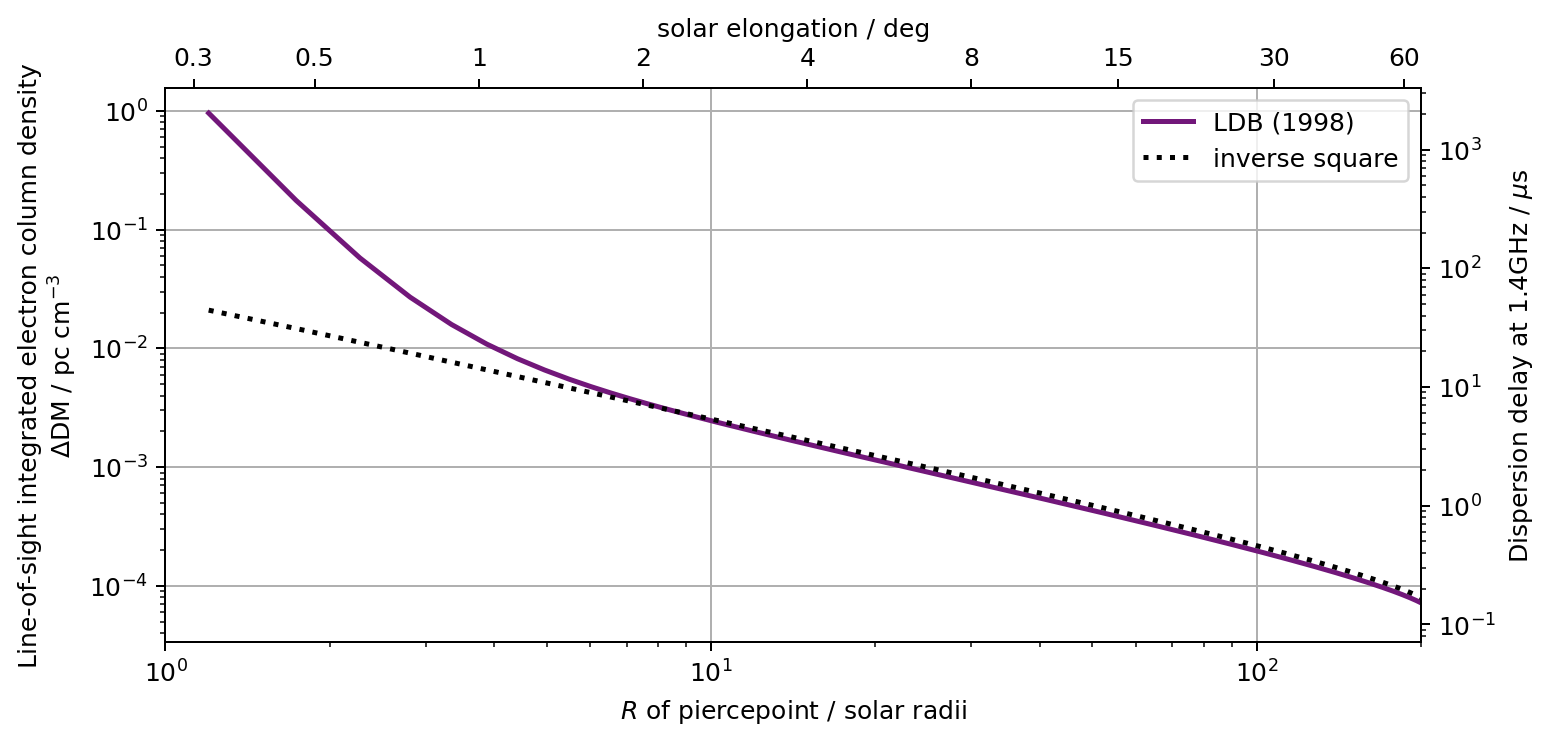}
    \caption{\small{A typical LDB model integrate along the line of sight to produce a delta DM as a function of elongation. The upper x-axis shows the solar elongation for a line of sight which reaches the corresponding `r' at the point of closest approach (the pierce point). The delta DM values are plotted relative to a simple inverse square model represented by the dashed line.
    }}
    \label{fig:solar_wind}
\end{figure}

As the solar wind expands into three-dimensional space, its density is expected to decrease inversely with distance from the Sun. However, this simple model is inaccurate, particularly close to the Sun, due to solar wind acceleration. A range of models for background solar wind density is available in the literature, a broad sample of which are collected by \cite{Kooi2022}. A typical example, that of \cite{Leblanc1998}, is illustrated in Figure~\ref{fig:solar_wind}, along with the simpler inverse square model. 

The background solar wind is also not radially symmetric. There is a very strong dependence on density with solar latitude with polar solar winds being less dense by a factor of $\sim$8 (e.g., \citealt{Tyler1977}). These effects change over the course of the solar cycle, with the slow, dense, equatorial solar wind increasing in coverage at solar maximum, and the fast, sparse, polar solar wind becoming more dominant at solar minimum (e.g., \citealt{Manoharan2012}).

More generally, the fast solar wind is associated with open field lines (coronal holes) close to the Sun's surface. This means that reconstructions of the Sun's magnetic field can be used directly to estimate the density of the solar wind emanating from each point on the Sun. This is the approach used by the popular ``Wang, Sheeley \& Arge'' model (\citealt{Arge2000}, and references therein). A similar, slightly simpler approach, has also been developed for correcting pulsar dispersion measures \citep{You2007}. Finally, solar wind density may change due to transients such as CMEs. These can be expected to cause an increase in column density of a factor of 4 or more \citep{Ontiveros2009}.

Measuring a pulsar's dispersion measure (DM) provides the column density of free electrons along the line-of-sight to the pulsar, and its rotation measure (RM) provides the component of the magnetic field parallel to the line-of-sight. Study of variations in these parameters have been conducted to specifically analyze the solar wind and also as a by-product of long-term pulsar monitoring projects. 

\begin{table*}
\centering
\label{tab2}
\caption{Properties of pulsars likely of use in the space weather experiments}
\begin{tabular}{lllllll}
\hline
PSR J & P0 & DM & S600 & S1400 & RM & $\rm E_{lat}$\\
      & (s) & (\DMunit) & (mJy) & (mJy) & (\RMunit) & (degrees) \\
\hline
J0034$-$0721 & 0.94 & 10.922(6) & 51 & 11 & 9.8(7) & $-$10.15 \\
J0534$+$2200 & 0.033 & 56.771(3) & 211 & 14 & $-$45.4(8) & $-$1.29 \\
J0826$+$2637 & 0.53 & 19.476 & 236.800 & 10 & 5.3(6) & 7.24 \\
J0837$+$0610 & 1.27 & 12.8640(4) & 441.047 & 5.0 & 25.3(7) & $-$11.98 \\
J0922$+$0638 & 0.43 & 27.299 & 200.305 & 10 & 29.(3) & $-$8.33 \\
J0953$+$0755 & 0.25 & 2.96927(8) & 403 & 100 & $-$0.6(4) & $-$4.62 \\
J1136$+$1551 & 1.18 & 4.8407(4) & 612.989 & 20 & 3.9(7) & 12.17 \\
J1720$-$2933 & 0.62 & 42.64(3) & 309.035 & 1.69 & 10 & $-$6.41 \\
J1722$-$3712 & 0.236 & 99.49(3) & 166.852 & 3.8 & 104 & $-$14.01 \\
J1744$-$2335 & 1.68 & 96.66(2) & 220.565 & 0.2 & -  & $-$0.21 \\
J1745$-$3040 & 0.36 & 88.373(4) & 279.608 & 21 & 97.(5) & $-$7.27 \\
J1752$-$2806 & 0.56 & 50.372(8) & 335 & 48 & 96.(2) & $-$4.68 \\
J1824$-$1945 & 0.18 & 224.38(6) & 36 & 7.8 & $-$302.(7) & 3.55 \\
J1900$-$2600 & 0.612 & 37.994(5) & 407.756 & 15 & $-$9.(2) & $-$3.29 \\
J2145$-$0750 & 0.016 & 9.000(1) & 19 & 5.5 & $-$4.40(3) & 5.3 \\
\hline
\end{tabular}
\end{table*}

\begin{figure}[hbt!]
    \centering
    \includegraphics[height=0.25\textheight, width=0.49\textwidth]{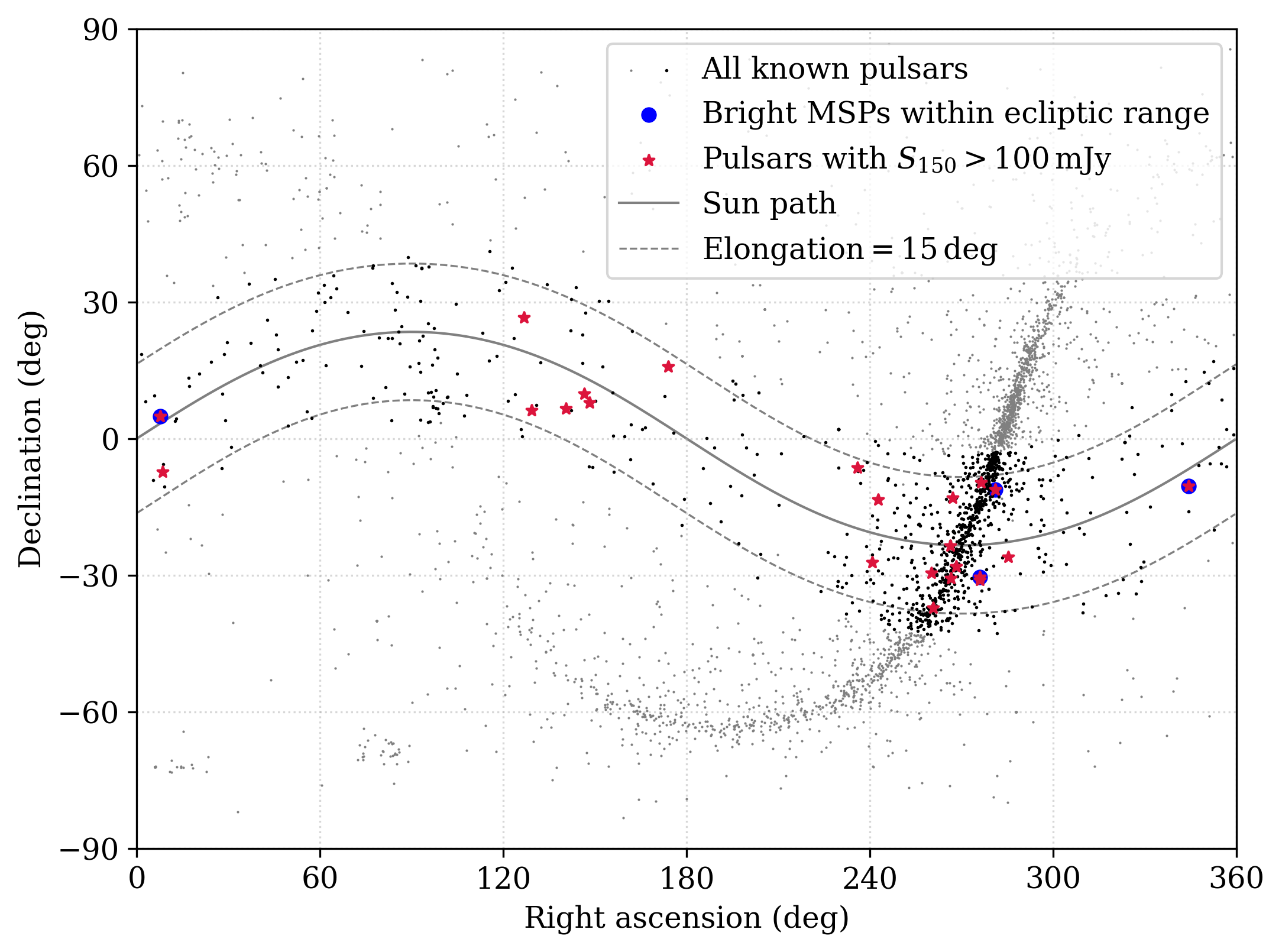}
    \caption{\small{All known pulsars to date are presented by grey dots. The dashed lines present the $\pm$\,15 degree from the ecliptic plane. Solid line presents the path followed by the Sun. Bright pulsars with mean flux density \textgreater 100\,mJy are highlighted in red. Blue circles represent millisecond pulsars within the given ecliptic latitude range.}}
    \label{fig:skymap}
\end{figure}

For example, \cite{mca+2019} showed the impact of the solar wind on their 11-year pulsar timing data. They also noted that a high-cadence monitoring campaign of pulsars near the ecliptic plane would enable the latitudinal structure in the solar wind to be mapped. \cite{tmt+20} used low-frequency observations of giant pulses from a single pulsar, PSR~B0531$+$21, to determine variations in the DM as the line-of-sight to the pulsar went close to the Sun. \cite{Tiburzi2019, Tiburzi2021} also demonstrated the effects of the solar wind on the pulsar timing data. \cite{Howard2016} successfully observed a pulsar (PSR~B0950$+$08) through a CME and managed to estimate an upper bound on the magnetic field strength of the CME. Because the pulsar observations have only probed a very small number of CMEs and for only a few lines-of-sight, the measurements of RM and DM have not been compared in detail with modeling of the CMEs. Instead, the majority of pulsar studies have primarily been used to determine whether existing solar wind models are sufficient for high-precision timing experiments and to demonstrate that DMs and RMs can be measured as the line of sight to a pulsar goes close to the Sun. 

Of the more than 3300 known pulsars, approximately 800 are within 15 degrees of the ecliptic plane, $\sim$\,300 within 5 degrees and $\sim$\,50, which are so close that the line of sight to the pulsar can go through the Sun. In Figure~\ref{fig:skymap} we plot the equatorial coordinates of these pulsars, and highlight the path of the Sun through the galactic plane, bright pulsars with flux densities below 600\,MHz and the pulsars within the 15 degree ecliptic (within dashed gray dashed lines) are also presented. Every year in December, the Sun passes through the galactic plane, close to the galactic centre, where the majority of pulsars are found. Such opportunities have not been explored yet with dedicated radio pulsar observations, hence, such instances can be used to measure DM and RM from a relatively large number of pulsars.

However, the ability to measure pulsar DMs and RMs depends on the observing frequency, bandwidth, S/N of the pulse profile, linear polarization fraction, and the achievable timing precision. Telescopes such as the MWA can measure DMs to a precision in the range $10^{-3}$ to $10^{-6}$\,\DMunit and RMs to 0.09\,\RMunit, whereas higher frequency telescopes such as the Parkes Murriyang telescope which has a wider bandwidth can obtain $10^{-2}$ to $10^{-5}$\,\DMunit\, DM precision for normal (long period) pulsars, $10^{-3}$ to $10^{-5}$\,\DMunit\, for millisecond pulsars, and RMs to 0.01 to 0.1\,\RMunit\, respectively \footnote{Noting that these typical values are not close to the Sun}.

It remains challenging to accurately determine the electron density within CMEs, as it varies within CMEs and is subject to measurement discrepancies due to different observational methodologies. Moreover, as a CME evolves, its electron density at the core and leading edge undergoes significant changes. The location of measurement plays a crucial role, with regions closer to the Sun exhibiting higher densities and those farther away displaying lower densities. Consequently, total mass emerges as a more reliable metric than density. Typical values of the electron density within CMEs vary between different regions. At the leading front, electron densities range from $ \rm 10^{4}\, to\,10^{6}\,cm^{-3}$ within the solar elongation range of 1.3 to 10 solar radii, while at the core, densities span from $\rm 1.4\times10^{6}\,to\,7.0\times10^{8}\,cm^{-3}$ at 1.3 solar radii \citep{Jyoti2023}, and a typical magnetic field strength of $\sim$30\,mG at 10 solar radii \citep{Kooi2021, Oberoi2012}.

If the line of sight to a pulsar went directly through such an example CME then this would correspond to a change in DM of $\rm (1.60\,\pm\,0.06)\,\times\,10^{-3}$\,\DMunit and RM of $\sim$ 0.5,-,0.8\,\RMunit according to the values measured by \cite{Howard2016}. Such changes will be easily measured with Parkes/MWA. For DM estimates we will achieve the required precision easily with Parkes and MWA for millisecond pulsars. In Figure~\ref{fig:skymap} we highlight the millisecond pulsars with blue squares, but note that there are only 200-300. For RM measurements, we need a high S/N pulse profile and significant linear polarization in the pulse profile. From the existed database, for the brightest pulsars in the ecliptic plane Table~\ref{tab2} presents a list of some pulsars that are good targets for such studies. However, others such as PSR~J2145$-$0750, J1909$-$3744 are ideal sources, while pulsars such as PSR~J1939$+$2134, J1740$+$1000, J1829$+$ 2456 and J1918$-$0642 do not have the required amount of linear polarization (below 20$\%$).

As shown in Figure~\ref{fig:skymap} there are not an enormous number of pulsars at any given time whose line of sight is in the vicinity of the Sun, except for the time every year when the Sun passes through the centre of the galactic plane. At other times of the year, as compared to the pulsars, there is a high density of extra-galactic radio sources \citep{Morgan2022}, which can be used to measure RMs \citep{Kooi2022}, and to compare results with simulations. There will never be the same density of pulsars, but the pulsar observations provide a direct simultaneous measurement of the DM and the RM, which is not possible from a continuum extra-Galactic source measurement. The simultaneous measurement of the DM and RM directly provides a model-independent integrated magnetic field strength of the CME event.

Both pulsars and many extra-Galactic sources also scintillate. Interplanetary scintillation can be used as a tool for coronal and heliospheric studies. Distant galaxies provide a high density of sources, but pulsars are uniquely point sources, and hence the scintillation properties are not affected by any source structure.

In the future, more pulsars in the ecliptic plane will be discovered (e.g., \citealt{Bhat2023, Spiewak2020}). Multibeam instruments such as the Parkes cryogenically cooled phased array feed and multibeams with low-frequency telescopes will allow us to observe multiple pulsars simultaneously, allowing multiple lines-of-sight through the solar corona simultaneously. We can already do this for pulsars in globular clusters, but existing telescopes do not have the sensitivity to detect many pulsars without extremely long integration times.

\subsection{The ionosphere}
\label{ionosphere}

Observations of pulsars provide information about the structure and dynamics of the ionosphere, while offering independent measurements that can be compared with predictions from ionospheric models. Deviations between the modeled effects and those observed, such as pulse delays or scintillation patterns, indicate where the model needs to be refined. Iteratively adjusting the model parameters, based on comparisons between observations and model output, the researchers can obtain a better agreement of the ionospheric models. Furthermore, pulsar and other radio sources (e.g., GPS satellites and transmitters) provide ionospheric tomography: by measuring the differential delays between signals from the multiple sources arriving at different locations on Earth's surface, scientists can reconstruct the three-dimensional structure of the ionosphere and map the distribution of its electron density.

The work of \cite{Burgin2024} examined the ionosphere by conducting low-frequency pulsar observations with ground-space VLBI data using the RadioAstron project archive. \cite{Blaszkiewicz2020} studied the influence of ionospheric scintillations on pulsar observations made with the LOFAR low-frequency radio telescope. They were able to show that ionospheric scintillations of pulsars induce time and frequency modulations of the signal, with further consequences such as variations in DM and scintillation. These results are important in showing the relevance of taking into consideration ionospheric effects during pulsar observations and provide a general overview of the pulsar-ionosphere interaction.

In \cite{Zhuravlev2020}, it had been investigated with the scattering of pulses coming from PSR~B0950$+$08, measured by the 10\,m RadioAstron Space Radio Telescope, the 300\,m Arecibo Radio Telescope and the $14\times25$,m Westerbork Synthesis Radio Telescope (WSRT) in a frequency band of 316 to 332\,MHz. Changes in DM, scattering features, and scintillation patterns characterize the propagation effects generated by ionospheric disturbances, e.g., plasma bubbles, ionospheric storms, and irregularities, which can easily be identified and studied in pulsar signals. These insights help to draw valuable inferences about ionospheric dynamics and their interaction with radio wave propagation, communication, and navigation systems. They also provide improved ionospheric models and help develop forecasting techniques associated with space weather.

\section{Randomness}
\label{randomness}

\begin{figure*}[hbt!]
\centering
\includegraphics[angle=-90,width=8cm]{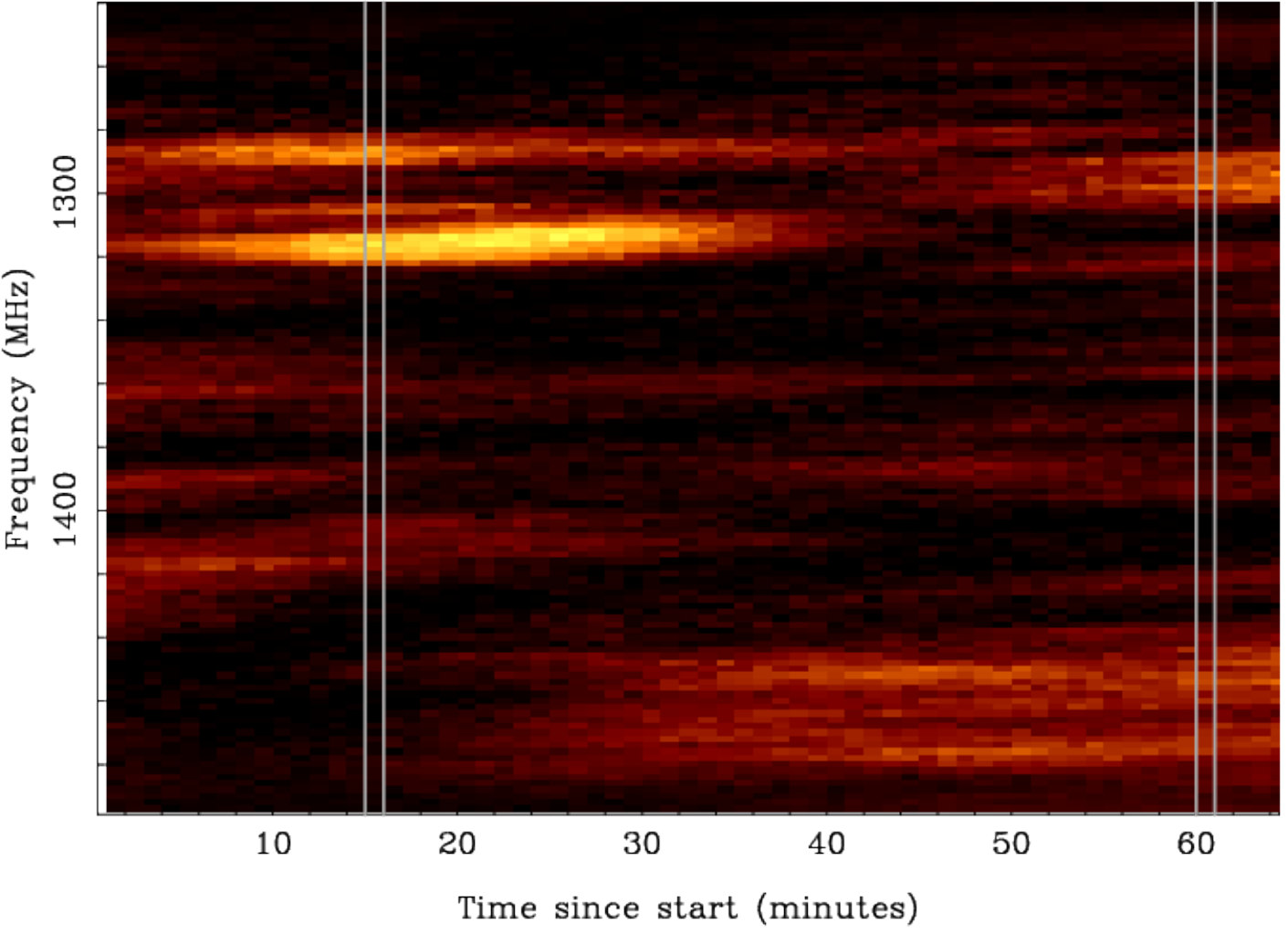}
\includegraphics[angle=-90,width=8cm]{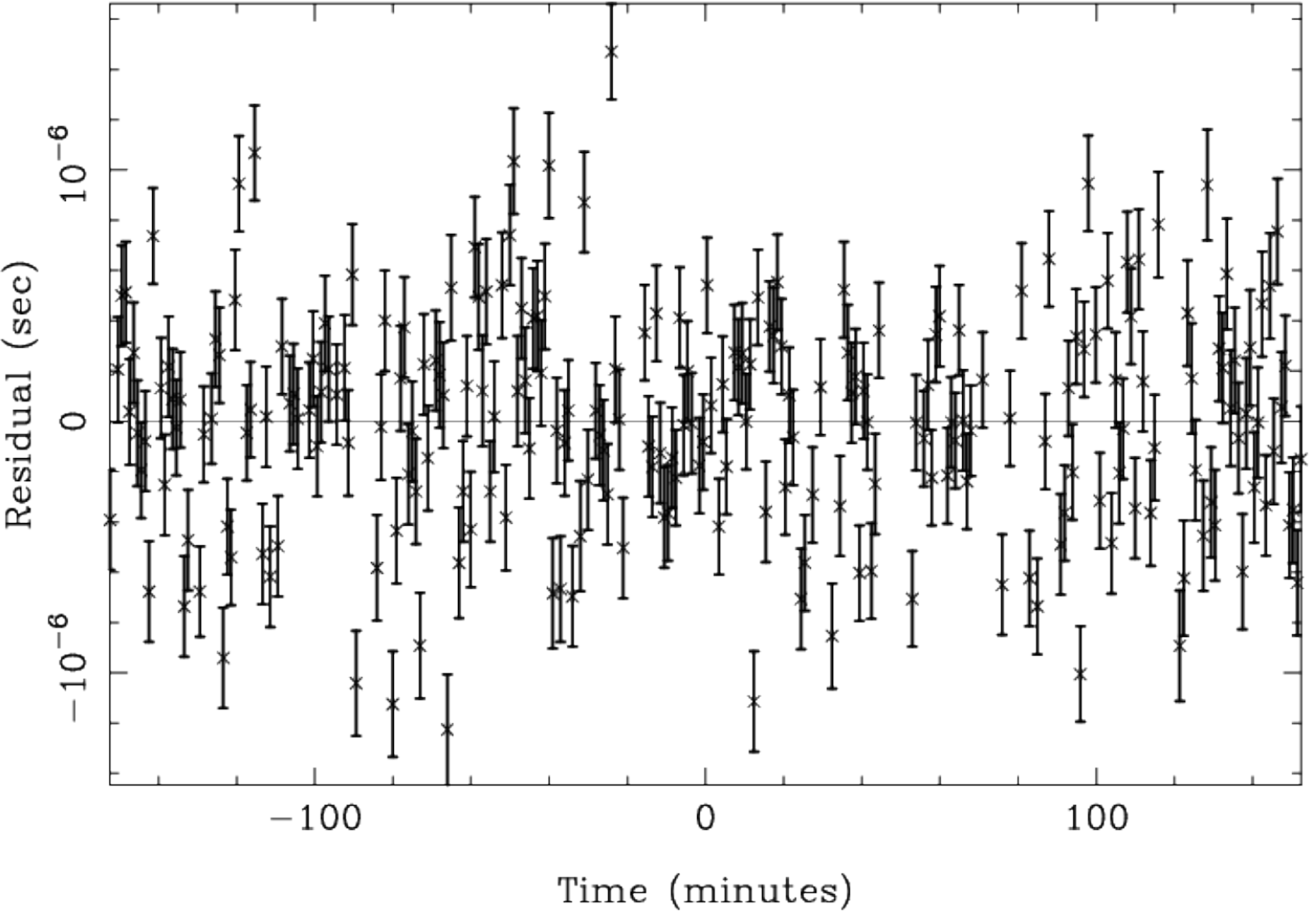}
\caption{(left) Dynamic spectrum of PSR~J1603$-$7202 observed with the Parkes radio telescope on 2018-06-03 as part of the Parkes Pulsar Timing Array project. The vertical lines indicate slices through the dynamic spectrum used for extracting random number sequences. (right) Timing residuals obtained from PSR~J0437$-$4715 on 2010-02-16, which exhibit the jitter phenomenon.}
\label{fg:randomFigure}
\end{figure*}

\begin{table*}
\centering
\caption{Properties of a representative set of pulsars likely of use in producing random number sequences. This includes the brightest pulsars, those exhibiting easily detectable giant pulses, pulsars that glitch regularly, and those with large nulling fractions.}\label{tb:randomnessPulsars}
\begin{tabular}{lllll}
\hline
PSR J & P0 & DM  & S1400 & Reason \\
      & (s) & (cm$^{-3}$pc)  & (mJy) \\
\hline
J0332$+$5434 & 0.714 & 27 & 203 & Single pulse variability \\
J0437$-$4715 & 0.006 & 3 & 150 & Single pulse variability, jitter, scintillation \\
J0534$+$2200 & 0.033 & 56.8 & 14 & Giant pulses, glitch events\\
J0738$-$4042 & 0.375 & 161 & 112 & Single pulse variability \\
J0835$-$4510 & 0.089 & 68 & 1050 & Single pulse variability, Glitch events, self noise\\
J0953$+$0755 & 0.253 & 3 & 100 & Single pulse variability, scintillation and giant pulses \\
J1022$+$1001 & 0.016 & 10 & 4 & Jitter \\
J1049$-$5833 & 2.202 & 447& 1 &  Nulling \\
J1341$-$6220 & 0.193 & 720 & 3 & Glitch events \\
J1456$-$6843 & 0.263 & 9 & 64 & Single pulse variability \\
J1502$-$5653 & 0.536 & 194 & 0.4 &  Nulling \\
J1603$-$7202 & 0.015 & 38 & 3 & Scintillation \\
J1644$-$4559 & 0.455 & 479 & 300 & Single pulse variability \\
J1717$-$4054 & 0.888 & 307 & 1 & Intermittency \\
J1740$-$3015 & 0.607 & 152 & 9 & Glitch events \\
J1935$+$0755 & 0.359 & 159 & 58 & Single pulse variability \\
\hline
\end{tabular}
\end{table*}

The regularity of the pulses detected from pulsars forms the basis of many of the societal uses described in this article so far. However, the intensity of each individual pulse often varies and \cite{dhg+22} showed how measurements of these intensity fluctuations could potentially be used to produce random number sequences. The primary use-cases relate to applications where the same random sequence needs to be obtained simultaneously at multiple, independent geographic locations and/or for scenarios in which it is essential that no human can bias the random values without detection (as no human can influence the emission from a pulsar). 

Other astronomical sources have also been explored for generating random number sequences. For instance, \cite{chapman16} considered random number generation from a methanol maser and an $\rm H_{II}$ region. However, they only used a single telescope and so their method is not directly viable for publicly verifiable random sequences especially as they do not distinguish in their analysis between the radiometer noise from the telescope and the noise from the astronomical signals. \cite{lc2015} considered using the cosmic microwave background radiation (and also mentioned other astronomical sources of radiation, but not pulsars) as a random bit generator for cryptography. Random numbers have also been derived from the direction of incident events. For instance, \cite{pkb05} used incident cosmic rays and a similar analysis could potentially now be carried out using fast radio burst event times or directions. \cite{cps07} considered the use of sunspots in cryptographic applications. 

Pulsars have the advantage over many of these other astronomical sources as (1) there are a large number of sources observable during the day and night, (2) in contrast to e.g., maser observations, they are broadband sources and so difficult to jam or spoof, (3) the data volumes that require to be shared amongst parties can be relatively small, and (4) pulsars can produce random numbers with a millisecond-scale cadence. 

\cite{dhg+22} explored the use of pulsars for generating publicly verifiable randomness (PVR). The expected properties of PVR include public accessibility, unpredictability, non-malleability, public verifiability, and the absence of a need for a trusted server. Pulsars are freely accessible; the only limitation is the geographic location of observers, which may restrict access to certain pulsars. The random sequences extracted from the pulsar signals successfully passed the NIST statistical tests. Non-malleability is guaranteed because no one can influence the behavior of pulsar emissions, which originate millions of light-years away. Public verifiability is achieved when parties, synchronized in time and observing the same pulsar, apply the same algorithm for randomness extraction, resulting in identical outputs. There is clearly no need for a trusted server to access or interact with pulsars. 

PVR has numerous applications, including cryptography, scientific simulations, electoral audits, and monitoring international treaties. As reported in \cite{dhg+22}, this concept has been considered in real-world scenarios, such as generating PVR sequences for "enforcing Soviet and United States strategic arms limitation agreements in the late 1970s."

\cite{Rocha2022} explored how pulsar observations from geographically-separated telescopes could be used for establishing a common cryptographic key over huge distances as well as methods that reduce the bit error rate between the two observatories. The authors consider parties separated by thousands of kilometres on Earth, but we note that these ideas could also be applied to much larger, Solar-system scales.

Challenges in using pulsars for such applications include -- unlike the averaged pulse profiles discussed in the previous sections of this work -- individual pulses require relatively large (and highly sensitive) telescopes to detect, and that properties such as pulse intensities in adjacent pulses can be correlated. Accounting for such correlations will reduce the bit rate of the output data stream. Most methods to extract individual random bits from pulsars (or other natural sources) do not lead to uniform random values. For instance, the flux densities of individual pulse intensities can be log-normal-distributed. Post-processing steps are therefore required to convert the pulsar datasets into data streams useful for most randomness applications.

The fastest rotating pulsars would produce around 1000 values per second, with typical pulsars providing around 1 value per second, if a single random value is extracted from each pulse. Confirming the random sequence with another telescope requires that the telescopes are synchronized in time to a precision better than the pulse period. Previous research into obtaining random sequences from pulsars has assumed large observatories that can detect individual pulses from the brightest pulsars with high S/N. We can estimate the minimum size of a dish antenna that could be used by considering the brightest pulsar, Vela, which has an average flux density of $\sim$1\,Jy in the 20\,cm observing band and a pulse width of $\sim 2$\,ms. The typical peak flux density for a single pulse is $\sim 50$\,Jy. A small antenna is likely to be fitted with a receiving system that can produce around 64\,MHz of bandwidth and may have a system temperature as high as 300\,K. With such a system a 10\,m-diameter telescope would be needed to detect individual pulses from the Vela pulsar with reasonable S/N.

\cite{dhg+22} concentrated on the flux density variations of individual pulses, but other properties of pulsars could be used for randomness studies including (1) pulse jitter \citep{pbv21}, (2) scintillation (e.g., \cite{lv+22} and references therein) or (3) the time between (or of) glitch, nulling, moding or giant pulse events. Glitch events refer to sudden spin-up events whose event times can follow a Poisson process \citep{fmh17}. Nulling refers to the emission switching off and one or more pulses being missing. The randomness of null events was explored in \cite{rsr09}, where they noted that the nulling events in only some pulsars are random. Moding refers to a sudden change in the shape of the time-averaged pulse profile. Giant pulses refer to the phenomenon whereby some pulsars exhibit individual pulses that exceed the average flux density by a factor of ten or more (see, for example, \cite{kb22}). The ToA distribution of giant pulses was found by \cite{gsa21} to follow a Poisson distribution on most time scales.
Whereas any event originating from the pulsar itself (e.g., a giant pulse event) would be detectable by using any sufficiently-sensitive telescope observing the pulsar, ISM-related events (such as \#2 above) would have a spatial scale of 1000s of km \citep{nar92} and hence, using these properties of a pulsar, widely-separated observatories would measure different random sequences, whereas closer-spaced observatories would obtain the same. 

In Figure~\ref{fg:randomFigure} we show two examples how random number sequences could potentially be formed based on \#1 and \#2 above, but the verification does not require exact time or frequency synchronization. In the left-hand panel we show a dynamic spectrum (the pulsar signal strength as a function of observing frequency and time) for PSR~J1603$-$7202 as a representation of \#2 above. A slice in the frequency direction for a given time would produce a random sequence correlated on the diffractive bandwidth scale. If we simply ask whether every 10th frequency channel (approximately representing the diffractive bandwidth) is lower (a zero) or higher (a one) than the previous one we obtain a sequence of random numbers. If the verifying telescope was not perfectly time-synchronized and observed a few minutes later then we recover an almost identical random sequence. An independent random sequence would be obtained after a delay corresponding to the diffractive timescale\footnote{In Figure~\ref{fg:randomFigure} using this method on the left-most slice gives $100100101000$, the next slice gives the almost identical $100101101000$, whereas the rightmost slices give 110011000101 and $110010100101$ respectively. Follow-up work is required to demonstrate whether such sequences would pass standardized randomness tests for randomness and how to reduce the bit-error-rate further.}.

In the right-hand panel of Figure~\ref{fg:randomFigure} we show the timing residuals observed for single pulses from PSR~J0437$-$4715 centred on 1369\,MHz (in black). Simply obtaining the median and determining a sequence from values above the median (a one) or below (a zero) produces a bit sequence\footnote{In this case if we choose values where the residuals deviate by more than 4$\sigma$ we obtain the sequence $01100001111000101$}. The jitter phenomenon is broad-band \citep{pbv21} and so a telescope operating in a near, but non-identical observing band would be able to verify the sequence.

There are therefore a range of methods to extract randomness from pulsars. The choice of method would depend on the sensitivity of the telescopes being used and the bit rate required. For verifiable randomness the algorithm will depend on the geographic distance between the telescopes and how well synchronized in time and frequency they are. A list of pulsars that potentially could be used in producing PVR sequences are listed in Table~\ref{tb:randomnessPulsars}. For a moderate-sized telescope only the very brightest pulsars can be detected through their individual pulses and hence only pulsars with an average flux density of $>$50\,mJy at 20\,cm are included\footnote{This cutoff is arbitrary, sensitive radio telescopes will be able to detect individual pulses from pulsars that are weaker than this cutoff value.}.

\section{Making use of pulsar data sets and algorithm development}
\label{algorithm development}

Searches for new pulsars with radio telescopes have led to massive (PB-scale) data sets that are primarily a low bit-quantization of noise along with rare events of interest. Individual pulses from pulsars in such data sets often have a S/N$<1$, making them detectable only after a search for their periodic signal. Searching for pulsars in these datasets typically relies on Fourier transforms or a fast-folding algorithm (FFA; \citealt{staelin69}). The FFA can detect `weak, noisy pulse trains of unknown period and phase', which has applications outside of pulsar astronomy. In particular \cite{ph18} described how an FFA can be used to detect and identify a beacon acting as a `license plate with a unique identification number' on a spacecraft. \cite{stanciu11} noted that determining the period of noisy discrete events (for instance, using an FFA) is relevant in numerous aspects of digital communication research.

Large datasets containing rare events of interest have attracted machine learning communities because they offer large-scale, labelled data that are unusual (e.g., low-bit quantized) yet retain discovery potential. A selection of pulsar candidates from the Parkes High Time Resolution Universe (HTRU) survey were labelled and made available as the HTRU-2 pulsar data set\footnote{\url{https://www.kaggle.com/datasets/charitarth/pulsar-dataset-htru2}} with the primary purpose of training and comparing machine learning algorithms. One of the most exciting uses of this data set was by \cite{kas21} who used the data set to train a quantum machine learning algorithm. The small number of parameters for each candidate (only eight) makes this dataset ideal for studying the effectiveness of quantum computers. This pioneering work is being further developed, for instance by \cite{slp23} and the concepts are now being developed into more areas of radio astronomy \citep{btc+23}.

\section{Calibrating science observations and commissioning new instrumentation}
\label{calibrating instrumentation}
The averaged pulse profile for most pulsars is stable in terms of pulse shape and polarized, exhibiting both linear and circular components. Pulsars are also point sources in which we record both the pulsed emission and the off-pulse signal. This provides an ideal tool for calibrating Epoch of Reionisation (EoR) fields and radio telescope systems in terms of beam shape, polarization, and timing.

\cite{Paciga2011} demonstrated an experiment of removing EoR foreground using pulsar observations. The experiment utilized PSR~B0823$+$26 as a calibrator due to its stable flux and short period compared to ionospheric fluctuations. By comparing the pulsar signal received by each antenna, the relative system gain of each antenna was calibrated in real-time for phase and polarization.

\cite{straten2013} described how observations of a stable, and well-parameterized, pulsar (such as PSR~J0437$-$4715) can be used to calibrate antenna polarization and \cite{Hobbs+2020} demonstrated how the beam pattern can be determined through scanning the beam over the Vela pulsar. These concepts have not, to our knowledge, been applied more widely to antenna calibration methods. As mentioned earlier, pulsars are highly stable cosmic clocks, emitting radio pulses at extremely regular intervals. By observing pulsars with known positions and precisely measuring the time delay of their signals as they arrive at different telescopes in the array, astronomers can accurately calibrate the timing and phase of the observations. Using pulsars as phase calibrators enables the KKO CHIME/FRB outrigger to achieve precise synchronization between different telescopes in the VLBI array \citep{Lanman2024}. This synchronization is essential for combining the signals from multiple telescopes to form high-resolution images of celestial sources and to study transient phenomena such as fast radio bursts with unprecedented detail.

\section{Outreach and education}
\label{outreach}

Pulsar observations have educational, outreach and artistic value. \cite{craft1970} showed a single pulse stack for PSR~J1921$+$2153 (B1919$+$21) in his PhD thesis to study the shape, intensity and drifting of the pulses (as linked with the randomness studies described in Section~\ref{randomness}). This image captured the attention of a member of the \textsc{Joy Division} rock band, becoming iconic as their album cover\footnote{\url{https://archive.is/IGJjF}} and subsequently included on numerous fashion accessories. Similar data sets have been converted into sound\footnote{The standard method is to modulate white noise by the pulsar's de-dispersed signal intensity, but there are numerous ways for the sonification of a pulsar data stream.}. Such sound files are often used in outreach and education presentations to showcase pulsar's rotational speed, stability, and variability in pulse emission. Such recordings (in particular for PSR~B0359$+$54) are used as part of the composition, \emph{Le Noir De L'Etoile}, by G. Grisey for six percussionists and the pulsar.  

With a sensitive antenna, pulsars can be detected with short integration times and hence provide the means for real-time student observing sessions. Pulsar data sets also provide a large discovery space allowing members of the public (or high school students) to make actual scientific discoveries. \cite{fhr+14} reviewed 22 student (in the classroom) research projects of which three are directly related to pulsar observing. The projects describe include the Arecibo Remote Command Center (ARCC; \citealt{mjzs09}) in which students were able to discover pulsars using archival data sets, the Pulsar Search Collaboratory (PSC; \citealt{rsm+13}) again relating to the discovery of new pulsars and PULSE@Parkes \citep{hhc+09} in which high school students use the Parkes telescope to monitor known pulsars. Their paper did not explicitly describe the pros and cons of using pulsars for outreach. However, it is clear that the challenges relate to pulsars not being directly recognizable to students (in contrast, for instance to optical images of the planets) and require more background knowledge than students typically have. However, the pros include the ability to observe during the day, to make significant discoveries and to observe astronomical variability on timescales of seconds to minutes.

Over 500,000 individuals have been introduced to pulsars through the Einstein at Home project\footnote{\url{https://einsteinathome.org/science/brp}} in which a screensaver is used to use a large number of individual computers in order to search for pulsars in datasets from the MeerKAT, Fermi and Arecibo telescopes. The majority of the outreach/education activities with pulsars have come from observing with large radio telescopes, using public data sets or through data products. As described in Section~\ref{detecting pulsars} it is challenging to detect pulsars with small-scale antennas and so currently remains inaccessible to small-scale projects.

\section{Summary and Conclusions}
The unique characteristics of pulsars render them invaluable tools in both fundamental astrophysical inquiries and practical applications such as navigation and timekeeping. Our discussion has only provided a glimpse into the diverse applications of pulsars and their distinctive phenomena. Nonetheless, their remarkable and sometimes unexpected attributes are evident. Pulsars facilitate the development of resilient positioning and navigation systems, mitigating the vulnerabilities of existing systems like GNSS. They also enable the creation of highly precise and stable timing mechanisms, provide innovative tools for studying space weather and calibrating radio telescopes, and serve as potential sources of genuine random numbers. These examples merely scratch the surface of the extensive applicability of pulsars and their counterparts.

Looking towards the future, prospects for pulsar research and applications appear promising. Advancements in observational methodologies, novel algorithms, and sophisticated technologies promise new ways to detect and study pulsars while harnessing their signals for diverse applications. Large-scale observatories such as the Square Kilometre Array (SKA) hold the potential to revolutionize pulsar astronomy, ushering in a new era of discovery.

In conclusion, pulsars serve as beacons of exploration and innovation, illuminating pathways to deeper understanding of the Universe and offering practical solutions to contemporary challenges. The future of pulsar research is bright, with opportunities to uncover the mysteries of both known and newly discovered pulsars. Through continued research, discovery, and outreach efforts, the pulsar narrative will continue to captivate our imaginations, spur innovation, and expand the boundaries of human knowledge.

\section*{Acknowledgments}
Murriyang, the Parkes radio telescope, is part of the Australia Telescope National Facility (https://ror.org/05qajvd42), funded by the Australian Government for operation as a National Facility managed by CSIRO. This paper includes archived data obtained through the Parkes Pulsar Data archive on the CSIRO Data Access Portal (http://data.csiro.au).

\section*{Data Availability}

The data used in the left-hand panel of Figure~\ref{fg:randomFigure} are available from the CSIRO DAP (\url{https://data.csiro.au/domain/atnf}) by searching for the filename t180603\_152524.rf and frequency averaging by a factor of 8. Similarly the data for the right-hand panel can be obtained by searching for the filename s100216\_053518.rf.

\bibliographystyle{elsarticle-harv} 
\bibliography{ver1}

\begin{thebibliography}{104}
\expandafter\ifx\csname natexlab\endcsname\relax\def\natexlab#1{#1}\fi
\providecommand{\url}[1]{\texttt{#1}}
\providecommand{\href}[2]{#2}
\providecommand{\path}[1]{#1}
\providecommand{\DOIprefix}{doi:}
\providecommand{\ArXivprefix}{arXiv:}
\providecommand{\URLprefix}{URL: }
\providecommand{\Pubmedprefix}{pmid:}
\providecommand{\doi}[1]{\href{http://dx.doi.org/#1}{\path{#1}}}
\providecommand{\Pubmed}[1]{\href{pmid:#1}{\path{#1}}}
\providecommand{\bibinfo}[2]{#2}
\ifx\xfnm\relax \def\xfnm[#1]{\unskip,\space#1}\fi
\bibitem[{{Agazie} et~al.(2023){Agazie}, {Anumarlapudi}, {Archibald}, {Arzoumanian}, {Baker}, {B{\'e}csy}, {Blecha}, {Brazier}, {Brook}, {Burke-Spolaor}, {Burnette}, {Case}, {Charisi}, {Chatterjee}, {Chatziioannou}, {Cheeseboro}, {Chen}, {Cohen}, {Cordes}, {Cornish}, {Crawford}, {Cromartie}, {Crowter}, {Cutler}, {Decesar}, {Degan}, {Demorest}, {Deng}, {Dolch}, {Drachler}, {Ellis}, {Ferrara}, {Fiore}, {Fonseca}, {Freedman}, {Garver-Daniels}, {Gentile}, {Gersbach}, {Glaser}, {Good}, {G{\"u}ltekin}, {Hazboun}, {Hourihane}, {Islo}, {Jennings}, {Johnson}, {Jones}, {Kaiser}, {Kaplan}, {Kelley}, {Kerr}, {Key}, {Klein}, {Laal}, {Lam}, {Lamb}, {Lazio}, {Lewandowska}, {Littenberg}, {Liu}, {Lommen}, {Lorimer}, {Luo}, {Lynch}, {Ma}, {Madison}, {Mattson}, {McEwen}, {McKee}, {McLaughlin}, {McMann}, {Meyers}, {Meyers}, {Mingarelli}, {Mitridate}, {Natarajan}, {Ng}, {Nice}, {Ocker}, {Olum}, {Pennucci}, {Perera}, {Petrov}, {Pol}, {Radovan}, {Ransom}, {Ray}, {Romano}, {Sardesai}, {Schmiedekamp}, {Schmiedekamp}, {Schmitz},
  {Schult}, {Shapiro-Albert}, {Siemens}, {Simon}, {Siwek}, {Stairs}, {Stinebring}, {Stovall}, {Sun}, {Susobhanan}, {Swiggum}, {Taylor}, {Taylor}, {Turner}, {Unal}, {Vallisneri}, {van Haasteren}, {Vigeland}, {Wahl}, {Wang}, {Witt}, {Young} and {Nanograv Collaboration}}]{Agazie2023}
\bibinfo{author}{{Agazie}, G.}, \bibinfo{author}{{Anumarlapudi}, A.}, \bibinfo{author}{{Archibald}, A.M.}, \bibinfo{author}{{Arzoumanian}, Z.}, \bibinfo{author}{{Baker}, P.T.}, \bibinfo{author}{{B{\'e}csy}, B.}, \bibinfo{author}{{Blecha}, L.}, \bibinfo{author}{{Brazier}, A.}, \bibinfo{author}{{Brook}, P.R.}, \bibinfo{author}{{Burke-Spolaor}, S.}, \bibinfo{author}{{Burnette}, R.}, \bibinfo{author}{{Case}, R.}, \bibinfo{author}{{Charisi}, M.}, \bibinfo{author}{{Chatterjee}, S.}, \bibinfo{author}{{Chatziioannou}, K.}, \bibinfo{author}{{Cheeseboro}, B.D.}, \bibinfo{author}{{Chen}, S.}, \bibinfo{author}{{Cohen}, T.}, \bibinfo{author}{{Cordes}, J.M.}, \bibinfo{author}{{Cornish}, N.J.}, \bibinfo{author}{{Crawford}, F.}, \bibinfo{author}{{Cromartie}, H.T.}, \bibinfo{author}{{Crowter}, K.}, \bibinfo{author}{{Cutler}, C.J.}, \bibinfo{author}{{Decesar}, M.E.}, \bibinfo{author}{{Degan}, D.}, \bibinfo{author}{{Demorest}, P.B.}, \bibinfo{author}{{Deng}, H.}, \bibinfo{author}{{Dolch}, T.}, \bibinfo{author}{{Drachler}, B.},
  \bibinfo{author}{{Ellis}, J.A.}, \bibinfo{author}{{Ferrara}, E.C.}, \bibinfo{author}{{Fiore}, W.}, \bibinfo{author}{{Fonseca}, E.}, \bibinfo{author}{{Freedman}, G.E.}, \bibinfo{author}{{Garver-Daniels}, N.}, \bibinfo{author}{{Gentile}, P.A.}, \bibinfo{author}{{Gersbach}, K.A.}, \bibinfo{author}{{Glaser}, J.}, \bibinfo{author}{{Good}, D.C.}, \bibinfo{author}{{G{\"u}ltekin}, K.}, \bibinfo{author}{{Hazboun}, J.S.}, \bibinfo{author}{{Hourihane}, S.}, \bibinfo{author}{{Islo}, K.}, \bibinfo{author}{{Jennings}, R.J.}, \bibinfo{author}{{Johnson}, A.D.}, \bibinfo{author}{{Jones}, M.L.}, \bibinfo{author}{{Kaiser}, A.R.}, \bibinfo{author}{{Kaplan}, D.L.}, \bibinfo{author}{{Kelley}, L.Z.}, \bibinfo{author}{{Kerr}, M.}, \bibinfo{author}{{Key}, J.S.}, \bibinfo{author}{{Klein}, T.C.}, \bibinfo{author}{{Laal}, N.}, \bibinfo{author}{{Lam}, M.T.}, \bibinfo{author}{{Lamb}, W.G.}, \bibinfo{author}{{Lazio}, T.J.W.}, \bibinfo{author}{{Lewandowska}, N.}, \bibinfo{author}{{Littenberg}, T.B.}, \bibinfo{author}{{Liu}, T.},
  \bibinfo{author}{{Lommen}, A.}, \bibinfo{author}{{Lorimer}, D.R.}, \bibinfo{author}{{Luo}, J.}, \bibinfo{author}{{Lynch}, R.S.}, \bibinfo{author}{{Ma}, C.P.}, \bibinfo{author}{{Madison}, D.R.}, \bibinfo{author}{{Mattson}, M.A.}, \bibinfo{author}{{McEwen}, A.}, \bibinfo{author}{{McKee}, J.W.}, \bibinfo{author}{{McLaughlin}, M.A.}, \bibinfo{author}{{McMann}, N.}, \bibinfo{author}{{Meyers}, B.W.}, \bibinfo{author}{{Meyers}, P.M.}, \bibinfo{author}{{Mingarelli}, C.M.F.}, \bibinfo{author}{{Mitridate}, A.}, \bibinfo{author}{{Natarajan}, P.}, \bibinfo{author}{{Ng}, C.}, \bibinfo{author}{{Nice}, D.J.}, \bibinfo{author}{{Ocker}, S.K.}, \bibinfo{author}{{Olum}, K.D.}, \bibinfo{author}{{Pennucci}, T.T.}, \bibinfo{author}{{Perera}, B.B.P.}, \bibinfo{author}{{Petrov}, P.}, \bibinfo{author}{{Pol}, N.S.}, \bibinfo{author}{{Radovan}, H.A.}, \bibinfo{author}{{Ransom}, S.M.}, \bibinfo{author}{{Ray}, P.S.}, \bibinfo{author}{{Romano}, J.D.}, \bibinfo{author}{{Sardesai}, S.C.}, \bibinfo{author}{{Schmiedekamp}, A.},
  \bibinfo{author}{{Schmiedekamp}, C.}, \bibinfo{author}{{Schmitz}, K.}, \bibinfo{author}{{Schult}, L.}, \bibinfo{author}{{Shapiro-Albert}, B.J.}, \bibinfo{author}{{Siemens}, X.}, \bibinfo{author}{{Simon}, J.}, \bibinfo{author}{{Siwek}, M.S.}, \bibinfo{author}{{Stairs}, I.H.}, \bibinfo{author}{{Stinebring}, D.R.}, \bibinfo{author}{{Stovall}, K.}, \bibinfo{author}{{Sun}, J.P.}, \bibinfo{author}{{Susobhanan}, A.}, \bibinfo{author}{{Swiggum}, J.K.}, \bibinfo{author}{{Taylor}, J.}, \bibinfo{author}{{Taylor}, S.R.}, \bibinfo{author}{{Turner}, J.E.}, \bibinfo{author}{{Unal}, C.}, \bibinfo{author}{{Vallisneri}, M.}, \bibinfo{author}{{van Haasteren}, R.}, \bibinfo{author}{{Vigeland}, S.J.}, \bibinfo{author}{{Wahl}, H.M.}, \bibinfo{author}{{Wang}, Q.}, \bibinfo{author}{{Witt}, C.A.}, \bibinfo{author}{{Young}, O.}, \bibinfo{author}{{Nanograv Collaboration}}, \bibinfo{year}{2023}.
\newblock \bibinfo{title}{{The NANOGrav 15 yr Data Set: Evidence for a Gravitational-wave Background}}.
\newblock \bibinfo{journal}{APJL} \bibinfo{volume}{951}, \bibinfo{pages}{L8}.
\newblock \DOIprefix\doi{10.3847/2041-8213/acdac6}, \href{http://arxiv.org/abs/2306.16213}{{\tt arXiv:2306.16213}}.
\bibitem[{Arge and Pizzo(2000)}]{Arge2000}
\bibinfo{author}{Arge, C.N.}, \bibinfo{author}{Pizzo, V.J.}, \bibinfo{year}{2000}.
\newblock \bibinfo{title}{Improvement in the prediction of solar wind conditions using near-real time solar magnetic field updates}.
\newblock \bibinfo{journal}{Journal of Geophysical Research: Space Physics} \bibinfo{volume}{105}, \bibinfo{pages}{10465--10479}.
\newblock \DOIprefix\doi{https://doi.org/10.1029/1999JA000262}.
\bibitem[{Avramenko(2007)}]{avramenko2007}
\bibinfo{author}{Avramenko, A.}, \bibinfo{year}{2007}.
\newblock \bibinfo{title}{Synchronization on the basis of pulsar time}.
\newblock \bibinfo{journal}{Measurement Techniques} \bibinfo{volume}{50}, \bibinfo{pages}{725}.
\bibitem[{{Bates} et~al.(2013){Bates}, {Lorimer} and {Verbiest}}]{Bates2013}
\bibinfo{author}{{Bates}, S.D.}, \bibinfo{author}{{Lorimer}, D.R.}, \bibinfo{author}{{Verbiest}, J.P.W.}, \bibinfo{year}{2013}.
\newblock \bibinfo{title}{{The pulsar spectral index distribution}}.
\newblock \bibinfo{journal}{MNRAS} \bibinfo{volume}{431}, \bibinfo{pages}{1352--1358}.
\newblock \DOIprefix\doi{10.1093/mnras/stt257}, \href{http://arxiv.org/abs/1302.2053}{{\tt arXiv:1302.2053}}.
\bibitem[{{Bhat} et~al.(2023){Bhat}, {Swainston}, {McSweeney}, {Xue}, {Meyers}, {Kudale}, {Dai}, {Tremblay}, {van Straten}, {Shannon}, {Smith}, {Sokolowski}, {Ord}, {Sleap}, {Williams}, {Hancock}, {Lange}, {Tocknell}, {Johnston-Hollitt}, {Kaplan}, {Tingay} and {Walker}}]{Bhat2023}
\bibinfo{author}{{Bhat}, N.D.R.}, \bibinfo{author}{{Swainston}, N.A.}, \bibinfo{author}{{McSweeney}, S.J.}, \bibinfo{author}{{Xue}, M.}, \bibinfo{author}{{Meyers}, B.W.}, \bibinfo{author}{{Kudale}, S.}, \bibinfo{author}{{Dai}, S.}, \bibinfo{author}{{Tremblay}, S.E.}, \bibinfo{author}{{van Straten}, W.}, \bibinfo{author}{{Shannon}, R.M.}, \bibinfo{author}{{Smith}, K.R.}, \bibinfo{author}{{Sokolowski}, M.}, \bibinfo{author}{{Ord}, S.M.}, \bibinfo{author}{{Sleap}, G.}, \bibinfo{author}{{Williams}, A.}, \bibinfo{author}{{Hancock}, P.J.}, \bibinfo{author}{{Lange}, R.}, \bibinfo{author}{{Tocknell}, J.}, \bibinfo{author}{{Johnston-Hollitt}, M.}, \bibinfo{author}{{Kaplan}, D.L.}, \bibinfo{author}{{Tingay}, S.J.}, \bibinfo{author}{{Walker}, M.}, \bibinfo{year}{2023}.
\newblock \bibinfo{title}{{The Southern-sky MWA Rapid Two-metre (SMART) pulsar survey{\textemdash}II. Survey status, pulsar census, and first pulsar discoveries}}.
\newblock \bibinfo{journal}{PASA} \bibinfo{volume}{40}, \bibinfo{pages}{e020}.
\newblock \DOIprefix\doi{10.1017/pasa.2023.18}, \href{http://arxiv.org/abs/2302.11920}{{\tt arXiv:2302.11920}}.
\bibitem[{{Bhat} et~al.(2007){Bhat}, {Wayth}, {Knight}, {Bowman}, {Oberoi}, {Barnes}, {Briggs}, {Cappallo}, {Herne}, {Kocz}, {Lonsdale}, {Lynch}, {Stansby}, {Stevens}, {Torr}, {Webster} and {Wyithe}}]{bwk+07}
\bibinfo{author}{{Bhat}, N.D.R.}, \bibinfo{author}{{Wayth}, R.B.}, \bibinfo{author}{{Knight}, H.S.}, \bibinfo{author}{{Bowman}, J.D.}, \bibinfo{author}{{Oberoi}, D.}, \bibinfo{author}{{Barnes}, D.G.}, \bibinfo{author}{{Briggs}, F.H.}, \bibinfo{author}{{Cappallo}, R.J.}, \bibinfo{author}{{Herne}, D.}, \bibinfo{author}{{Kocz}, J.}, \bibinfo{author}{{Lonsdale}, C.J.}, \bibinfo{author}{{Lynch}, M.J.}, \bibinfo{author}{{Stansby}, B.}, \bibinfo{author}{{Stevens}, J.}, \bibinfo{author}{{Torr}, G.}, \bibinfo{author}{{Webster}, R.L.}, \bibinfo{author}{{Wyithe}, J.S.B.}, \bibinfo{year}{2007}.
\newblock \bibinfo{title}{{Detection of Crab Giant Pulses Using the Mileura Widefield Array Low Frequency Demonstrator Field Prototype System}}.
\newblock \bibinfo{journal}{APJ} \bibinfo{volume}{665}, \bibinfo{pages}{618--627}.
\newblock \DOIprefix\doi{10.1086/519444}, \href{http://arxiv.org/abs/0705.0404}{{\tt arXiv:0705.0404}}.
\bibitem[{{Billings}(1966)}]{Billings1966}
\bibinfo{author}{{Billings}, D.E.}, \bibinfo{year}{1966}.
\newblock \bibinfo{title}{{A guide to the solar corona}}.
\bibitem[{{B{\l}aszkiewicz} et~al.(2020){B{\l}aszkiewicz}, {Flisek}, {Kotulak}, {Krankowski}, {Lewandowski}, {Kijak} and {Fro{\'n}}}]{Blaszkiewicz2020}
\bibinfo{author}{{B{\l}aszkiewicz}, L.P.}, \bibinfo{author}{{Flisek}, P.}, \bibinfo{author}{{Kotulak}, K.}, \bibinfo{author}{{Krankowski}, A.}, \bibinfo{author}{{Lewandowski}, W.}, \bibinfo{author}{{Kijak}, J.}, \bibinfo{author}{{Fro{\'n}}, A.}, \bibinfo{year}{2020}.
\newblock \bibinfo{title}{{Finding the Ionospheric Fluctuations Reflection in the Pulsar Signals' Characteristics Observed with LOFAR}}.
\newblock \bibinfo{journal}{Sensors} \bibinfo{volume}{21}, \bibinfo{pages}{51}.
\newblock \DOIprefix\doi{10.3390/s21010051}.
\bibitem[{{Brunet} et~al.(2023){Brunet}, {Tolley}, {Corda}, {Ilic}, {Broekema} and {Kneib}}]{btc+23}
\bibinfo{author}{{Brunet}, T.}, \bibinfo{author}{{Tolley}, E.}, \bibinfo{author}{{Corda}, S.}, \bibinfo{author}{{Ilic}, R.}, \bibinfo{author}{{Broekema}, P.C.}, \bibinfo{author}{{Kneib}, J.P.}, \bibinfo{year}{2023}.
\newblock \bibinfo{title}{{Quantum Radio Astronomy: Data Encodings and Quantum Image Processing}}.
\newblock \bibinfo{journal}{arXiv e-prints} , \bibinfo{pages}{arXiv:2310.12084}\DOIprefix\doi{10.48550/arXiv.2310.12084}, \href{http://arxiv.org/abs/2310.12084}{{\tt arXiv:2310.12084}}.
\bibitem[{Buist et~al.(2014)Buist, Hesselink, Gibbs, Keuning, Gaubitch, Noroozi, Bentum, Verhoeven, Heusdens, Fernandes, Kabakchiev and Kestil{\"a}}]{bhg+14}
\bibinfo{author}{Buist, P.}, \bibinfo{author}{Hesselink, H.}, \bibinfo{author}{Gibbs, A.}, \bibinfo{author}{Keuning, M.}, \bibinfo{author}{Gaubitch, N.}, \bibinfo{author}{Noroozi, A.}, \bibinfo{author}{Bentum, M.}, \bibinfo{author}{Verhoeven, C.}, \bibinfo{author}{Heusdens, R.}, \bibinfo{author}{Fernandes, J.}, \bibinfo{author}{Kabakchiev, H.}, \bibinfo{author}{Kestil{\"a}, A.}, \bibinfo{year}{2014}.
\newblock \bibinfo{title}{Pulsarplane: a feasibility study for millisecond radio pulsar navigation}, in: \bibinfo{booktitle}{Proceedings of the 65th International Astronautical Congress (IAC2014)}, \bibinfo{publisher}{International Astronautical Federation (IAF)}. pp. \bibinfo{pages}{1--10}.
\newblock \bibinfo{note}{Eemcs-eprint-25633 ; null ; Conference date: 29-09-2014 Through 03-10-2014}.
\bibitem[{Burgin and Popov(2024)}]{Burgin2024}
\bibinfo{author}{Burgin, M.S.}, \bibinfo{author}{Popov, M.V.}, \bibinfo{year}{2024}.
\newblock \bibinfo{title}{{Probing the Ionosphere with Pulses from the Pulsar B2016+28 at a Frequency of 324 MHz}}.
\newblock \bibinfo{journal}{Astronomy Reports} \bibinfo{volume}{68}, \bibinfo{pages}{257--267}.
\newblock \URLprefix \url{https://doi.org/10.1134/S1063772924700276 https://doi.org/10.1134/S1063772924700276}, \DOIprefix\doi{10.1134/S1063772924700276}.
\bibitem[{{Caballero} et~al.(2018){Caballero}, {Guo}, {Lee}, {Lazarus}, {Champion}, {Desvignes}, {Kramer}, {Plant}, {Arzoumanian}, {Bailes}, {Bassa}, {Bhat}, {Brazier}, {Burgay}, {Burke-Spolaor}, {Chamberlin}, {Chatterjee}, {Cognard}, {Cordes}, {Dai}, {Demorest}, {Dolch}, {Ferdman}, {Fonseca}, {Gair}, {Garver-Daniels}, {Gentile}, {Gonzalez}, {Graikou}, {Guillemot}, {Hobbs}, {Janssen}, {Karuppusamy}, {Keith}, {Kerr}, {Lam}, {Lasky}, {Lazio}, {Levin}, {Liu}, {Lommen}, {Lorimer}, {Lynch}, {Madison}, {Manchester}, {McKee}, {McLaughlin}, {McWilliams}, {Mingarelli}, {Nice}, {Os{\l}owski}, {Palliyaguru}, {Pennucci}, {Perera}, {Perrodin}, {Possenti}, {Ransom}, {Reardon}, {Sanidas}, {Sesana}, {Shaifullah}, {Shannon}, {Siemens}, {Simon}, {Spiewak}, {Stairs}, {Stappers}, {Stinebring}, {Stovall}, {Swiggum}, {Taylor}, {Theureau}, {Tiburzi}, {Toomey}, {van Haasteren}, {van Straten}, {Verbiest}, {Wang}, {Zhu} and {Zhu}}]{cgl+18}
\bibinfo{author}{{Caballero}, R.N.}, \bibinfo{author}{{Guo}, Y.J.}, \bibinfo{author}{{Lee}, K.J.}, \bibinfo{author}{{Lazarus}, P.}, \bibinfo{author}{{Champion}, D.J.}, \bibinfo{author}{{Desvignes}, G.}, \bibinfo{author}{{Kramer}, M.}, \bibinfo{author}{{Plant}, K.}, \bibinfo{author}{{Arzoumanian}, Z.}, \bibinfo{author}{{Bailes}, M.}, \bibinfo{author}{{Bassa}, C.G.}, \bibinfo{author}{{Bhat}, N.D.R.}, \bibinfo{author}{{Brazier}, A.}, \bibinfo{author}{{Burgay}, M.}, \bibinfo{author}{{Burke-Spolaor}, S.}, \bibinfo{author}{{Chamberlin}, S.J.}, \bibinfo{author}{{Chatterjee}, S.}, \bibinfo{author}{{Cognard}, I.}, \bibinfo{author}{{Cordes}, J.M.}, \bibinfo{author}{{Dai}, S.}, \bibinfo{author}{{Demorest}, P.}, \bibinfo{author}{{Dolch}, T.}, \bibinfo{author}{{Ferdman}, R.D.}, \bibinfo{author}{{Fonseca}, E.}, \bibinfo{author}{{Gair}, J.R.}, \bibinfo{author}{{Garver-Daniels}, N.}, \bibinfo{author}{{Gentile}, P.}, \bibinfo{author}{{Gonzalez}, M.E.}, \bibinfo{author}{{Graikou}, E.}, \bibinfo{author}{{Guillemot}, L.},
  \bibinfo{author}{{Hobbs}, G.}, \bibinfo{author}{{Janssen}, G.H.}, \bibinfo{author}{{Karuppusamy}, R.}, \bibinfo{author}{{Keith}, M.J.}, \bibinfo{author}{{Kerr}, M.}, \bibinfo{author}{{Lam}, M.T.}, \bibinfo{author}{{Lasky}, P.D.}, \bibinfo{author}{{Lazio}, T.J.W.}, \bibinfo{author}{{Levin}, L.}, \bibinfo{author}{{Liu}, K.}, \bibinfo{author}{{Lommen}, A.N.}, \bibinfo{author}{{Lorimer}, D.R.}, \bibinfo{author}{{Lynch}, R.S.}, \bibinfo{author}{{Madison}, D.R.}, \bibinfo{author}{{Manchester}, R.N.}, \bibinfo{author}{{McKee}, J.W.}, \bibinfo{author}{{McLaughlin}, M.A.}, \bibinfo{author}{{McWilliams}, S.T.}, \bibinfo{author}{{Mingarelli}, C.M.F.}, \bibinfo{author}{{Nice}, D.J.}, \bibinfo{author}{{Os{\l}owski}, S.}, \bibinfo{author}{{Palliyaguru}, N.T.}, \bibinfo{author}{{Pennucci}, T.T.}, \bibinfo{author}{{Perera}, B.B.P.}, \bibinfo{author}{{Perrodin}, D.}, \bibinfo{author}{{Possenti}, A.}, \bibinfo{author}{{Ransom}, S.M.}, \bibinfo{author}{{Reardon}, D.J.}, \bibinfo{author}{{Sanidas}, S.A.},
  \bibinfo{author}{{Sesana}, A.}, \bibinfo{author}{{Shaifullah}, G.}, \bibinfo{author}{{Shannon}, R.M.}, \bibinfo{author}{{Siemens}, X.}, \bibinfo{author}{{Simon}, J.}, \bibinfo{author}{{Spiewak}, R.}, \bibinfo{author}{{Stairs}, I.}, \bibinfo{author}{{Stappers}, B.}, \bibinfo{author}{{Stinebring}, D.R.}, \bibinfo{author}{{Stovall}, K.}, \bibinfo{author}{{Swiggum}, J.K.}, \bibinfo{author}{{Taylor}, S.R.}, \bibinfo{author}{{Theureau}, G.}, \bibinfo{author}{{Tiburzi}, C.}, \bibinfo{author}{{Toomey}, L.}, \bibinfo{author}{{van Haasteren}, R.}, \bibinfo{author}{{van Straten}, W.}, \bibinfo{author}{{Verbiest}, J.P.W.}, \bibinfo{author}{{Wang}, J.B.}, \bibinfo{author}{{Zhu}, X.J.}, \bibinfo{author}{{Zhu}, W.W.}, \bibinfo{year}{2018}.
\newblock \bibinfo{title}{{Studying the Solar system with the International Pulsar Timing Array}}.
\newblock \bibinfo{journal}{MNRAS} \bibinfo{volume}{481}, \bibinfo{pages}{5501--5516}.
\newblock \DOIprefix\doi{10.1093/mnras/sty2632}, \href{http://arxiv.org/abs/1809.10744}{{\tt arXiv:1809.10744}}.
\bibitem[{{Campbell-Wilson} et~al.(2024){Campbell-Wilson}, {Flynn} and {Bateman}}]{ATel2024}
\bibinfo{author}{{Campbell-Wilson}, D.}, \bibinfo{author}{{Flynn}, C.}, \bibinfo{author}{{Bateman}, T.}, \bibinfo{year}{2024}.
\newblock \bibinfo{title}{{Confirmation of glitch event observed in PSR J0835-4510 (Vela pulsar)}}.
\newblock \bibinfo{journal}{The Astronomer's Telegram} \bibinfo{volume}{16610}, \bibinfo{pages}{1}.
\bibitem[{Canetti et~al.(2007)Canetti, Pass and Shelat}]{cps07}
\bibinfo{author}{Canetti, R.}, \bibinfo{author}{Pass, R.}, \bibinfo{author}{Shelat, A.}, \bibinfo{year}{2007}.
\newblock \bibinfo{title}{Cryptography from sunspots: How to use an imperfect reference string}, in: \bibinfo{booktitle}{Proceedings of the 48th Annual IEEE Symposium on Foundations of Computer Science}, \bibinfo{publisher}{IEEE Computer Society}, \bibinfo{address}{USA}. p. \bibinfo{pages}{249–259}.
\newblock \URLprefix \url{https://doi.org/10.1109/FOCS.2007.22}, \DOIprefix\doi{10.1109/FOCS.2007.22}.
\bibitem[{{Carley} et~al.(2017){Carley}, {Vilmer}, {Sim{\~o}es} and {{\'O} Fearraigh}}]{Carley+2017}
\bibinfo{author}{{Carley}, E.P.}, \bibinfo{author}{{Vilmer}, N.}, \bibinfo{author}{{Sim{\~o}es}, P.J.A.}, \bibinfo{author}{{{\'O} Fearraigh}, B.}, \bibinfo{year}{2017}.
\newblock \bibinfo{title}{{Estimation of a coronal mass ejection magnetic field strength using radio observations of gyrosynchrotron radiation}}.
\newblock \bibinfo{journal}{AAP} \bibinfo{volume}{608}, \bibinfo{pages}{A137}.
\newblock \DOIprefix\doi{10.1051/0004-6361/201731368}, \href{http://arxiv.org/abs/1709.05184}{{\tt arXiv:1709.05184}}.
\bibitem[{{Champion} et~al.(2010){Champion}, {Hobbs}, {Manchester}, {Edwards}, {Backer}, {Bailes}, {Bhat}, {Burke-Spolaor}, {Coles}, {Demorest}, {Ferdman}, {Folkner}, {Hotan}, {Kramer}, {Lommen}, {Nice}, {Purver}, {Sarkissian}, {Stairs}, {van Straten}, {Verbiest} and {Yardley}}]{chm+10}
\bibinfo{author}{{Champion}, D.J.}, \bibinfo{author}{{Hobbs}, G.B.}, \bibinfo{author}{{Manchester}, R.N.}, \bibinfo{author}{{Edwards}, R.T.}, \bibinfo{author}{{Backer}, D.C.}, \bibinfo{author}{{Bailes}, M.}, \bibinfo{author}{{Bhat}, N.D.R.}, \bibinfo{author}{{Burke-Spolaor}, S.}, \bibinfo{author}{{Coles}, W.}, \bibinfo{author}{{Demorest}, P.B.}, \bibinfo{author}{{Ferdman}, R.D.}, \bibinfo{author}{{Folkner}, W.M.}, \bibinfo{author}{{Hotan}, A.W.}, \bibinfo{author}{{Kramer}, M.}, \bibinfo{author}{{Lommen}, A.N.}, \bibinfo{author}{{Nice}, D.J.}, \bibinfo{author}{{Purver}, M.B.}, \bibinfo{author}{{Sarkissian}, J.M.}, \bibinfo{author}{{Stairs}, I.H.}, \bibinfo{author}{{van Straten}, W.}, \bibinfo{author}{{Verbiest}, J.P.W.}, \bibinfo{author}{{Yardley}, D.R.B.}, \bibinfo{year}{2010}.
\newblock \bibinfo{title}{{Measuring the Mass of Solar System Planets Using Pulsar Timing}}.
\newblock \bibinfo{journal}{APJL} \bibinfo{volume}{720}, \bibinfo{pages}{L201--L205}.
\newblock \DOIprefix\doi{10.1088/2041-8205/720/2/L201}, \href{http://arxiv.org/abs/1008.3607}{{\tt arXiv:1008.3607}}.
\bibitem[{Chapman~E.(2016)}]{chapman16}
\bibinfo{author}{Chapman~E., Grewar~J., N.T.}, \bibinfo{year}{2016}.
\newblock \bibinfo{title}{{Celestial sources for random number generation}}.
\newblock \bibinfo{journal}{The Proceedings of 14th Australian Information Security Management Conference} \DOIprefix\doi{10.4225/75/58a6975133e06}.
\bibitem[{Chen et~al.(2018)Chen, Zhan, Liu and Yuan}]{czly18}
\bibinfo{author}{Chen, M.}, \bibinfo{author}{Zhan, X.}, \bibinfo{author}{Liu, B.}, \bibinfo{author}{Yuan, W.}, \bibinfo{year}{2018}.
\newblock \bibinfo{title}{Gnss vulnerability network risk assessment and alleviation strategies considering efficiency cost}.
\newblock \bibinfo{journal}{Journal of Aeronautics, Astronautics and Aviation, Series A} \bibinfo{volume}{52}.
\newblock \DOIprefix\doi{10.6125/JoAAA.201806_50(2).05}.
\bibitem[{{Chen}(2018)}]{poting2018}
\bibinfo{author}{{Chen}, P.T.}, \bibinfo{year}{2018}.
\newblock \bibinfo{title}{{Pulsar-Based Navigation and Timing: Analysis and Estimation}}.
\newblock Ph.D. thesis. University of California, Los Angeles.
\bibitem[{{Craft}(1970)}]{craft1970}
\bibinfo{author}{{Craft}, Harold~Dumont, J.}, \bibinfo{year}{1970}.
\newblock \bibinfo{title}{{Radio Observations of the Pulse Profiles and Dispersion Measures of Twelve Pulsars.}}
\newblock Ph.D. thesis. Cornell University, New York.
\bibitem[{{Dawson} et~al.(2022){Dawson}, {Hobbs}, {Gao}, {Camtepe}, {Pieprzyk}, {Feng}, {Tranfa}, {Bradbury}, {Zhu} and {Li}}]{dhg+22}
\bibinfo{author}{{Dawson}, J.R.}, \bibinfo{author}{{Hobbs}, G.}, \bibinfo{author}{{Gao}, Y.}, \bibinfo{author}{{Camtepe}, S.}, \bibinfo{author}{{Pieprzyk}, J.}, \bibinfo{author}{{Feng}, Y.}, \bibinfo{author}{{Tranfa}, L.}, \bibinfo{author}{{Bradbury}, S.}, \bibinfo{author}{{Zhu}, W.}, \bibinfo{author}{{Li}, D.}, \bibinfo{year}{2022}.
\newblock \bibinfo{title}{{Physical publicly verifiable randomness from pulsars}}.
\newblock \bibinfo{journal}{Astronomy and Computing} \bibinfo{volume}{38}, \bibinfo{pages}{100549}.
\newblock \DOIprefix\doi{10.1016/j.ascom.2022.100549}, \href{http://arxiv.org/abs/2201.05763}{{\tt arXiv:2201.05763}}.
\bibitem[{{Deng} et~al.(2013){Deng}, {Hobbs}, {You}, {Li}, {Keith}, {Shannon}, {Coles}, {Manchester}, {Zheng}, {Yu}, {Gao}, {Wu} and {Chen}}]{dhy+13}
\bibinfo{author}{{Deng}, X.P.}, \bibinfo{author}{{Hobbs}, G.}, \bibinfo{author}{{You}, X.P.}, \bibinfo{author}{{Li}, M.T.}, \bibinfo{author}{{Keith}, M.J.}, \bibinfo{author}{{Shannon}, R.M.}, \bibinfo{author}{{Coles}, W.}, \bibinfo{author}{{Manchester}, R.N.}, \bibinfo{author}{{Zheng}, J.H.}, \bibinfo{author}{{Yu}, X.Z.}, \bibinfo{author}{{Gao}, D.}, \bibinfo{author}{{Wu}, X.}, \bibinfo{author}{{Chen}, D.}, \bibinfo{year}{2013}.
\newblock \bibinfo{title}{{Interplanetary spacecraft navigation using pulsars}}.
\newblock \bibinfo{journal}{Advances in Space Research} \bibinfo{volume}{52}, \bibinfo{pages}{1602--1621}.
\newblock \DOIprefix\doi{10.1016/j.asr.2013.07.025}, \href{http://arxiv.org/abs/1307.5375}{{\tt arXiv:1307.5375}}.
\bibitem[{{Desvignes} et~al.(2016){Desvignes}, {Caballero}, {Lentati}, {Verbiest}, {Champion}, {Stappers}, {Janssen}, {Lazarus}, {Os{\l}owski}, {Babak}, {Bassa}, {Brem}, {Burgay}, {Cognard}, {Gair}, {Graikou}, {Guillemot}, {Hessels}, {Jessner}, {Jordan}, {Karuppusamy}, {Kramer}, {Lassus}, {Lazaridis}, {Lee}, {Liu}, {Lyne}, {McKee}, {Mingarelli}, {Perrodin}, {Petiteau}, {Possenti}, {Purver}, {Rosado}, {Sanidas}, {Sesana}, {Shaifullah}, {Smits}, {Taylor}, {Theureau}, {Tiburzi}, {van Haasteren} and {Vecchio}}]{dcl+16}
\bibinfo{author}{{Desvignes}, G.}, \bibinfo{author}{{Caballero}, R.N.}, \bibinfo{author}{{Lentati}, L.}, \bibinfo{author}{{Verbiest}, J.P.W.}, \bibinfo{author}{{Champion}, D.J.}, \bibinfo{author}{{Stappers}, B.W.}, \bibinfo{author}{{Janssen}, G.H.}, \bibinfo{author}{{Lazarus}, P.}, \bibinfo{author}{{Os{\l}owski}, S.}, \bibinfo{author}{{Babak}, S.}, \bibinfo{author}{{Bassa}, C.G.}, \bibinfo{author}{{Brem}, P.}, \bibinfo{author}{{Burgay}, M.}, \bibinfo{author}{{Cognard}, I.}, \bibinfo{author}{{Gair}, J.R.}, \bibinfo{author}{{Graikou}, E.}, \bibinfo{author}{{Guillemot}, L.}, \bibinfo{author}{{Hessels}, J.W.T.}, \bibinfo{author}{{Jessner}, A.}, \bibinfo{author}{{Jordan}, C.}, \bibinfo{author}{{Karuppusamy}, R.}, \bibinfo{author}{{Kramer}, M.}, \bibinfo{author}{{Lassus}, A.}, \bibinfo{author}{{Lazaridis}, K.}, \bibinfo{author}{{Lee}, K.J.}, \bibinfo{author}{{Liu}, K.}, \bibinfo{author}{{Lyne}, A.G.}, \bibinfo{author}{{McKee}, J.}, \bibinfo{author}{{Mingarelli}, C.M.F.}, \bibinfo{author}{{Perrodin}, D.},
  \bibinfo{author}{{Petiteau}, A.}, \bibinfo{author}{{Possenti}, A.}, \bibinfo{author}{{Purver}, M.B.}, \bibinfo{author}{{Rosado}, P.A.}, \bibinfo{author}{{Sanidas}, S.}, \bibinfo{author}{{Sesana}, A.}, \bibinfo{author}{{Shaifullah}, G.}, \bibinfo{author}{{Smits}, R.}, \bibinfo{author}{{Taylor}, S.R.}, \bibinfo{author}{{Theureau}, G.}, \bibinfo{author}{{Tiburzi}, C.}, \bibinfo{author}{{van Haasteren}, R.}, \bibinfo{author}{{Vecchio}, A.}, \bibinfo{year}{2016}.
\newblock \bibinfo{title}{{High-precision timing of 42 millisecond pulsars with the European Pulsar Timing Array}}.
\newblock \bibinfo{journal}{MNRAS} \bibinfo{volume}{458}, \bibinfo{pages}{3341--3380}.
\newblock \DOIprefix\doi{10.1093/mnras/stw483}, \href{http://arxiv.org/abs/1602.08511}{{\tt arXiv:1602.08511}}.
\bibitem[{{EPTA Collaboration} et~al.(2023){EPTA Collaboration}, {InPTA Collaboration}, {Antoniadis}, {Arumugam}, {Arumugam}, {Babak}, {Bagchi}, {Bak Nielsen}, {Bassa}, {Bathula}, {Berthereau}, {Bonetti}, {Bortolas}, {Brook}, {Burgay}, {Caballero}, {Chalumeau}, {Champion}, {Chanlaridis}, {Chen}, {Cognard}, {Dandapat}, {Deb}, {Desai}, {Desvignes}, {Dhanda-Batra}, {Dwivedi}, {Falxa}, {Ferdman}, {Franchini}, {Gair}, {Goncharov}, {Gopakumar}, {Graikou}, {Grie{\ss}meier}, {Guillemot}, {Guo}, {Gupta}, {Hisano}, {Hu}, {Iraci}, {Izquierdo-Villalba}, {Jang}, {Jawor}, {Janssen}, {Jessner}, {Joshi}, {Kareem}, {Karuppusamy}, {Keane}, {Keith}, {Kharbanda}, {Kikunaga}, {Kolhe}, {Kramer}, {Krishnakumar}, {Lackeos}, {Lee}, {Liu}, {Liu}, {Lyne}, {McKee}, {Maan}, {Main}, {Mickaliger}, {Ni{\c{t}}u}, {Nobleson}, {Paladi}, {Parthasarathy}, {Perera}, {Perrodin}, {Petiteau}, {Porayko}, {Possenti}, {Prabu}, {Quelquejay Leclere}, {Rana}, {Samajdar}, {Sanidas}, {Sesana}, {Shaifullah}, {Singha}, {Speri}, {Spiewak}, {Srivastava},
  {Stappers}, {Surnis}, {Susarla}, {Susobhanan}, {Takahashi}, {Tarafdar}, {Theureau}, {Tiburzi}, {van der Wateren}, {Vecchio}, {Venkatraman Krishnan}, {Verbiest}, {Wang}, {Wang} and {Wu}}]{EPTA2023}
\bibinfo{author}{{EPTA Collaboration}}, \bibinfo{author}{{InPTA Collaboration}}, \bibinfo{author}{{Antoniadis}, J.}, \bibinfo{author}{{Arumugam}, P.}, \bibinfo{author}{{Arumugam}, S.}, \bibinfo{author}{{Babak}, S.}, \bibinfo{author}{{Bagchi}, M.}, \bibinfo{author}{{Bak Nielsen}, A.S.}, \bibinfo{author}{{Bassa}, C.G.}, \bibinfo{author}{{Bathula}, A.}, \bibinfo{author}{{Berthereau}, A.}, \bibinfo{author}{{Bonetti}, M.}, \bibinfo{author}{{Bortolas}, E.}, \bibinfo{author}{{Brook}, P.R.}, \bibinfo{author}{{Burgay}, M.}, \bibinfo{author}{{Caballero}, R.N.}, \bibinfo{author}{{Chalumeau}, A.}, \bibinfo{author}{{Champion}, D.J.}, \bibinfo{author}{{Chanlaridis}, S.}, \bibinfo{author}{{Chen}, S.}, \bibinfo{author}{{Cognard}, I.}, \bibinfo{author}{{Dandapat}, S.}, \bibinfo{author}{{Deb}, D.}, \bibinfo{author}{{Desai}, S.}, \bibinfo{author}{{Desvignes}, G.}, \bibinfo{author}{{Dhanda-Batra}, N.}, \bibinfo{author}{{Dwivedi}, C.}, \bibinfo{author}{{Falxa}, M.}, \bibinfo{author}{{Ferdman}, R.D.}, \bibinfo{author}{{Franchini},
  A.}, \bibinfo{author}{{Gair}, J.R.}, \bibinfo{author}{{Goncharov}, B.}, \bibinfo{author}{{Gopakumar}, A.}, \bibinfo{author}{{Graikou}, E.}, \bibinfo{author}{{Grie{\ss}meier}, J.M.}, \bibinfo{author}{{Guillemot}, L.}, \bibinfo{author}{{Guo}, Y.J.}, \bibinfo{author}{{Gupta}, Y.}, \bibinfo{author}{{Hisano}, S.}, \bibinfo{author}{{Hu}, H.}, \bibinfo{author}{{Iraci}, F.}, \bibinfo{author}{{Izquierdo-Villalba}, D.}, \bibinfo{author}{{Jang}, J.}, \bibinfo{author}{{Jawor}, J.}, \bibinfo{author}{{Janssen}, G.H.}, \bibinfo{author}{{Jessner}, A.}, \bibinfo{author}{{Joshi}, B.C.}, \bibinfo{author}{{Kareem}, F.}, \bibinfo{author}{{Karuppusamy}, R.}, \bibinfo{author}{{Keane}, E.F.}, \bibinfo{author}{{Keith}, M.J.}, \bibinfo{author}{{Kharbanda}, D.}, \bibinfo{author}{{Kikunaga}, T.}, \bibinfo{author}{{Kolhe}, N.}, \bibinfo{author}{{Kramer}, M.}, \bibinfo{author}{{Krishnakumar}, M.A.}, \bibinfo{author}{{Lackeos}, K.}, \bibinfo{author}{{Lee}, K.J.}, \bibinfo{author}{{Liu}, K.}, \bibinfo{author}{{Liu}, Y.},
  \bibinfo{author}{{Lyne}, A.G.}, \bibinfo{author}{{McKee}, J.W.}, \bibinfo{author}{{Maan}, Y.}, \bibinfo{author}{{Main}, R.A.}, \bibinfo{author}{{Mickaliger}, M.B.}, \bibinfo{author}{{Ni{\c{t}}u}, I.C.}, \bibinfo{author}{{Nobleson}, K.}, \bibinfo{author}{{Paladi}, A.K.}, \bibinfo{author}{{Parthasarathy}, A.}, \bibinfo{author}{{Perera}, B.B.P.}, \bibinfo{author}{{Perrodin}, D.}, \bibinfo{author}{{Petiteau}, A.}, \bibinfo{author}{{Porayko}, N.K.}, \bibinfo{author}{{Possenti}, A.}, \bibinfo{author}{{Prabu}, T.}, \bibinfo{author}{{Quelquejay Leclere}, H.}, \bibinfo{author}{{Rana}, P.}, \bibinfo{author}{{Samajdar}, A.}, \bibinfo{author}{{Sanidas}, S.A.}, \bibinfo{author}{{Sesana}, A.}, \bibinfo{author}{{Shaifullah}, G.}, \bibinfo{author}{{Singha}, J.}, \bibinfo{author}{{Speri}, L.}, \bibinfo{author}{{Spiewak}, R.}, \bibinfo{author}{{Srivastava}, A.}, \bibinfo{author}{{Stappers}, B.W.}, \bibinfo{author}{{Surnis}, M.}, \bibinfo{author}{{Susarla}, S.C.}, \bibinfo{author}{{Susobhanan}, A.},
  \bibinfo{author}{{Takahashi}, K.}, \bibinfo{author}{{Tarafdar}, P.}, \bibinfo{author}{{Theureau}, G.}, \bibinfo{author}{{Tiburzi}, C.}, \bibinfo{author}{{van der Wateren}, E.}, \bibinfo{author}{{Vecchio}, A.}, \bibinfo{author}{{Venkatraman Krishnan}, V.}, \bibinfo{author}{{Verbiest}, J.P.W.}, \bibinfo{author}{{Wang}, J.}, \bibinfo{author}{{Wang}, L.}, \bibinfo{author}{{Wu}, Z.}, \bibinfo{year}{2023}.
\newblock \bibinfo{title}{{The second data release from the European Pulsar Timing Array. III. Search for gravitational wave signals}}.
\newblock \bibinfo{journal}{AAP} \bibinfo{volume}{678}, \bibinfo{pages}{A50}.
\newblock \DOIprefix\doi{10.1051/0004-6361/202346844}, \href{http://arxiv.org/abs/2306.16214}{{\tt arXiv:2306.16214}}.
\bibitem[{{Fitzgerald} et~al.(2014){Fitzgerald}, {Hollow}, {Rebull}, {Danaia} and {McKinnon}}]{fhr+14}
\bibinfo{author}{{Fitzgerald}, M.T.}, \bibinfo{author}{{Hollow}, R.}, \bibinfo{author}{{Rebull}, L.M.}, \bibinfo{author}{{Danaia}, L.}, \bibinfo{author}{{McKinnon}, D.H.}, \bibinfo{year}{2014}.
\newblock \bibinfo{title}{{A Review of High School Level Astronomy Student Research Projects Over the Last Two Decades}}.
\newblock \bibinfo{journal}{PASA} \bibinfo{volume}{31}, \bibinfo{pages}{e037}.
\newblock \DOIprefix\doi{10.1017/pasa.2014.30}, \href{http://arxiv.org/abs/1407.6586}{{\tt arXiv:1407.6586}}.
\bibitem[{{Fulgenzi} et~al.(2017){Fulgenzi}, {Melatos} and {Hughes}}]{fmh17}
\bibinfo{author}{{Fulgenzi}, W.}, \bibinfo{author}{{Melatos}, A.}, \bibinfo{author}{{Hughes}, B.D.}, \bibinfo{year}{2017}.
\newblock \bibinfo{title}{{Radio pulsar glitches as a state-dependent Poisson process}}.
\newblock \bibinfo{journal}{MNRAS} \bibinfo{volume}{470}, \bibinfo{pages}{4307--4329}.
\newblock \DOIprefix\doi{10.1093/mnras/stx1353}.
\bibitem[{{Geyer} et~al.(2021){Geyer}, {Serylak}, {Abbate}, {Bailes}, {Buchner}, {Chilufya}, {Johnston}, {Karastergiou}, {Main}, {van Straten} and {Shamohammadi}}]{gsa21}
\bibinfo{author}{{Geyer}, M.}, \bibinfo{author}{{Serylak}, M.}, \bibinfo{author}{{Abbate}, F.}, \bibinfo{author}{{Bailes}, M.}, \bibinfo{author}{{Buchner}, S.}, \bibinfo{author}{{Chilufya}, J.}, \bibinfo{author}{{Johnston}, S.}, \bibinfo{author}{{Karastergiou}, A.}, \bibinfo{author}{{Main}, R.}, \bibinfo{author}{{van Straten}, W.}, \bibinfo{author}{{Shamohammadi}, M.}, \bibinfo{year}{2021}.
\newblock \bibinfo{title}{{The Thousand-Pulsar-Array programme on MeerKAT - III. Giant pulse characteristics of PSR J0540-6919}}.
\newblock \bibinfo{journal}{MNRAS} \bibinfo{volume}{505}, \bibinfo{pages}{4468--4482}.
\newblock \DOIprefix\doi{10.1093/mnras/stab1501}, \href{http://arxiv.org/abs/2105.09096}{{\tt arXiv:2105.09096}}.
\bibitem[{{Guo} et~al.(2019){Guo}, {Li}, {Lee} and {Caballero}}]{gllc19}
\bibinfo{author}{{Guo}, Y.J.}, \bibinfo{author}{{Li}, G.Y.}, \bibinfo{author}{{Lee}, K.J.}, \bibinfo{author}{{Caballero}, R.N.}, \bibinfo{year}{2019}.
\newblock \bibinfo{title}{{Studying the Solar system dynamics using pulsar timing arrays and the LINIMOSS dynamical model}}.
\newblock \bibinfo{journal}{MNRAS} \bibinfo{volume}{489}, \bibinfo{pages}{5573--5581}.
\newblock \DOIprefix\doi{10.1093/mnras/stz2515}, \href{http://arxiv.org/abs/1909.04507}{{\tt arXiv:1909.04507}}.
\bibitem[{{Han} et~al.(2020){Han}, {Wang}, {Wang}, {Sun} and {He}}]{hww+20}
\bibinfo{author}{{Han}, W.}, \bibinfo{author}{{Wang}, J.}, \bibinfo{author}{{Wang}, N.}, \bibinfo{author}{{Sun}, G.}, \bibinfo{author}{{He}, D.}, \bibinfo{year}{2020}.
\newblock \bibinfo{title}{{A method of ground target positioning by observing radio pulsars}}.
\newblock \bibinfo{journal}{Experimental Astronomy} \bibinfo{volume}{49}, \bibinfo{pages}{43--60}.
\newblock \DOIprefix\doi{10.1007/s10686-020-09651-2}.
\bibitem[{{Hewish} and {Dennison}(1967)}]{Hewish/Dennison1967}
\bibinfo{author}{{Hewish}, A.}, \bibinfo{author}{{Dennison}, P.A.}, \bibinfo{year}{1967}.
\newblock \bibinfo{title}{{Measurements of the Solar Wind and the Small-Scale Structure of the Interplanetary Medium}}.
\newblock \bibinfo{journal}{Journal of Geophysical Research} \bibinfo{volume}{72}, \bibinfo{pages}{1977}.
\newblock \DOIprefix\doi{10.1029/JZ072i007p01977}.
\bibitem[{{Hobbs} et~al.(2012){Hobbs}, {Coles}, {Manchester}, {Keith}, {Shannon}, {Chen}, {Bailes}, {Bhat}, {Burke-Spolaor}, {Champion}, {Chaudhary}, {Hotan}, {Khoo}, {Kocz}, {Levin}, {Oslowski}, {Preisig}, {Ravi}, {Reynolds}, {Sarkissian}, {van Straten}, {Verbiest}, {Yardley} and {You}}]{hcm12}
\bibinfo{author}{{Hobbs}, G.}, \bibinfo{author}{{Coles}, W.}, \bibinfo{author}{{Manchester}, R.N.}, \bibinfo{author}{{Keith}, M.J.}, \bibinfo{author}{{Shannon}, R.M.}, \bibinfo{author}{{Chen}, D.}, \bibinfo{author}{{Bailes}, M.}, \bibinfo{author}{{Bhat}, N.D.R.}, \bibinfo{author}{{Burke-Spolaor}, S.}, \bibinfo{author}{{Champion}, D.}, \bibinfo{author}{{Chaudhary}, A.}, \bibinfo{author}{{Hotan}, A.}, \bibinfo{author}{{Khoo}, J.}, \bibinfo{author}{{Kocz}, J.}, \bibinfo{author}{{Levin}, Y.}, \bibinfo{author}{{Oslowski}, S.}, \bibinfo{author}{{Preisig}, B.}, \bibinfo{author}{{Ravi}, V.}, \bibinfo{author}{{Reynolds}, J.E.}, \bibinfo{author}{{Sarkissian}, J.}, \bibinfo{author}{{van Straten}, W.}, \bibinfo{author}{{Verbiest}, J.P.W.}, \bibinfo{author}{{Yardley}, D.}, \bibinfo{author}{{You}, X.P.}, \bibinfo{year}{2012}.
\newblock \bibinfo{title}{{Development of a pulsar-based time-scale}}.
\newblock \bibinfo{journal}{MNRAS} \bibinfo{volume}{427}, \bibinfo{pages}{2780--2787}.
\newblock \DOIprefix\doi{10.1111/j.1365-2966.2012.21946.x}, \href{http://arxiv.org/abs/1208.3560}{{\tt arXiv:1208.3560}}.
\bibitem[{{Hobbs} et~al.(2020){Hobbs}, {Guo}, {Caballero}, {Coles}, {Lee}, {Manchester}, {Reardon}, {Matsakis}, {Tong}, {Arzoumanian}, {Bailes}, {Bassa}, {Bhat}, {Brazier}, {Burke-Spolaor}, {Champion}, {Chatterjee}, {Cognard}, {Dai}, {Desvignes}, {Dolch}, {Ferdman}, {Graikou}, {Guillemot}, {Janssen}, {Keith}, {Kerr}, {Kramer}, {Lam}, {Liu}, {Lyne}, {Lazio}, {Lynch}, {McKee}, {McLaughlin}, {Mingarelli}, {Nice}, {Os{\l}owski}, {Pennucci}, {Perera}, {Perrodin}, {Possenti}, {Russell}, {Sanidas}, {Sesana}, {Shaifullah}, {Shannon}, {Simon}, {Spiewak}, {Stairs}, {Stappers}, {Swiggum}, {Taylor}, {Theureau}, {Toomey}, {van Haasteren}, {Wang}, {Wang} and {Zhu}}]{hgc+20}
\bibinfo{author}{{Hobbs}, G.}, \bibinfo{author}{{Guo}, L.}, \bibinfo{author}{{Caballero}, R.N.}, \bibinfo{author}{{Coles}, W.}, \bibinfo{author}{{Lee}, K.J.}, \bibinfo{author}{{Manchester}, R.N.}, \bibinfo{author}{{Reardon}, D.J.}, \bibinfo{author}{{Matsakis}, D.}, \bibinfo{author}{{Tong}, M.L.}, \bibinfo{author}{{Arzoumanian}, Z.}, \bibinfo{author}{{Bailes}, M.}, \bibinfo{author}{{Bassa}, C.G.}, \bibinfo{author}{{Bhat}, N.D.R.}, \bibinfo{author}{{Brazier}, A.}, \bibinfo{author}{{Burke-Spolaor}, S.}, \bibinfo{author}{{Champion}, D.J.}, \bibinfo{author}{{Chatterjee}, S.}, \bibinfo{author}{{Cognard}, I.}, \bibinfo{author}{{Dai}, S.}, \bibinfo{author}{{Desvignes}, G.}, \bibinfo{author}{{Dolch}, T.}, \bibinfo{author}{{Ferdman}, R.D.}, \bibinfo{author}{{Graikou}, E.}, \bibinfo{author}{{Guillemot}, L.}, \bibinfo{author}{{Janssen}, G.H.}, \bibinfo{author}{{Keith}, M.J.}, \bibinfo{author}{{Kerr}, M.}, \bibinfo{author}{{Kramer}, M.}, \bibinfo{author}{{Lam}, M.T.}, \bibinfo{author}{{Liu}, K.}, \bibinfo{author}{{Lyne},
  A.}, \bibinfo{author}{{Lazio}, T.J.W.}, \bibinfo{author}{{Lynch}, R.}, \bibinfo{author}{{McKee}, J.W.}, \bibinfo{author}{{McLaughlin}, M.A.}, \bibinfo{author}{{Mingarelli}, C.M.F.}, \bibinfo{author}{{Nice}, D.J.}, \bibinfo{author}{{Os{\l}owski}, S.}, \bibinfo{author}{{Pennucci}, T.T.}, \bibinfo{author}{{Perera}, B.B.P.}, \bibinfo{author}{{Perrodin}, D.}, \bibinfo{author}{{Possenti}, A.}, \bibinfo{author}{{Russell}, C.J.}, \bibinfo{author}{{Sanidas}, S.}, \bibinfo{author}{{Sesana}, A.}, \bibinfo{author}{{Shaifullah}, G.}, \bibinfo{author}{{Shannon}, R.M.}, \bibinfo{author}{{Simon}, J.}, \bibinfo{author}{{Spiewak}, R.}, \bibinfo{author}{{Stairs}, I.H.}, \bibinfo{author}{{Stappers}, B.W.}, \bibinfo{author}{{Swiggum}, J.K.}, \bibinfo{author}{{Taylor}, S.R.}, \bibinfo{author}{{Theureau}, G.}, \bibinfo{author}{{Toomey}, L.}, \bibinfo{author}{{van Haasteren}, R.}, \bibinfo{author}{{Wang}, J.B.}, \bibinfo{author}{{Wang}, Y.}, \bibinfo{author}{{Zhu}, X.J.}, \bibinfo{year}{2020}.
\newblock \bibinfo{title}{{A pulsar-based time-scale from the International Pulsar Timing Array}}.
\newblock \bibinfo{journal}{MNRAS} \bibinfo{volume}{491}, \bibinfo{pages}{5951--5965}.
\newblock \DOIprefix\doi{10.1093/mnras/stz3071}, \href{http://arxiv.org/abs/1910.13628}{{\tt arXiv:1910.13628}}.
\bibitem[{{Hobbs} et~al.(2009){Hobbs}, {Hollow}, {Champion}, {Khoo}, {Yardley}, {Carr}, {Keith}, {Jenet}, {Amy}, {Burgay}, {Burke-Spolaor}, {Chapman}, {Danaia}, {Homewood}, {Kovacevic}, {Mao}, {McKinnon}, {Mulcahy}, {Oslowski} and {van Straten}}]{hhc+09}
\bibinfo{author}{{Hobbs}, G.}, \bibinfo{author}{{Hollow}, R.}, \bibinfo{author}{{Champion}, D.}, \bibinfo{author}{{Khoo}, J.}, \bibinfo{author}{{Yardley}, D.}, \bibinfo{author}{{Carr}, M.}, \bibinfo{author}{{Keith}, M.}, \bibinfo{author}{{Jenet}, F.}, \bibinfo{author}{{Amy}, S.}, \bibinfo{author}{{Burgay}, M.}, \bibinfo{author}{{Burke-Spolaor}, S.}, \bibinfo{author}{{Chapman}, J.}, \bibinfo{author}{{Danaia}, L.}, \bibinfo{author}{{Homewood}, B.}, \bibinfo{author}{{Kovacevic}, A.}, \bibinfo{author}{{Mao}, M.}, \bibinfo{author}{{McKinnon}, D.}, \bibinfo{author}{{Mulcahy}, M.}, \bibinfo{author}{{Oslowski}, S.}, \bibinfo{author}{{van Straten}, W.}, \bibinfo{year}{2009}.
\newblock \bibinfo{title}{{The PULSE@Parkes Project: a New Observing Technique for Long-Term Pulsar Monitoring}}.
\newblock \bibinfo{journal}{PASA} \bibinfo{volume}{26}, \bibinfo{pages}{468--475}.
\newblock \DOIprefix\doi{10.1071/AS09021}, \href{http://arxiv.org/abs/0907.4847}{{\tt arXiv:0907.4847}}.
\bibitem[{Hobbs et~al.(2020)Hobbs, Manchester, Dunning, Jameson, Roberts, George, Green, Tuthill, Toomey, Kaczmarek, Mader, Marquarding, Ahmed, Amy, Bailes, Beresford, Bhat, Bock, Bourne, Bowen, Brothers, Cameron, Carretti, Carter, Castillo, Chekkala, Cheng, Chung, Craig, Dai, Dawson, Dempsey, Doherty, Dong, Edwards, Ergesh, Gao, Han, Hayman, Indermuehle, Jeganathan, Johnston, Kanoniuk, Kesteven, Kramer, Leach, Mcintyre, Moss, Os{\l}owski, Phillips, Pope, Preisig, Price, Reeves, Reilly, Reynolds, Robishaw, Roush, Ruckley, Sadler, Sarkissian, Severs, Shannon, Smart, Smith, Smith, Sobey, Staveley-Smith, Tzioumis, van Straten, Wang, Wen and Whiting}]{Hobbs+2020}
\bibinfo{author}{Hobbs, G.}, \bibinfo{author}{Manchester, R.N.}, \bibinfo{author}{Dunning, A.}, \bibinfo{author}{Jameson, A.}, \bibinfo{author}{Roberts, P.}, \bibinfo{author}{George, D.}, \bibinfo{author}{Green, J.A.}, \bibinfo{author}{Tuthill, J.}, \bibinfo{author}{Toomey, L.}, \bibinfo{author}{Kaczmarek, J.F.}, \bibinfo{author}{Mader, S.}, \bibinfo{author}{Marquarding, M.}, \bibinfo{author}{Ahmed, A.}, \bibinfo{author}{Amy, S.W.}, \bibinfo{author}{Bailes, M.}, \bibinfo{author}{Beresford, R.}, \bibinfo{author}{Bhat, N.D.R.}, \bibinfo{author}{Bock, D.C.J.}, \bibinfo{author}{Bourne, M.}, \bibinfo{author}{Bowen, M.}, \bibinfo{author}{Brothers, M.}, \bibinfo{author}{Cameron, A.D.}, \bibinfo{author}{Carretti, E.}, \bibinfo{author}{Carter, N.}, \bibinfo{author}{Castillo, S.}, \bibinfo{author}{Chekkala, R.}, \bibinfo{author}{Cheng, W.}, \bibinfo{author}{Chung, Y.}, \bibinfo{author}{Craig, D.A.}, \bibinfo{author}{Dai, S.}, \bibinfo{author}{Dawson, J.}, \bibinfo{author}{Dempsey, J.}, \bibinfo{author}{Doherty, P.},
  \bibinfo{author}{Dong, B.}, \bibinfo{author}{Edwards, P.}, \bibinfo{author}{Ergesh, T.}, \bibinfo{author}{Gao, X.}, \bibinfo{author}{Han, J.}, \bibinfo{author}{Hayman, D.}, \bibinfo{author}{Indermuehle, B.}, \bibinfo{author}{Jeganathan, K.}, \bibinfo{author}{Johnston, S.}, \bibinfo{author}{Kanoniuk, H.}, \bibinfo{author}{Kesteven, M.}, \bibinfo{author}{Kramer, M.}, \bibinfo{author}{Leach, M.}, \bibinfo{author}{Mcintyre, V.}, \bibinfo{author}{Moss, V.}, \bibinfo{author}{Os{\l}owski, S.}, \bibinfo{author}{Phillips, C.}, \bibinfo{author}{Pope, N.}, \bibinfo{author}{Preisig, B.}, \bibinfo{author}{Price, D.}, \bibinfo{author}{Reeves, K.}, \bibinfo{author}{Reilly, L.}, \bibinfo{author}{Reynolds, J.}, \bibinfo{author}{Robishaw, T.}, \bibinfo{author}{Roush, P.}, \bibinfo{author}{Ruckley, T.}, \bibinfo{author}{Sadler, E.}, \bibinfo{author}{Sarkissian, J.}, \bibinfo{author}{Severs, S.}, \bibinfo{author}{Shannon, R.}, \bibinfo{author}{Smart, K.}, \bibinfo{author}{Smith, M.}, \bibinfo{author}{Smith, S.},
  \bibinfo{author}{Sobey, C.}, \bibinfo{author}{Staveley-Smith, L.}, \bibinfo{author}{Tzioumis, A.}, \bibinfo{author}{van Straten, W.}, \bibinfo{author}{Wang, N.}, \bibinfo{author}{Wen, L.}, \bibinfo{author}{Whiting, M.}, \bibinfo{year}{2020}.
\newblock \bibinfo{title}{An ultra-wide bandwidth (704 to 4~032~{MHz}) receiver for the parkes radio telescope}.
\newblock \bibinfo{journal}{Publications of the Astronomical Society of Australia} \bibinfo{volume}{37}.
\newblock \URLprefix \url{https://doi.org/10.1017%2Fpasa.2020.2}, \DOIprefix\doi{10.1017/pasa.2020.2}.
\bibitem[{{Hobbs} et~al.(2011){Hobbs}, {Miller}, {Manchester}, {Dempsey}, {Chapman}, {Khoo}, {Applegate}, {Bailes}, {Bhat}, {Bridle}, {Borg}, {Brown}, {Burnett}, {Camilo}, {Cattalini}, {Chaudhary}, {Chen}, {D'Amico}, {Kedziora-Chudczer}, {Cornwell}, {George}, {Hampson}, {Hepburn}, {Jameson}, {Keith}, {Kelly}, {Kosmynin}, {Lenc}, {Lorimer}, {Love}, {Lyne}, {McIntyre}, {Morrissey}, {Pienaar}, {Reynolds}, {Ryder}, {Sarkissian}, {Stevenson}, {Treloar}, {van Straten}, {Whiting} and {Wilson}}]{hmm+11}
\bibinfo{author}{{Hobbs}, G.}, \bibinfo{author}{{Miller}, D.}, \bibinfo{author}{{Manchester}, R.N.}, \bibinfo{author}{{Dempsey}, J.}, \bibinfo{author}{{Chapman}, J.M.}, \bibinfo{author}{{Khoo}, J.}, \bibinfo{author}{{Applegate}, J.}, \bibinfo{author}{{Bailes}, M.}, \bibinfo{author}{{Bhat}, N.D.R.}, \bibinfo{author}{{Bridle}, R.}, \bibinfo{author}{{Borg}, A.}, \bibinfo{author}{{Brown}, A.}, \bibinfo{author}{{Burnett}, C.}, \bibinfo{author}{{Camilo}, F.}, \bibinfo{author}{{Cattalini}, C.}, \bibinfo{author}{{Chaudhary}, A.}, \bibinfo{author}{{Chen}, R.}, \bibinfo{author}{{D'Amico}, N.}, \bibinfo{author}{{Kedziora-Chudczer}, L.}, \bibinfo{author}{{Cornwell}, T.}, \bibinfo{author}{{George}, R.}, \bibinfo{author}{{Hampson}, G.}, \bibinfo{author}{{Hepburn}, M.}, \bibinfo{author}{{Jameson}, A.}, \bibinfo{author}{{Keith}, M.}, \bibinfo{author}{{Kelly}, T.}, \bibinfo{author}{{Kosmynin}, A.}, \bibinfo{author}{{Lenc}, E.}, \bibinfo{author}{{Lorimer}, D.}, \bibinfo{author}{{Love}, C.}, \bibinfo{author}{{Lyne}, A.},
  \bibinfo{author}{{McIntyre}, V.}, \bibinfo{author}{{Morrissey}, J.}, \bibinfo{author}{{Pienaar}, M.}, \bibinfo{author}{{Reynolds}, J.}, \bibinfo{author}{{Ryder}, G.}, \bibinfo{author}{{Sarkissian}, J.}, \bibinfo{author}{{Stevenson}, A.}, \bibinfo{author}{{Treloar}, A.}, \bibinfo{author}{{van Straten}, W.}, \bibinfo{author}{{Whiting}, M.}, \bibinfo{author}{{Wilson}, G.}, \bibinfo{year}{2011}.
\newblock \bibinfo{title}{{The Parkes Observatory Pulsar Data Archive}}.
\newblock \bibinfo{journal}{PASA} \bibinfo{volume}{28}, \bibinfo{pages}{202--214}.
\newblock \DOIprefix\doi{10.1071/AS11016}, \href{http://arxiv.org/abs/1105.5746}{{\tt arXiv:1105.5746}}.
\bibitem[{{Hobbs} et~al.(2006){Hobbs}, {Edwards} and {Manchester}}]{hem06}
\bibinfo{author}{{Hobbs}, G.B.}, \bibinfo{author}{{Edwards}, R.T.}, \bibinfo{author}{{Manchester}, R.N.}, \bibinfo{year}{2006}.
\newblock \bibinfo{title}{{TEMPO2, a new pulsar-timing package - I. An overview}}.
\newblock \bibinfo{journal}{MNRAS} \bibinfo{volume}{369}, \bibinfo{pages}{655--672}.
\newblock \DOIprefix\doi{10.1111/j.1365-2966.2006.10302.x}, \href{http://arxiv.org/abs/astro-ph/0603381}{{\tt arXiv:astro-ph/0603381}}.
\bibitem[{{Hong} et~al.(2021){Hong}, {Park} and {Ryoo}}]{hpr21}
\bibinfo{author}{{Hong}, J.H.}, \bibinfo{author}{{Park}, W.}, \bibinfo{author}{{Ryoo}, C.}, \bibinfo{year}{2021}.
\newblock \bibinfo{title}{{An Autonomous Space Navigation System Using Image Sensors.}}
\newblock \bibinfo{journal}{Int. J. Control Autom. Syst.} \bibinfo{volume}{19}, \bibinfo{pages}{2122–2133}.
\bibitem[{{Howard} et~al.(2016){Howard}, {Stovall}, {Dowell}, {Taylor} and {White}}]{Howard2016}
\bibinfo{author}{{Howard}, T.A.}, \bibinfo{author}{{Stovall}, K.}, \bibinfo{author}{{Dowell}, J.}, \bibinfo{author}{{Taylor}, G.B.}, \bibinfo{author}{{White}, S.M.}, \bibinfo{year}{2016}.
\newblock \bibinfo{title}{{Measuring the Magnetic Field of Coronal Mass Ejections Near the Sun Using Pulsars}}.
\newblock \bibinfo{journal}{APJ} \bibinfo{volume}{831}, \bibinfo{pages}{208}.
\newblock \DOIprefix\doi{10.3847/0004-637X/831/2/208}.
\bibitem[{{Hurd}(1974)}]{hurd1974}
\bibinfo{author}{{Hurd}, W.J.}, \bibinfo{year}{1974}.
\newblock \bibinfo{title}{{An Analysis and Demonstration of Clock Synchronization by VLBI}}.
\newblock \bibinfo{journal}{IEEE Transactions on Instrumentation Measurement} \bibinfo{volume}{23}, \bibinfo{pages}{80--89}.
\newblock \DOIprefix\doi{10.1109/TIM.1974.4314222}.
\bibitem[{{Kazantsev} and {Basalaeva}(2022)}]{kb22}
\bibinfo{author}{{Kazantsev}, A.N.}, \bibinfo{author}{{Basalaeva}, M.Y.}, \bibinfo{year}{2022}.
\newblock \bibinfo{title}{{Low-frequency observations of giant pulses from ordinary pulsars}}.
\newblock \bibinfo{journal}{MNRAS} \bibinfo{volume}{513}, \bibinfo{pages}{4332--4340}.
\newblock \DOIprefix\doi{10.1093/mnras/stac473}.
\bibitem[{{Kooi} et~al.(2021){Kooi}, {Ascione}, {Reyes-Rosa}, {Rier} and {Ashas}}]{Kooi2021}
\bibinfo{author}{{Kooi}, J.E.}, \bibinfo{author}{{Ascione}, M.L.}, \bibinfo{author}{{Reyes-Rosa}, L.V.}, \bibinfo{author}{{Rier}, S.K.}, \bibinfo{author}{{Ashas}, M.}, \bibinfo{year}{2021}.
\newblock \bibinfo{title}{{VLA Measurements of Faraday Rotation Through a Coronal Mass Ejection Using Multiple Lines of Sight}}.
\newblock \bibinfo{journal}{Solar Physics} \bibinfo{volume}{296}, \bibinfo{pages}{11}.
\newblock \DOIprefix\doi{10.1007/s11207-020-01755-4}.
\bibitem[{{Kooi} et~al.(2022){Kooi}, {Wexler}, {Jensen}, {Kenny}, {Nieves-Chinchilla}, {Wilson}, {Wood}, {Jian}, {Fung}, {Pevtsov}, {Gopalswamy} and {Manchester}}]{Kooi2022}
\bibinfo{author}{{Kooi}, J.E.}, \bibinfo{author}{{Wexler}, D.B.}, \bibinfo{author}{{Jensen}, E.A.}, \bibinfo{author}{{Kenny}, M.N.}, \bibinfo{author}{{Nieves-Chinchilla}, T.}, \bibinfo{author}{{Wilson}, Lynn~B., I.}, \bibinfo{author}{{Wood}, B.E.}, \bibinfo{author}{{Jian}, L.K.}, \bibinfo{author}{{Fung}, S.F.}, \bibinfo{author}{{Pevtsov}, A.}, \bibinfo{author}{{Gopalswamy}, N.}, \bibinfo{author}{{Manchester}, W.B.}, \bibinfo{year}{2022}.
\newblock \bibinfo{title}{{Modern Faraday Rotation Studies to Probe the Solar Wind}}.
\newblock \bibinfo{journal}{Frontiers in Astronomy and Space Sciences} \bibinfo{volume}{9}, \bibinfo{pages}{841866}.
\newblock \DOIprefix\doi{10.3389/fspas.2022.841866}.
\bibitem[{{Kordzanganeh} et~al.(2021){Kordzanganeh}, {Utting} and {Scaife}}]{kas21}
\bibinfo{author}{{Kordzanganeh}, M.}, \bibinfo{author}{{Utting}, A.}, \bibinfo{author}{{Scaife}, A.}, \bibinfo{year}{2021}.
\newblock \bibinfo{title}{{Quantum Machine Learning for Radio Astronomy}}.
\newblock \bibinfo{journal}{arXiv e-prints} , \bibinfo{pages}{arXiv:2112.02655}\DOIprefix\doi{10.48550/arXiv.2112.02655}, \href{http://arxiv.org/abs/2112.02655}{{\tt arXiv:2112.02655}}.
\bibitem[{{Kramer} et~al.(2021){Kramer}, {Stairs}, {Manchester}, {Wex}, {Deller}, {Coles}, {Ali}, {Burgay}, {Camilo}, {Cognard}, {Damour}, {Desvignes}, {Ferdman}, {Freire}, {Grondin}, {Guillemot}, {Hobbs}, {Janssen}, {Karuppusamy}, {Lorimer}, {Lyne}, {McKee}, {McLaughlin}, {M{\"u}nch}, {Perera}, {Pol}, {Possenti}, {Sarkissian}, {Stappers} and {Theureau}}]{Kramer2021}
\bibinfo{author}{{Kramer}, M.}, \bibinfo{author}{{Stairs}, I.H.}, \bibinfo{author}{{Manchester}, R.N.}, \bibinfo{author}{{Wex}, N.}, \bibinfo{author}{{Deller}, A.T.}, \bibinfo{author}{{Coles}, W.A.}, \bibinfo{author}{{Ali}, M.}, \bibinfo{author}{{Burgay}, M.}, \bibinfo{author}{{Camilo}, F.}, \bibinfo{author}{{Cognard}, I.}, \bibinfo{author}{{Damour}, T.}, \bibinfo{author}{{Desvignes}, G.}, \bibinfo{author}{{Ferdman}, R.D.}, \bibinfo{author}{{Freire}, P.C.C.}, \bibinfo{author}{{Grondin}, S.}, \bibinfo{author}{{Guillemot}, L.}, \bibinfo{author}{{Hobbs}, G.B.}, \bibinfo{author}{{Janssen}, G.}, \bibinfo{author}{{Karuppusamy}, R.}, \bibinfo{author}{{Lorimer}, D.R.}, \bibinfo{author}{{Lyne}, A.G.}, \bibinfo{author}{{McKee}, J.W.}, \bibinfo{author}{{McLaughlin}, M.}, \bibinfo{author}{{M{\"u}nch}, L.E.}, \bibinfo{author}{{Perera}, B.B.P.}, \bibinfo{author}{{Pol}, N.}, \bibinfo{author}{{Possenti}, A.}, \bibinfo{author}{{Sarkissian}, J.}, \bibinfo{author}{{Stappers}, B.W.}, \bibinfo{author}{{Theureau}, G.},
  \bibinfo{year}{2021}.
\newblock \bibinfo{title}{{Strong-Field Gravity Tests with the Double Pulsar}}.
\newblock \bibinfo{journal}{Physical Review X} \bibinfo{volume}{11}, \bibinfo{pages}{041050}.
\newblock \DOIprefix\doi{10.1103/PhysRevX.11.041050}, \href{http://arxiv.org/abs/2112.06795}{{\tt arXiv:2112.06795}}.
\bibitem[{{Lanman} et~al.(2024){Lanman}, {Andrew}, {Lazda}, {Shah}, {Amiri}, {Balasubramanian}, {Bandura}, {Boyle}, {Brar}, {Carlson}, {Cliche}, {Gusinskaia}, {Hendricksen}, {Kaczmarek}, {Landecker}, {Leung}, {Mckinven}, {Mena-Parra}, {Milutinovic}, {Nimmo}, {Pearlman}, {Renard}, {Rahman}, {Shaw}, {Siegel}, {Smegal}, {Cassanelli}, {Chatterjee}, {Curtin}, {Dobbs}, {Dong}, {Halpern}, {Hopkins}, {Kaspi}, {Khairy}, {Masui}, {Meyers}, {Michilli}, {Petroff}, {Pinsonneault-Marotte}, {Pleunis}, {Rafiei-Ravandi}, {Shin}, {Smith}, {Vanderlinde} and {Zegmott}}]{Lanman2024}
\bibinfo{author}{{Lanman}, A.E.}, \bibinfo{author}{{Andrew}, S.}, \bibinfo{author}{{Lazda}, M.}, \bibinfo{author}{{Shah}, V.}, \bibinfo{author}{{Amiri}, M.}, \bibinfo{author}{{Balasubramanian}, A.}, \bibinfo{author}{{Bandura}, K.}, \bibinfo{author}{{Boyle}, P.J.}, \bibinfo{author}{{Brar}, C.}, \bibinfo{author}{{Carlson}, M.}, \bibinfo{author}{{Cliche}, J.F.}, \bibinfo{author}{{Gusinskaia}, N.}, \bibinfo{author}{{Hendricksen}, I.T.}, \bibinfo{author}{{Kaczmarek}, J.F.}, \bibinfo{author}{{Landecker}, T.}, \bibinfo{author}{{Leung}, C.}, \bibinfo{author}{{Mckinven}, R.}, \bibinfo{author}{{Mena-Parra}, J.}, \bibinfo{author}{{Milutinovic}, N.}, \bibinfo{author}{{Nimmo}, K.}, \bibinfo{author}{{Pearlman}, A.B.}, \bibinfo{author}{{Renard}, A.}, \bibinfo{author}{{Rahman}, M.}, \bibinfo{author}{{Shaw}, J.R.}, \bibinfo{author}{{Siegel}, S.R.}, \bibinfo{author}{{Smegal}, R.J.}, \bibinfo{author}{{Cassanelli}, T.}, \bibinfo{author}{{Chatterjee}, S.}, \bibinfo{author}{{Curtin}, A.P.}, \bibinfo{author}{{Dobbs}, M.},
  \bibinfo{author}{{Dong}, F.A.}, \bibinfo{author}{{Halpern}, M.}, \bibinfo{author}{{Hopkins}, H.}, \bibinfo{author}{{Kaspi}, V.M.}, \bibinfo{author}{{Khairy}, K.}, \bibinfo{author}{{Masui}, K.W.}, \bibinfo{author}{{Meyers}, B.W.}, \bibinfo{author}{{Michilli}, D.}, \bibinfo{author}{{Petroff}, E.}, \bibinfo{author}{{Pinsonneault-Marotte}, T.}, \bibinfo{author}{{Pleunis}, Z.}, \bibinfo{author}{{Rafiei-Ravandi}, M.}, \bibinfo{author}{{Shin}, K.}, \bibinfo{author}{{Smith}, K.}, \bibinfo{author}{{Vanderlinde}, K.}, \bibinfo{author}{{Zegmott}, T.J.}, \bibinfo{year}{2024}.
\newblock \bibinfo{title}{{CHIME/FRB Outriggers: KKO Station System and Commissioning Results}}.
\newblock \bibinfo{journal}{arXiv e-prints} , \bibinfo{pages}{arXiv:2402.07898}\DOIprefix\doi{10.48550/arXiv.2402.07898}, \href{http://arxiv.org/abs/2402.07898}{{\tt arXiv:2402.07898}}.
\bibitem[{{Leblanc} et~al.(1998){Leblanc}, {Dulk} and {Bougeret}}]{Leblanc1998}
\bibinfo{author}{{Leblanc}, Y.}, \bibinfo{author}{{Dulk}, G.A.}, \bibinfo{author}{{Bougeret}, J.L.}, \bibinfo{year}{1998}.
\newblock \bibinfo{title}{{Tracing the Electron Density from the Corona to 1 au}}.
\newblock \bibinfo{journal}{Solar Physics} \bibinfo{volume}{183}, \bibinfo{pages}{165--180}.
\newblock \DOIprefix\doi{10.1023/A:1005049730506}.
\bibitem[{{Lee} and {Cleaver}(2015)}]{lc2015}
\bibinfo{author}{{Lee}, J.S.}, \bibinfo{author}{{Cleaver}, G.B.}, \bibinfo{year}{2015}.
\newblock \bibinfo{title}{{The Cosmic Microwave Background Radiation Power Spectrum as a Random Bit Generator for Symmetric and Asymmetric-Key Cryptography}}.
\newblock \bibinfo{journal}{arXiv e-prints} , \bibinfo{pages}{arXiv:1511.02511}\DOIprefix\doi{10.48550/arXiv.1511.02511}, \href{http://arxiv.org/abs/1511.02511}{{\tt arXiv:1511.02511}}.
\bibitem[{{Liu} et~al.(2019){Liu}, {Young}, {Wharton}, {Blackburn}, {Cappallo}, {Chatterjee}, {Cordes}, {Crew}, {Desvignes}, {Doeleman}, {Eatough}, {Falcke}, {Goddi}, {Johnson}, {Johnston}, {Karuppusamy}, {Kramer}, {Matthews}, {Ransom}, {Rezzolla}, {Rottmann}, {Tilanus} and {Torne}}]{Liu+2019}
\bibinfo{author}{{Liu}, K.}, \bibinfo{author}{{Young}, A.}, \bibinfo{author}{{Wharton}, R.}, \bibinfo{author}{{Blackburn}, L.}, \bibinfo{author}{{Cappallo}, R.}, \bibinfo{author}{{Chatterjee}, S.}, \bibinfo{author}{{Cordes}, J.M.}, \bibinfo{author}{{Crew}, G.B.}, \bibinfo{author}{{Desvignes}, G.}, \bibinfo{author}{{Doeleman}, S.S.}, \bibinfo{author}{{Eatough}, R.P.}, \bibinfo{author}{{Falcke}, H.}, \bibinfo{author}{{Goddi}, C.}, \bibinfo{author}{{Johnson}, M.D.}, \bibinfo{author}{{Johnston}, S.}, \bibinfo{author}{{Karuppusamy}, R.}, \bibinfo{author}{{Kramer}, M.}, \bibinfo{author}{{Matthews}, L.D.}, \bibinfo{author}{{Ransom}, S.M.}, \bibinfo{author}{{Rezzolla}, L.}, \bibinfo{author}{{Rottmann}, H.}, \bibinfo{author}{{Tilanus}, R.P.J.}, \bibinfo{author}{{Torne}, P.}, \bibinfo{year}{2019}.
\newblock \bibinfo{title}{{Detection of Pulses from the Vela Pulsar at Millimeter Wavelengths with Phased ALMA}}.
\newblock \bibinfo{journal}{APJL} \bibinfo{volume}{885}, \bibinfo{pages}{L10}.
\newblock \DOIprefix\doi{10.3847/2041-8213/ab4da8}, \href{http://arxiv.org/abs/1910.07974}{{\tt arXiv:1910.07974}}.
\bibitem[{{Liu} et~al.(2022a){Liu}, {Zhu}, {Antoniadis}, {Liu}, {Zhang} and {Jiang}}]{lzl+22}
\bibinfo{author}{{Liu}, N.}, \bibinfo{author}{{Zhu}, Z.}, \bibinfo{author}{{Antoniadis}, J.}, \bibinfo{author}{{Liu}, J.C.}, \bibinfo{author}{{Zhang}, H.}, \bibinfo{author}{{Jiang}, N.}, \bibinfo{year}{2022}a.
\newblock \bibinfo{title}{{Comparison of dynamical and kinematic reference frames via pulsar positions from timing, Gaia, and interferometric astrometry}}.
\newblock \bibinfo{journal}{arXiv e-prints} , \bibinfo{pages}{arXiv:2212.07178}\DOIprefix\doi{10.48550/arXiv.2212.07178}, \href{http://arxiv.org/abs/2212.07178}{{\tt arXiv:2212.07178}}.
\bibitem[{{Liu} et~al.(2022b){Liu}, {Verbiest}, {Main}, {Wu}, {Ambalappat}, {Champion}, {Cognard}, {Guillemot}, {Gaikwad}, {Janssen}, {Kramer}, {Keith}, {Karuppusamy}, {K{\"u}nkel}, {Liu}, {McKee}, {Mickaliger}, {Stappers}, {Shaifullah} and {Theureau}}]{lv+22}
\bibinfo{author}{{Liu}, Y.}, \bibinfo{author}{{Verbiest}, J.P.W.}, \bibinfo{author}{{Main}, R.A.}, \bibinfo{author}{{Wu}, Z.}, \bibinfo{author}{{Ambalappat}, K.M.}, \bibinfo{author}{{Champion}, D.J.}, \bibinfo{author}{{Cognard}, I.}, \bibinfo{author}{{Guillemot}, L.}, \bibinfo{author}{{Gaikwad}, M.}, \bibinfo{author}{{Janssen}, G.H.}, \bibinfo{author}{{Kramer}, M.}, \bibinfo{author}{{Keith}, M.J.}, \bibinfo{author}{{Karuppusamy}, R.}, \bibinfo{author}{{K{\"u}nkel}, L.}, \bibinfo{author}{{Liu}, K.}, \bibinfo{author}{{McKee}, J.W.}, \bibinfo{author}{{Mickaliger}, M.B.}, \bibinfo{author}{{Stappers}, B.W.}, \bibinfo{author}{{Shaifullah}, G.M.}, \bibinfo{author}{{Theureau}, G.}, \bibinfo{year}{2022}b.
\newblock \bibinfo{title}{{Long-term scintillation studies of EPTA pulsars. I. Observations and basic results}}.
\newblock \bibinfo{journal}{AAP} \bibinfo{volume}{664}, \bibinfo{pages}{A116}.
\newblock \DOIprefix\doi{10.1051/0004-6361/202142552}, \href{http://arxiv.org/abs/2203.16950}{{\tt arXiv:2203.16950}}.
\bibitem[{{Lorimer} and {Kramer}(2012)}]{lk12}
\bibinfo{author}{{Lorimer}, D.R.}, \bibinfo{author}{{Kramer}, M.}, \bibinfo{year}{2012}.
\newblock \bibinfo{title}{{Handbook of Pulsar Astronomy}}.
\bibitem[{{Madison} et~al.(2013){Madison}, {Chatterjee} and {Cordes}}]{mcc13}
\bibinfo{author}{{Madison}, D.R.}, \bibinfo{author}{{Chatterjee}, S.}, \bibinfo{author}{{Cordes}, J.M.}, \bibinfo{year}{2013}.
\newblock \bibinfo{title}{{The Benefits of VLBI Astrometry to Pulsar Timing Array Searches for Gravitational Radiation}}.
\newblock \bibinfo{journal}{APJ} \bibinfo{volume}{777}, \bibinfo{pages}{104}.
\newblock \DOIprefix\doi{10.1088/0004-637X/777/2/104}, \href{http://arxiv.org/abs/1210.2469}{{\tt arXiv:1210.2469}}.
\bibitem[{{Madison} et~al.(2019){Madison}, {Cordes}, {Arzoumanian}, {Chatterjee}, {Crowter}, {DeCesar}, {Demorest}, {Dolch}, {Ellis}, {Ferdman}, {Ferrara}, {Fonseca}, {Gentile}, {Jones}, {Jones}, {Lam}, {Levin}, {Lorimer}, {Lynch}, {McLaughlin}, {Mingarelli}, {Ng}, {Nice}, {Pennucci}, {Ransom}, {Ray}, {Spiewak}, {Stairs}, {Stovall}, {Swiggum} and {Zhu}}]{mca+2019}
\bibinfo{author}{{Madison}, D.R.}, \bibinfo{author}{{Cordes}, J.M.}, \bibinfo{author}{{Arzoumanian}, Z.}, \bibinfo{author}{{Chatterjee}, S.}, \bibinfo{author}{{Crowter}, K.}, \bibinfo{author}{{DeCesar}, M.E.}, \bibinfo{author}{{Demorest}, P.B.}, \bibinfo{author}{{Dolch}, T.}, \bibinfo{author}{{Ellis}, J.A.}, \bibinfo{author}{{Ferdman}, R.D.}, \bibinfo{author}{{Ferrara}, E.C.}, \bibinfo{author}{{Fonseca}, E.}, \bibinfo{author}{{Gentile}, P.A.}, \bibinfo{author}{{Jones}, G.}, \bibinfo{author}{{Jones}, M.L.}, \bibinfo{author}{{Lam}, M.T.}, \bibinfo{author}{{Levin}, L.}, \bibinfo{author}{{Lorimer}, D.R.}, \bibinfo{author}{{Lynch}, R.S.}, \bibinfo{author}{{McLaughlin}, M.A.}, \bibinfo{author}{{Mingarelli}, C.M.F.}, \bibinfo{author}{{Ng}, C.}, \bibinfo{author}{{Nice}, D.J.}, \bibinfo{author}{{Pennucci}, T.T.}, \bibinfo{author}{{Ransom}, S.M.}, \bibinfo{author}{{Ray}, P.S.}, \bibinfo{author}{{Spiewak}, R.}, \bibinfo{author}{{Stairs}, I.H.}, \bibinfo{author}{{Stovall}, K.}, \bibinfo{author}{{Swiggum}, J.K.},
  \bibinfo{author}{{Zhu}, W.W.}, \bibinfo{year}{2019}.
\newblock \bibinfo{title}{{The NANOGrav 11 yr Data Set: Solar Wind Sounding through Pulsar Timing}}.
\newblock \bibinfo{journal}{APJ} \bibinfo{volume}{872}, \bibinfo{pages}{150}.
\newblock \DOIprefix\doi{10.3847/1538-4357/ab01fd}, \href{http://arxiv.org/abs/1808.07078}{{\tt arXiv:1808.07078}}.
\bibitem[{{Mahrous} et~al.(2018){Mahrous}, {Alielden}, {Vr{\v{s}}nak} and {Youssef}}]{Mahrous2018}
\bibinfo{author}{{Mahrous}, A.}, \bibinfo{author}{{Alielden}, K.}, \bibinfo{author}{{Vr{\v{s}}nak}, B.}, \bibinfo{author}{{Youssef}, M.}, \bibinfo{year}{2018}.
\newblock \bibinfo{title}{{Type II solar radio burst band-splitting: Measure of coronal magnetic field strength}}.
\newblock \bibinfo{journal}{Journal of Atmospheric and Solar-Terrestrial Physics} \bibinfo{volume}{172}, \bibinfo{pages}{75--82}.
\newblock \DOIprefix\doi{10.1016/j.jastp.2018.03.018}.
\bibitem[{{Manchester}(2017)}]{man2017}
\bibinfo{author}{{Manchester}, R.N.}, \bibinfo{year}{2017}.
\newblock \bibinfo{title}{{Millisecond Pulsars, their Evolution and Applications}}.
\newblock \bibinfo{journal}{Journal of Astrophysics and Astronomy} \bibinfo{volume}{38}, \bibinfo{pages}{42}.
\newblock \DOIprefix\doi{10.1007/s12036-017-9469-2}, \href{http://arxiv.org/abs/1709.09434}{{\tt arXiv:1709.09434}}.
\bibitem[{{Manchester} et~al.(2005){Manchester}, {Hobbs}, {Teoh} and {Hobbs}}]{mhth05}
\bibinfo{author}{{Manchester}, R.N.}, \bibinfo{author}{{Hobbs}, G.B.}, \bibinfo{author}{{Teoh}, A.}, \bibinfo{author}{{Hobbs}, M.}, \bibinfo{year}{2005}.
\newblock \bibinfo{title}{{The Australia Telescope National Facility Pulsar Catalogue}}.
\newblock \bibinfo{journal}{AJ} \bibinfo{volume}{129}, \bibinfo{pages}{1993--2006}.
\newblock \DOIprefix\doi{10.1086/428488}, \href{http://arxiv.org/abs/astro-ph/0412641}{{\tt arXiv:astro-ph/0412641}}.
\bibitem[{Manoharan(2012)}]{Manoharan2012}
\bibinfo{author}{Manoharan, P.K.}, \bibinfo{year}{2012}.
\newblock \bibinfo{title}{Three-dimensional evolution of solar wind during solar cycles 22–24}.
\newblock \bibinfo{journal}{The Astrophysical Journal} \bibinfo{volume}{751}, \bibinfo{pages}{128}.
\newblock \URLprefix \url{https://dx.doi.org/10.1088/0004-637X/751/2/128}, \DOIprefix\doi{10.1088/0004-637X/751/2/128}.
\bibitem[{{Miles} et~al.(2023){Miles}, {Shannon}, {Bailes}, {Reardon}, {Keith}, {Cameron}, {Parthasarathy}, {Shamohammadi}, {Spiewak}, {van Straten}, {Buchner}, {Camilo}, {Geyer}, {Karastergiou}, {Kramer}, {Serylak}, {Theureau} and {Venkatraman Krishnan}}]{msb+23}
\bibinfo{author}{{Miles}, M.T.}, \bibinfo{author}{{Shannon}, R.M.}, \bibinfo{author}{{Bailes}, M.}, \bibinfo{author}{{Reardon}, D.J.}, \bibinfo{author}{{Keith}, M.J.}, \bibinfo{author}{{Cameron}, A.D.}, \bibinfo{author}{{Parthasarathy}, A.}, \bibinfo{author}{{Shamohammadi}, M.}, \bibinfo{author}{{Spiewak}, R.}, \bibinfo{author}{{van Straten}, W.}, \bibinfo{author}{{Buchner}, S.}, \bibinfo{author}{{Camilo}, F.}, \bibinfo{author}{{Geyer}, M.}, \bibinfo{author}{{Karastergiou}, A.}, \bibinfo{author}{{Kramer}, M.}, \bibinfo{author}{{Serylak}, M.}, \bibinfo{author}{{Theureau}, G.}, \bibinfo{author}{{Venkatraman Krishnan}, V.}, \bibinfo{year}{2023}.
\newblock \bibinfo{title}{{The MeerKAT Pulsar Timing Array: first data release}}.
\newblock \bibinfo{journal}{MNRAS} \bibinfo{volume}{519}, \bibinfo{pages}{3976--3991}.
\newblock \DOIprefix\doi{10.1093/mnras/stac3644}, \href{http://arxiv.org/abs/2212.04648}{{\tt arXiv:2212.04648}}.
\bibitem[{{Miller} et~al.(2009){Miller}, {Jenet}, {Zermeno} and {Stovall}}]{mjzs09}
\bibinfo{author}{{Miller}, A.F.}, \bibinfo{author}{{Jenet}, F.A.}, \bibinfo{author}{{Zermeno}, A.}, \bibinfo{author}{{Stovall}, K.}, \bibinfo{year}{2009}.
\newblock \bibinfo{title}{{The Arecibo Remote Command Center: Creating an Inspiring Environment for Astrophysics}}, in: \bibinfo{booktitle}{American Astronomical Society Meeting Abstracts \#213}, p. \bibinfo{pages}{431.04}.
\bibitem[{{Morgan} et~al.(2022){Morgan}, {Chhetri} and {Ekers}}]{Morgan2022}
\bibinfo{author}{{Morgan}, J.S.}, \bibinfo{author}{{Chhetri}, R.}, \bibinfo{author}{{Ekers}, R.}, \bibinfo{year}{2022}.
\newblock \bibinfo{title}{{A census of compact sources at 162 MHz: First data release from the MWA Phase II IPS Survey}}.
\newblock \bibinfo{journal}{PASA} \bibinfo{volume}{39}, \bibinfo{pages}{e063}.
\newblock \DOIprefix\doi{10.1017/pasa.2022.56}, \href{http://arxiv.org/abs/2210.05400}{{\tt arXiv:2210.05400}}.
\bibitem[{{Narayan}(1992)}]{nar92}
\bibinfo{author}{{Narayan}, R.}, \bibinfo{year}{1992}.
\newblock \bibinfo{title}{{The Physics of Pulsar Scintillation}}.
\newblock \bibinfo{journal}{Philosophical Transactions of the Royal Society of London Series A} \bibinfo{volume}{341}, \bibinfo{pages}{151--165}.
\newblock \DOIprefix\doi{10.1098/rsta.1992.0090}.
\bibitem[{{Oberoi} and {Lonsdale}(2012)}]{Oberoi2012}
\bibinfo{author}{{Oberoi}, D.}, \bibinfo{author}{{Lonsdale}, C.J.}, \bibinfo{year}{2012}.
\newblock \bibinfo{title}{{Media responsible for Faraday rotation: A review}}.
\newblock \bibinfo{journal}{Radio Science} \bibinfo{volume}{47}, \bibinfo{pages}{RS0K08}.
\newblock \DOIprefix\doi{10.1029/2012RS004992}.
\bibitem[{Ontiveros and Vourlidas(2009)}]{Ontiveros2009}
\bibinfo{author}{Ontiveros, V.}, \bibinfo{author}{Vourlidas, A.}, \bibinfo{year}{2009}.
\newblock \bibinfo{title}{Quantitative measurements of coronal mass ejection-driven shocks from lasco observations}.
\newblock \bibinfo{journal}{The Astrophysical Journal} \bibinfo{volume}{693}, \bibinfo{pages}{267}.
\newblock \URLprefix \url{https://dx.doi.org/10.1088/0004-637X/693/1/267}, \DOIprefix\doi{10.1088/0004-637X/693/1/267}.
\bibitem[{Paciga et~al.(2011)Paciga, Chang, Gupta, Nityanada, Odegova, Pen, Peterson, Roy and Sigurdson}]{Paciga2011}
\bibinfo{author}{Paciga, G.}, \bibinfo{author}{Chang, T.C.}, \bibinfo{author}{Gupta, Y.}, \bibinfo{author}{Nityanada, R.}, \bibinfo{author}{Odegova, J.}, \bibinfo{author}{Pen, U.L.}, \bibinfo{author}{Peterson, J.B.}, \bibinfo{author}{Roy, J.}, \bibinfo{author}{Sigurdson, K.}, \bibinfo{year}{2011}.
\newblock \bibinfo{title}{{The GMRT Epoch of Reionization experiment: a new upper limit on the neutral hydrogen power spectrum at z$\approx$ 8.6}}.
\newblock \bibinfo{journal}{Monthly Notices of the Royal Astronomical Society} \bibinfo{volume}{413}, \bibinfo{pages}{1174--1183}.
\bibitem[{{Palmer} and {Holmes}(2018)}]{ph18}
\bibinfo{author}{{Palmer}, D.M.}, \bibinfo{author}{{Holmes}, R.M.}, \bibinfo{year}{2018}.
\newblock \bibinfo{title}{{ELROI: A License Plate For Your Satellite}}.
\newblock \bibinfo{journal}{arXiv e-prints} , \bibinfo{pages}{arXiv:1802.04820}\DOIprefix\doi{10.48550/arXiv.1802.04820}, \href{http://arxiv.org/abs/1802.04820}{{\tt arXiv:1802.04820}}.
\bibitem[{{Parthasarathy} et~al.(2021){Parthasarathy}, {Bailes}, {Shannon}, {van Straten}, {Os{\l}owski}, {Johnston}, {Spiewak}, {Reardon}, {Kramer}, {Venkatraman Krishnan}, {Pennucci}, {Abbate}, {Buchner}, {Camilo}, {Champion}, {Geyer}, {Hugo}, {Jameson}, {Karastergiou}, {Keith} and {Serylak}}]{pbv21}
\bibinfo{author}{{Parthasarathy}, A.}, \bibinfo{author}{{Bailes}, M.}, \bibinfo{author}{{Shannon}, R.M.}, \bibinfo{author}{{van Straten}, W.}, \bibinfo{author}{{Os{\l}owski}, S.}, \bibinfo{author}{{Johnston}, S.}, \bibinfo{author}{{Spiewak}, R.}, \bibinfo{author}{{Reardon}, D.J.}, \bibinfo{author}{{Kramer}, M.}, \bibinfo{author}{{Venkatraman Krishnan}, V.}, \bibinfo{author}{{Pennucci}, T.T.}, \bibinfo{author}{{Abbate}, F.}, \bibinfo{author}{{Buchner}, S.}, \bibinfo{author}{{Camilo}, F.}, \bibinfo{author}{{Champion}, D.J.}, \bibinfo{author}{{Geyer}, M.}, \bibinfo{author}{{Hugo}, B.}, \bibinfo{author}{{Jameson}, A.}, \bibinfo{author}{{Karastergiou}, A.}, \bibinfo{author}{{Keith}, M.J.}, \bibinfo{author}{{Serylak}, M.}, \bibinfo{year}{2021}.
\newblock \bibinfo{title}{{Measurements of pulse jitter and single-pulse variability in millisecond pulsars using MeerKAT}}.
\newblock \bibinfo{journal}{MNRAS} \bibinfo{volume}{502}, \bibinfo{pages}{407--422}.
\newblock \DOIprefix\doi{10.1093/mnras/stab037}, \href{http://arxiv.org/abs/2101.08531}{{\tt arXiv:2101.08531}}.
\bibitem[{{Pearlman} et~al.(2019){Pearlman}, {Majid} and {Prince}}]{pam+19}
\bibinfo{author}{{Pearlman}, A.B.}, \bibinfo{author}{{Majid}, W.A.}, \bibinfo{author}{{Prince}, T.A.}, \bibinfo{year}{2019}.
\newblock \bibinfo{title}{{Observations of Radio Magnetars with the Deep Space Network}}.
\newblock \bibinfo{journal}{Advances in Astronomy} \bibinfo{volume}{2019}, \bibinfo{pages}{6325183}.
\newblock \DOIprefix\doi{10.1155/2019/6325183}, \href{http://arxiv.org/abs/1902.10712}{{\tt arXiv:1902.10712}}.
\bibitem[{{Pimbblet} and {Bulmer}(2005)}]{pkb05}
\bibinfo{author}{{Pimbblet}, K.A.}, \bibinfo{author}{{Bulmer}, M.}, \bibinfo{year}{2005}.
\newblock \bibinfo{title}{{Random Numbers from Astronomical Imaging}}.
\newblock \bibinfo{journal}{PASA} \bibinfo{volume}{22}, \bibinfo{pages}{1--5}.
\newblock \DOIprefix\doi{10.1071/AS04043}, \href{http://arxiv.org/abs/astro-ph/0408281}{{\tt arXiv:astro-ph/0408281}}.
\bibitem[{Qiu et~al.(2021)Qiu, Yin, Zhang, Luo, Wang, Liu, Yao, Zhan, Fuhr and King}]{qiu2021pulsar}
\bibinfo{author}{Qiu, W.}, \bibinfo{author}{Yin, H.}, \bibinfo{author}{Zhang, L.}, \bibinfo{author}{Luo, X.}, \bibinfo{author}{Wang, W.}, \bibinfo{author}{Liu, Y.}, \bibinfo{author}{Yao, W.}, \bibinfo{author}{Zhan, L.}, \bibinfo{author}{Fuhr, P.L.}, \bibinfo{author}{King, T.J.}, \bibinfo{year}{2021}.
\newblock \bibinfo{title}{Pulsar based timing for grid synchronization}.
\newblock \bibinfo{journal}{IEEE Transactions on Industry Applications} \bibinfo{volume}{57}, \bibinfo{pages}{2067--2076}.
\bibitem[{{Ray} et~al.(2017){Ray}, {Wood} and {Wolff}}]{rww2017}
\bibinfo{author}{{Ray}, P.S.}, \bibinfo{author}{{Wood}, K.S.}, \bibinfo{author}{{Wolff}, M.T.}, \bibinfo{year}{2017}.
\newblock \bibinfo{title}{{Characterization of Pulsar Sources for X-ray Navigation}}.
\newblock \bibinfo{journal}{arXiv e-prints} , \bibinfo{pages}{arXiv:1711.08507}\href{http://arxiv.org/abs/1711.08507}{{\tt arXiv:1711.08507}}.
\bibitem[{{Reardon} et~al.(2023){Reardon}, {Zic}, {Shannon}, {Di Marco}, {Hobbs}, {Kapur}, {Lower}, {Mandow}, {Middleton}, {Miles}, {Rogers}, {Askew}, {Bailes}, {Bhat}, {Cameron}, {Kerr}, {Kulkarni}, {Manchester}, {Nathan}, {Russell}, {Os{\l}owski} and {Zhu}}]{Reardon2023}
\bibinfo{author}{{Reardon}, D.J.}, \bibinfo{author}{{Zic}, A.}, \bibinfo{author}{{Shannon}, R.M.}, \bibinfo{author}{{Di Marco}, V.}, \bibinfo{author}{{Hobbs}, G.B.}, \bibinfo{author}{{Kapur}, A.}, \bibinfo{author}{{Lower}, M.E.}, \bibinfo{author}{{Mandow}, R.}, \bibinfo{author}{{Middleton}, H.}, \bibinfo{author}{{Miles}, M.T.}, \bibinfo{author}{{Rogers}, A.F.}, \bibinfo{author}{{Askew}, J.}, \bibinfo{author}{{Bailes}, M.}, \bibinfo{author}{{Bhat}, N.D.R.}, \bibinfo{author}{{Cameron}, A.}, \bibinfo{author}{{Kerr}, M.}, \bibinfo{author}{{Kulkarni}, A.}, \bibinfo{author}{{Manchester}, R.N.}, \bibinfo{author}{{Nathan}, R.S.}, \bibinfo{author}{{Russell}, C.J.}, \bibinfo{author}{{Os{\l}owski}, S.}, \bibinfo{author}{{Zhu}, X.J.}, \bibinfo{year}{2023}.
\newblock \bibinfo{title}{{The Gravitational-wave Background Null Hypothesis: Characterizing Noise in Millisecond Pulsar Arrival Times with the Parkes Pulsar Timing Array}}.
\newblock \bibinfo{journal}{APJL} \bibinfo{volume}{951}, \bibinfo{pages}{L7}.
\newblock \DOIprefix\doi{10.3847/2041-8213/acdd03}, \href{http://arxiv.org/abs/2306.16229}{{\tt arXiv:2306.16229}}.
\bibitem[{{Redman} and {Rankin}(2009)}]{rsr09}
\bibinfo{author}{{Redman}, S.L.}, \bibinfo{author}{{Rankin}, J.M.}, \bibinfo{year}{2009}.
\newblock \bibinfo{title}{{On the randomness of pulsar nulls}}.
\newblock \bibinfo{journal}{MNRAS} \bibinfo{volume}{395}, \bibinfo{pages}{1529--1532}.
\newblock \DOIprefix\doi{10.1111/j.1365-2966.2009.14632.x}.
\bibitem[{Rocha et~al.(2022)Rocha, Mendonca and Ramos}]{Rocha2022}
\bibinfo{author}{Rocha, B.S.}, \bibinfo{author}{Mendonca, F.A.}, \bibinfo{author}{Ramos, R.V.}, \bibinfo{year}{2022}.
\newblock \bibinfo{title}{An algorithm to decrease the key distribution error rate using pulsars}, in: \bibinfo{booktitle}{2022 Workshop on Communication Networks and Power Systems (WCNPS)}, pp. \bibinfo{pages}{1--5}.
\newblock \DOIprefix\doi{10.1109/WCNPS56355.2022.9969705}.
\bibitem[{{Rosen} et~al.(2013){Rosen}, {Swiggum}, {McLaughlin}, {Lorimer}, {Yun}, {Heatherly}, {Boyles}, {Lynch}, {Kondratiev}, {Scoles}, {Ransom}, {Moniot}, {Cottrill}, {Weaver}, {Snider}, {Thompson}, {Raycraft}, {Dudenhoefer}, {Allphin}, {Thorley}, {Meadows}, {Marchiny}, {Liska}, {O'Dwyer}, {Butler}, {Bloxton}, {Mabry}, {Abate}, {Boothe}, {Pritt}, {Alberth}, {Green}, {Crowley}, {Agee}, {Nagley}, {Sargent}, {Hinson}, {Smith}, {McNeely}, {Quigley}, {Pennington}, {Chen}, {Maynard}, {Loope}, {Bielski}, {McGough}, {Gural}, {Colvin}, {Tso}, {Ewen}, {Zhang}, {Ciccarella}, {Bukowski}, {Novotny}, {Gore}, {Sarver}, {Johnson}, {Cunningham}, {Collins}, {Gardner}, {Monteleone}, {Hall}, {Schweinhagen}, {Ayers}, {Jay}, {Uosseph}, {Dunkum}, {Pal}, {Dydiw}, {Sterling} and {Phan}}]{rsm+13}
\bibinfo{author}{{Rosen}, R.}, \bibinfo{author}{{Swiggum}, J.}, \bibinfo{author}{{McLaughlin}, M.A.}, \bibinfo{author}{{Lorimer}, D.R.}, \bibinfo{author}{{Yun}, M.}, \bibinfo{author}{{Heatherly}, S.A.}, \bibinfo{author}{{Boyles}, J.}, \bibinfo{author}{{Lynch}, R.}, \bibinfo{author}{{Kondratiev}, V.I.}, \bibinfo{author}{{Scoles}, S.}, \bibinfo{author}{{Ransom}, S.M.}, \bibinfo{author}{{Moniot}, M.L.}, \bibinfo{author}{{Cottrill}, A.}, \bibinfo{author}{{Weaver}, M.}, \bibinfo{author}{{Snider}, A.}, \bibinfo{author}{{Thompson}, C.}, \bibinfo{author}{{Raycraft}, M.}, \bibinfo{author}{{Dudenhoefer}, J.}, \bibinfo{author}{{Allphin}, L.}, \bibinfo{author}{{Thorley}, J.}, \bibinfo{author}{{Meadows}, B.}, \bibinfo{author}{{Marchiny}, G.}, \bibinfo{author}{{Liska}, A.}, \bibinfo{author}{{O'Dwyer}, A.M.}, \bibinfo{author}{{Butler}, B.}, \bibinfo{author}{{Bloxton}, S.}, \bibinfo{author}{{Mabry}, H.}, \bibinfo{author}{{Abate}, H.}, \bibinfo{author}{{Boothe}, J.}, \bibinfo{author}{{Pritt}, S.}, \bibinfo{author}{{Alberth},
  J.}, \bibinfo{author}{{Green}, A.}, \bibinfo{author}{{Crowley}, R.J.}, \bibinfo{author}{{Agee}, A.}, \bibinfo{author}{{Nagley}, S.}, \bibinfo{author}{{Sargent}, N.}, \bibinfo{author}{{Hinson}, E.}, \bibinfo{author}{{Smith}, K.}, \bibinfo{author}{{McNeely}, R.}, \bibinfo{author}{{Quigley}, H.}, \bibinfo{author}{{Pennington}, A.}, \bibinfo{author}{{Chen}, S.}, \bibinfo{author}{{Maynard}, T.}, \bibinfo{author}{{Loope}, L.}, \bibinfo{author}{{Bielski}, N.}, \bibinfo{author}{{McGough}, J.R.}, \bibinfo{author}{{Gural}, J.C.}, \bibinfo{author}{{Colvin}, S.}, \bibinfo{author}{{Tso}, S.}, \bibinfo{author}{{Ewen}, Z.}, \bibinfo{author}{{Zhang}, M.}, \bibinfo{author}{{Ciccarella}, N.}, \bibinfo{author}{{Bukowski}, B.}, \bibinfo{author}{{Novotny}, C.B.}, \bibinfo{author}{{Gore}, J.}, \bibinfo{author}{{Sarver}, K.}, \bibinfo{author}{{Johnson}, S.}, \bibinfo{author}{{Cunningham}, H.}, \bibinfo{author}{{Collins}, D.}, \bibinfo{author}{{Gardner}, D.}, \bibinfo{author}{{Monteleone}, A.}, \bibinfo{author}{{Hall}, J.},
  \bibinfo{author}{{Schweinhagen}, R.}, \bibinfo{author}{{Ayers}, J.}, \bibinfo{author}{{Jay}, S.}, \bibinfo{author}{{Uosseph}, B.}, \bibinfo{author}{{Dunkum}, D.}, \bibinfo{author}{{Pal}, J.}, \bibinfo{author}{{Dydiw}, S.}, \bibinfo{author}{{Sterling}, M.}, \bibinfo{author}{{Phan}, E.}, \bibinfo{year}{2013}.
\newblock \bibinfo{title}{{The Pulsar Search Collaboratory: Discovery and Timing of Five New Pulsars}}.
\newblock \bibinfo{journal}{APJ} \bibinfo{volume}{768}, \bibinfo{pages}{85}.
\newblock \DOIprefix\doi{10.1088/0004-637X/768/1/85}, \href{http://arxiv.org/abs/1209.4108}{{\tt arXiv:1209.4108}}.
\bibitem[{{Sagan} et~al.(1972){Sagan}, {Salzman Sagan} and {Drake}}]{ssd+72}
\bibinfo{author}{{Sagan}, C.}, \bibinfo{author}{{Salzman Sagan}, L.}, \bibinfo{author}{{Drake}, F.}, \bibinfo{year}{1972}.
\newblock \bibinfo{title}{{A Message from Earth}}.
\newblock \bibinfo{journal}{Science} \bibinfo{volume}{175}, \bibinfo{pages}{881--884}.
\newblock \DOIprefix\doi{10.1126/science.175.4024.881}.
\bibitem[{{Sarkissian} et~al.(2017){Sarkissian}, {Reynolds}, {Hobbs} and {Harvey-Smith}}]{Sarkissian2017}
\bibinfo{author}{{Sarkissian}, J.M.}, \bibinfo{author}{{Reynolds}, J.E.}, \bibinfo{author}{{Hobbs}, G.}, \bibinfo{author}{{Harvey-Smith}, L.}, \bibinfo{year}{2017}.
\newblock \bibinfo{title}{{One Year of Monitoring the Vela Pulsar Using a Phased Array Feed}}.
\newblock \bibinfo{journal}{PASA} \bibinfo{volume}{34}, \bibinfo{pages}{e027}.
\newblock \DOIprefix\doi{10.1017/pasa.2017.19}, \href{http://arxiv.org/abs/1705.08355}{{\tt arXiv:1705.08355}}.
\bibitem[{Schmidt et~al.(2016)Schmidt, Radke, Camtepe, Foo and Ren}]{srcfr2016}
\bibinfo{author}{Schmidt, D.}, \bibinfo{author}{Radke, K.}, \bibinfo{author}{Camtepe, S.}, \bibinfo{author}{Foo, E.}, \bibinfo{author}{Ren, M.}, \bibinfo{year}{2016}.
\newblock \bibinfo{title}{A survey and analysis of the gnss spoofing threat and countermeasures}.
\newblock \bibinfo{journal}{ACM Comput. Surv.} \bibinfo{volume}{48}.
\newblock \URLprefix \url{https://doi.org/10.1145/2897166}, \DOIprefix\doi{10.1145/2897166}.
\bibitem[{{Schrijver} et~al.(2015){Schrijver}, {Kauristie}, {Aylward}, {Denardini}, {Gibson}, {Glover}, {Gopalswamy}, {Grande}, {Hapgood}, {Heynderickx}, {Jakowski}, {Kalegaev}, {Lapenta}, {Linker}, {Liu}, {Mandrini}, {Mann}, {Nagatsuma}, {Nandy}, {Obara}, {Paul O'Brien}, {Onsager}, {Opgenoorth}, {Terkildsen}, {Valladares} and {Vilmer}}]{Schrijver2015}
\bibinfo{author}{{Schrijver}, C.J.}, \bibinfo{author}{{Kauristie}, K.}, \bibinfo{author}{{Aylward}, A.D.}, \bibinfo{author}{{Denardini}, C.M.}, \bibinfo{author}{{Gibson}, S.E.}, \bibinfo{author}{{Glover}, A.}, \bibinfo{author}{{Gopalswamy}, N.}, \bibinfo{author}{{Grande}, M.}, \bibinfo{author}{{Hapgood}, M.}, \bibinfo{author}{{Heynderickx}, D.}, \bibinfo{author}{{Jakowski}, N.}, \bibinfo{author}{{Kalegaev}, V.V.}, \bibinfo{author}{{Lapenta}, G.}, \bibinfo{author}{{Linker}, J.A.}, \bibinfo{author}{{Liu}, S.}, \bibinfo{author}{{Mandrini}, C.H.}, \bibinfo{author}{{Mann}, I.R.}, \bibinfo{author}{{Nagatsuma}, T.}, \bibinfo{author}{{Nandy}, D.}, \bibinfo{author}{{Obara}, T.}, \bibinfo{author}{{Paul O'Brien}, T.}, \bibinfo{author}{{Onsager}, T.}, \bibinfo{author}{{Opgenoorth}, H.J.}, \bibinfo{author}{{Terkildsen}, M.}, \bibinfo{author}{{Valladares}, C.E.}, \bibinfo{author}{{Vilmer}, N.}, \bibinfo{year}{2015}.
\newblock \bibinfo{title}{{Understanding space weather to shield society: A global road map for 2015-2025 commissioned by COSPAR and ILWS}}.
\newblock \bibinfo{journal}{Advances in Space Research} \bibinfo{volume}{55}, \bibinfo{pages}{2745--2807}.
\newblock \DOIprefix\doi{10.1016/j.asr.2015.03.023}, \href{http://arxiv.org/abs/1503.06135}{{\tt arXiv:1503.06135}}.
\bibitem[{Sheoran et~al.(2023)Sheoran, Pant, Patel and Banerjee}]{Jyoti2023}
\bibinfo{author}{Sheoran, J.}, \bibinfo{author}{Pant, V.}, \bibinfo{author}{Patel, R.}, \bibinfo{author}{Banerjee, D.}, \bibinfo{year}{2023}.
\newblock \bibinfo{title}{Evolution of the thermodynamic properties of a coronal mass ejection in the inner corona}.
\newblock \bibinfo{journal}{Frontiers in Astronomy and Space Sciences} \bibinfo{volume}{10}.
\newblock \URLprefix \url{https://www.frontiersin.org/articles/10.3389/fspas.2023.1092881}, \DOIprefix\doi{10.3389/fspas.2023.1092881}.
\bibitem[{Singh et~al.(2022)Singh, Singh, Singh, Goyal, Raboaca, Verma and Suciu}]{sjs+22}
\bibinfo{author}{Singh, S.}, \bibinfo{author}{Singh, J.}, \bibinfo{author}{Singh, S.}, \bibinfo{author}{Goyal, S.B.}, \bibinfo{author}{Raboaca, M.S.}, \bibinfo{author}{Verma, C.}, \bibinfo{author}{Suciu, G.}, \bibinfo{year}{2022}.
\newblock \bibinfo{title}{Detection and mitigation of gnss spoofing attacks in maritime environments using a genetic algorithm}.
\newblock \bibinfo{journal}{Mathematics} \bibinfo{volume}{10}.
\newblock \URLprefix \url{https://www.mdpi.com/2227-7390/10/21/4097}, \DOIprefix\doi{10.3390/math10214097}.
\bibitem[{{Slabbert} et~al.(2023){Slabbert}, {Lourens} and {Petruccione}}]{slp23}
\bibinfo{author}{{Slabbert}, D.}, \bibinfo{author}{{Lourens}, M.}, \bibinfo{author}{{Petruccione}, F.}, \bibinfo{year}{2023}.
\newblock \bibinfo{title}{{Pulsar Classification: Comparing Quantum Convolutional Neural Networks and Quantum Support Vector Machines}}.
\newblock \bibinfo{journal}{arXiv e-prints} , \bibinfo{pages}{arXiv:2309.15592}\DOIprefix\doi{10.48550/arXiv.2309.15592}, \href{http://arxiv.org/abs/2309.15592}{{\tt arXiv:2309.15592}}.
\bibitem[{{Spiewak} et~al.(2020){Spiewak}, {Flynn}, {Johnston}, {Keane}, {Bailes}, {Barr}, {Bhandari}, {Burgay}, {Jankowski}, {Kramer}, {Morello}, {Possenti} and {Venkatraman Krishnan}}]{Spiewak2020}
\bibinfo{author}{{Spiewak}, R.}, \bibinfo{author}{{Flynn}, C.}, \bibinfo{author}{{Johnston}, S.}, \bibinfo{author}{{Keane}, E.F.}, \bibinfo{author}{{Bailes}, M.}, \bibinfo{author}{{Barr}, E.D.}, \bibinfo{author}{{Bhandari}, S.}, \bibinfo{author}{{Burgay}, M.}, \bibinfo{author}{{Jankowski}, F.}, \bibinfo{author}{{Kramer}, M.}, \bibinfo{author}{{Morello}, V.}, \bibinfo{author}{{Possenti}, A.}, \bibinfo{author}{{Venkatraman Krishnan}, V.}, \bibinfo{year}{2020}.
\newblock \bibinfo{title}{{The SUrvey for pulsars and extragalactic radio bursts V: recent discoveries and full timing solutions}}.
\newblock \bibinfo{journal}{MNRAS} \bibinfo{volume}{496}, \bibinfo{pages}{4836--4848}.
\newblock \DOIprefix\doi{10.1093/mnras/staa1869}, \href{http://arxiv.org/abs/2006.13637}{{\tt arXiv:2006.13637}}.
\bibitem[{{Sreeja}(2016)}]{sre16}
\bibinfo{author}{{Sreeja}, V.}, \bibinfo{year}{2016}.
\newblock \bibinfo{title}{{Impact and mitigation of space weather effects on GNSS receiver performance}}.
\newblock \bibinfo{journal}{Geoscience Letters} \bibinfo{volume}{3}, \bibinfo{pages}{24}.
\newblock \DOIprefix\doi{10.1186/s40562-016-0057-0}.
\bibitem[{{Staelin}(1969)}]{staelin69}
\bibinfo{author}{{Staelin}, D.H.}, \bibinfo{year}{1969}.
\newblock \bibinfo{title}{{Fast folding algorithm for detection of periodic pulse trains.}}
\newblock \bibinfo{journal}{IEEE Proceedings} \bibinfo{volume}{57}, \bibinfo{pages}{724--725}.
\newblock \DOIprefix\doi{10.1109/PROC.1969.7051}.
\bibitem[{{Stanciu} et~al.(2011){Stanciu}, {Azou} and {Serbanescu}}]{stanciu11}
\bibinfo{author}{{Stanciu}, M.I.}, \bibinfo{author}{{Azou}, S.}, \bibinfo{author}{{Serbanescu}, A.}, \bibinfo{year}{2011}.
\newblock \bibinfo{title}{{On the Blind Estimation of Chip Time of Time-Hopping Signals Through Minimization of a Multimodal Cost Function}}.
\newblock \bibinfo{journal}{IEEE Transactions on Signal Processing} \bibinfo{volume}{59}, \bibinfo{pages}{842--847}.
\newblock \DOIprefix\doi{10.1109/TSP.2010.2090872}.
\bibitem[{{Tiburzi} et~al.(2021){Tiburzi}, {Shaifullah}, {Bassa}, {Zucca}, {Verbiest}, {Porayko}, {van der Wateren}, {Fallows}, {Main}, {Janssen}, {Anderson}, {Bak Nielsen}, {Donner}, {Keane}, {K{\"u}nsem{\"o}ller}, {Os{\l}owski}, {Grie{\ss}meier}, {Serylak}, {Br{\"u}ggen}, {Ciardi}, {Dettmar}, {Hoeft}, {Kramer}, {Mann} and {Vocks}}]{Tiburzi2021}
\bibinfo{author}{{Tiburzi}, C.}, \bibinfo{author}{{Shaifullah}, G.M.}, \bibinfo{author}{{Bassa}, C.G.}, \bibinfo{author}{{Zucca}, P.}, \bibinfo{author}{{Verbiest}, J.P.W.}, \bibinfo{author}{{Porayko}, N.K.}, \bibinfo{author}{{van der Wateren}, E.}, \bibinfo{author}{{Fallows}, R.A.}, \bibinfo{author}{{Main}, R.A.}, \bibinfo{author}{{Janssen}, G.H.}, \bibinfo{author}{{Anderson}, J.M.}, \bibinfo{author}{{Bak Nielsen}, A.S.}, \bibinfo{author}{{Donner}, J.Y.}, \bibinfo{author}{{Keane}, E.F.}, \bibinfo{author}{{K{\"u}nsem{\"o}ller}, J.}, \bibinfo{author}{{Os{\l}owski}, S.}, \bibinfo{author}{{Grie{\ss}meier}, J.M.}, \bibinfo{author}{{Serylak}, M.}, \bibinfo{author}{{Br{\"u}ggen}, M.}, \bibinfo{author}{{Ciardi}, B.}, \bibinfo{author}{{Dettmar}, R.J.}, \bibinfo{author}{{Hoeft}, M.}, \bibinfo{author}{{Kramer}, M.}, \bibinfo{author}{{Mann}, G.}, \bibinfo{author}{{Vocks}, C.}, \bibinfo{year}{2021}.
\newblock \bibinfo{title}{{The impact of solar wind variability on pulsar timing}}.
\newblock \bibinfo{journal}{AAP} \bibinfo{volume}{647}, \bibinfo{pages}{A84}.
\newblock \DOIprefix\doi{10.1051/0004-6361/202039846}, \href{http://arxiv.org/abs/2012.11726}{{\tt arXiv:2012.11726}}.
\bibitem[{{Tiburzi} et~al.(2019){Tiburzi}, {Verbiest}, {Shaifullah}, {Janssen}, {Anderson}, {Horneffer}, {K{\"u}nsem{\"o}ller}, {Os{\l}owski}, {Donner}, {Kramer}, {Kumari}, {Porayko}, {Zucca}, {Ciardi}, {Dettmar}, {Grie{\ss}meier}, {Hoeft} and {Serylak}}]{Tiburzi2019}
\bibinfo{author}{{Tiburzi}, C.}, \bibinfo{author}{{Verbiest}, J.P.W.}, \bibinfo{author}{{Shaifullah}, G.M.}, \bibinfo{author}{{Janssen}, G.H.}, \bibinfo{author}{{Anderson}, J.M.}, \bibinfo{author}{{Horneffer}, A.}, \bibinfo{author}{{K{\"u}nsem{\"o}ller}, J.}, \bibinfo{author}{{Os{\l}owski}, S.}, \bibinfo{author}{{Donner}, J.Y.}, \bibinfo{author}{{Kramer}, M.}, \bibinfo{author}{{Kumari}, A.}, \bibinfo{author}{{Porayko}, N.K.}, \bibinfo{author}{{Zucca}, P.}, \bibinfo{author}{{Ciardi}, B.}, \bibinfo{author}{{Dettmar}, R.J.}, \bibinfo{author}{{Grie{\ss}meier}, J.M.}, \bibinfo{author}{{Hoeft}, M.}, \bibinfo{author}{{Serylak}, M.}, \bibinfo{year}{2019}.
\newblock \bibinfo{title}{{On the usefulness of existing solar wind models for pulsar timing corrections}}.
\newblock \bibinfo{journal}{MNRAS} \bibinfo{volume}{487}, \bibinfo{pages}{394--408}.
\newblock \DOIprefix\doi{10.1093/mnras/stz1278}, \href{http://arxiv.org/abs/1905.02989}{{\tt arXiv:1905.02989}}.
\bibitem[{{Tingay} et~al.(2013){Tingay}, {Goeke}, {Bowman}, {Emrich}, {Ord}, {Mitchell}, {Morales}, {Booler}, {Crosse}, {Wayth}, {Lonsdale}, {Tremblay}, {Pallot}, {Colegate}, {Wicenec}, {Kudryavtseva}, {Arcus}, {Barnes}, {Bernardi}, {Briggs}, {Burns}, {Bunton}, {Cappallo}, {Corey}, {Deshpande}, {Desouza}, {Gaensler}, {Greenhill}, {Hall}, {Hazelton}, {Herne}, {Hewitt}, {Johnston-Hollitt}, {Kaplan}, {Kasper}, {Kincaid}, {Koenig}, {Kratzenberg}, {Lynch}, {Mckinley}, {Mcwhirter}, {Morgan}, {Oberoi}, {Pathikulangara}, {Prabu}, {Remillard}, {Rogers}, {Roshi}, {Salah}, {Sault}, {Udaya-Shankar}, {Schlagenhaufer}, {Srivani}, {Stevens}, {Subrahmanyan}, {Waterson}, {Webster}, {Whitney}, {Williams}, {Williams} and {Wyithe}}]{Tingay2013}
\bibinfo{author}{{Tingay}, S.J.}, \bibinfo{author}{{Goeke}, R.}, \bibinfo{author}{{Bowman}, J.D.}, \bibinfo{author}{{Emrich}, D.}, \bibinfo{author}{{Ord}, S.M.}, \bibinfo{author}{{Mitchell}, D.A.}, \bibinfo{author}{{Morales}, M.F.}, \bibinfo{author}{{Booler}, T.}, \bibinfo{author}{{Crosse}, B.}, \bibinfo{author}{{Wayth}, R.B.}, \bibinfo{author}{{Lonsdale}, C.J.}, \bibinfo{author}{{Tremblay}, S.}, \bibinfo{author}{{Pallot}, D.}, \bibinfo{author}{{Colegate}, T.}, \bibinfo{author}{{Wicenec}, A.}, \bibinfo{author}{{Kudryavtseva}, N.}, \bibinfo{author}{{Arcus}, W.}, \bibinfo{author}{{Barnes}, D.}, \bibinfo{author}{{Bernardi}, G.}, \bibinfo{author}{{Briggs}, F.}, \bibinfo{author}{{Burns}, S.}, \bibinfo{author}{{Bunton}, J.D.}, \bibinfo{author}{{Cappallo}, R.J.}, \bibinfo{author}{{Corey}, B.E.}, \bibinfo{author}{{Deshpande}, A.}, \bibinfo{author}{{Desouza}, L.}, \bibinfo{author}{{Gaensler}, B.M.}, \bibinfo{author}{{Greenhill}, L.J.}, \bibinfo{author}{{Hall}, P.J.}, \bibinfo{author}{{Hazelton}, B.J.},
  \bibinfo{author}{{Herne}, D.}, \bibinfo{author}{{Hewitt}, J.N.}, \bibinfo{author}{{Johnston-Hollitt}, M.}, \bibinfo{author}{{Kaplan}, D.L.}, \bibinfo{author}{{Kasper}, J.C.}, \bibinfo{author}{{Kincaid}, B.B.}, \bibinfo{author}{{Koenig}, R.}, \bibinfo{author}{{Kratzenberg}, E.}, \bibinfo{author}{{Lynch}, M.J.}, \bibinfo{author}{{Mckinley}, B.}, \bibinfo{author}{{Mcwhirter}, S.R.}, \bibinfo{author}{{Morgan}, E.}, \bibinfo{author}{{Oberoi}, D.}, \bibinfo{author}{{Pathikulangara}, J.}, \bibinfo{author}{{Prabu}, T.}, \bibinfo{author}{{Remillard}, R.A.}, \bibinfo{author}{{Rogers}, A.E.E.}, \bibinfo{author}{{Roshi}, A.}, \bibinfo{author}{{Salah}, J.E.}, \bibinfo{author}{{Sault}, R.J.}, \bibinfo{author}{{Udaya-Shankar}, N.}, \bibinfo{author}{{Schlagenhaufer}, F.}, \bibinfo{author}{{Srivani}, K.S.}, \bibinfo{author}{{Stevens}, J.}, \bibinfo{author}{{Subrahmanyan}, R.}, \bibinfo{author}{{Waterson}, M.}, \bibinfo{author}{{Webster}, R.L.}, \bibinfo{author}{{Whitney}, A.R.}, \bibinfo{author}{{Williams}, A.},
  \bibinfo{author}{{Williams}, C.L.}, \bibinfo{author}{{Wyithe}, J.S.B.}, \bibinfo{year}{2013}.
\newblock \bibinfo{title}{{The Murchison Widefield Array: The Square Kilometre Array Precursor at Low Radio Frequencies}}.
\newblock \bibinfo{journal}{PASA} \bibinfo{volume}{30}, \bibinfo{pages}{e007}.
\newblock \DOIprefix\doi{10.1017/pasa.2012.007}, \href{http://arxiv.org/abs/1206.6945}{{\tt arXiv:1206.6945}}.
\bibitem[{{Tokumaru} et~al.(2020){Tokumaru}, {Tawara}, {Takefuji}, {Sekido} and {Terasawa}}]{tmt+20}
\bibinfo{author}{{Tokumaru}, M.}, \bibinfo{author}{{Tawara}, K.}, \bibinfo{author}{{Takefuji}, K.}, \bibinfo{author}{{Sekido}, M.}, \bibinfo{author}{{Terasawa}, T.}, \bibinfo{year}{2020}.
\newblock \bibinfo{title}{{Radio Sounding Measurements of the Solar Corona Using Giant Pulses of the Crab Pulsar in 2018}}.
\newblock \bibinfo{journal}{Solar Physics} \bibinfo{volume}{295}, \bibinfo{pages}{80}.
\newblock \DOIprefix\doi{10.1007/s11207-020-01644-w}.
\bibitem[{{Torne} et~al.(2017){Torne}, {Desvignes}, {Eatough}, {Karuppusamy}, {Paubert}, {Kramer}, {Cognard}, {Champion} and {Spitler}}]{Torne+2017}
\bibinfo{author}{{Torne}, P.}, \bibinfo{author}{{Desvignes}, G.}, \bibinfo{author}{{Eatough}, R.P.}, \bibinfo{author}{{Karuppusamy}, R.}, \bibinfo{author}{{Paubert}, G.}, \bibinfo{author}{{Kramer}, M.}, \bibinfo{author}{{Cognard}, I.}, \bibinfo{author}{{Champion}, D.J.}, \bibinfo{author}{{Spitler}, L.G.}, \bibinfo{year}{2017}.
\newblock \bibinfo{title}{{Detection of the magnetar SGR J1745-2900 up to 291 GHz with evidence of polarized millimetre emission}}.
\newblock \bibinfo{journal}{MNRAS} \bibinfo{volume}{465}, \bibinfo{pages}{242--247}.
\newblock \DOIprefix\doi{10.1093/mnras/stw2757}, \href{http://arxiv.org/abs/1610.07616}{{\tt arXiv:1610.07616}}.
\bibitem[{Tyler et~al.(1977)Tyler, Brenkle, Komarek and Zygielbaum}]{Tyler1977}
\bibinfo{author}{Tyler, G.L.}, \bibinfo{author}{Brenkle, J.P.}, \bibinfo{author}{Komarek, T.A.}, \bibinfo{author}{Zygielbaum, A.I.}, \bibinfo{year}{1977}.
\newblock \bibinfo{title}{The viking solar corona experiment}.
\newblock \bibinfo{journal}{Journal of Geophysical Research (1896-1977)} \bibinfo{volume}{82}, \bibinfo{pages}{4335--4340}.
\newblock \DOIprefix\doi{https://doi.org/10.1029/JS082i028p04335}.
\bibitem[{{van Haarlem} et~al.(2013){van Haarlem}, {Wise}, {Gunst}, {Heald}, {McKean}, {Hessels}, {de Bruyn}, {Nijboer}, {Swinbank}, {Fallows}, {Brentjens}, {Nelles}, {Beck}, {Falcke}, {Fender}, {H{\"o}randel}, {Koopmans}, {Mann}, {Miley}, {R{\"o}ttgering}, {Stappers}, {Wijers}, {Zaroubi}, {van den Akker}, {Alexov}, {Anderson}, {Anderson}, {van Ardenne}, {Arts}, {Asgekar}, {Avruch}, {Batejat}, {B{\"a}hren}, {Bell}, {Bell}, {van Bemmel}, {Bennema}, {Bentum}, {Bernardi}, {Best}, {B{\^i}rzan}, {Bonafede}, {Boonstra}, {Braun}, {Bregman}, {Breitling}, {van de Brink}, {Broderick}, {Broekema}, {Brouw}, {Br{\"u}ggen}, {Butcher}, {van Cappellen}, {Ciardi}, {Coenen}, {Conway}, {Coolen}, {Corstanje}, {Damstra}, {Davies}, {Deller}, {Dettmar}, {van Diepen}, {Dijkstra}, {Donker}, {Doorduin}, {Dromer}, {Drost}, {van Duin}, {Eisl{\"o}ffel}, {van Enst}, {Ferrari}, {Frieswijk}, {Gankema}, {Garrett}, {de Gasperin}, {Gerbers}, {de Geus}, {Grie{\ss}meier}, {Grit}, {Gruppen}, {Hamaker}, {Hassall}, {Hoeft}, {Holties}, {Horneffer},
  {van der Horst}, {van Houwelingen}, {Huijgen}, {Iacobelli}, {Intema}, {Jackson}, {Jelic}, {de Jong}, {Juette}, {Kant}, {Karastergiou}, {Koers}, {Kollen}, {Kondratiev}, {Kooistra}, {Koopman}, {Koster}, {Kuniyoshi}, {Kramer}, {Kuper}, {Lambropoulos}, {Law}, {van Leeuwen}, {Lemaitre}, {Loose}, {Maat}, {Macario}, {Markoff}, {Masters}, {McFadden}, {McKay-Bukowski}, {Meijering}, {Meulman}, {Mevius}, {Middelberg}, {Millenaar}, {Miller-Jones}, {Mohan}, {Mol}, {Morawietz}, {Morganti}, {Mulcahy}, {Mulder}, {Munk}, {Nieuwenhuis}, {van Nieuwpoort}, {Noordam}, {Norden}, {Noutsos}, {Offringa}, {Olofsson}, {Omar}, {Orr{\'u}}, {Overeem}, {Paas}, {Pandey-Pommier}, {Pandey}, {Pizzo}, {Polatidis}, {Rafferty}, {Rawlings}, {Reich}, {de Reijer}, {Reitsma}, {Renting}, {Riemers}, {Rol}, {Romein}, {Roosjen}, {Ruiter}, {Scaife}, {van der Schaaf}, {Scheers}, {Schellart}, {Schoenmakers}, {Schoonderbeek}, {Serylak}, {Shulevski}, {Sluman}, {Smirnov}, {Sobey}, {Spreeuw}, {Steinmetz}, {Sterks}, {Stiepel}, {Stuurwold}, {Tagger}, {Tang},
  {Tasse}, {Thomas}, {Thoudam}, {Toribio}, {van der Tol}, {Usov}, {van Veelen}, {van der Veen}, {ter Veen}, {Verbiest}, {Vermeulen}, {Vermaas}, {Vocks}, {Vogt}, {de Vos}, {van der Wal}, {van Weeren}, {Weggemans}, {Weltevrede}, {White}, {Wijnholds}, {Wilhelmsson}, {Wucknitz}, {Yatawatta}, {Zarka}, {Zensus} and {van Zwieten}}]{2013Van}
\bibinfo{author}{{van Haarlem}, M.P.}, \bibinfo{author}{{Wise}, M.W.}, \bibinfo{author}{{Gunst}, A.W.}, \bibinfo{author}{{Heald}, G.}, \bibinfo{author}{{McKean}, J.P.}, \bibinfo{author}{{Hessels}, J.W.T.}, \bibinfo{author}{{de Bruyn}, A.G.}, \bibinfo{author}{{Nijboer}, R.}, \bibinfo{author}{{Swinbank}, J.}, \bibinfo{author}{{Fallows}, R.}, \bibinfo{author}{{Brentjens}, M.}, \bibinfo{author}{{Nelles}, A.}, \bibinfo{author}{{Beck}, R.}, \bibinfo{author}{{Falcke}, H.}, \bibinfo{author}{{Fender}, R.}, \bibinfo{author}{{H{\"o}randel}, J.}, \bibinfo{author}{{Koopmans}, L.V.E.}, \bibinfo{author}{{Mann}, G.}, \bibinfo{author}{{Miley}, G.}, \bibinfo{author}{{R{\"o}ttgering}, H.}, \bibinfo{author}{{Stappers}, B.W.}, \bibinfo{author}{{Wijers}, R.A.M.J.}, \bibinfo{author}{{Zaroubi}, S.}, \bibinfo{author}{{van den Akker}, M.}, \bibinfo{author}{{Alexov}, A.}, \bibinfo{author}{{Anderson}, J.}, \bibinfo{author}{{Anderson}, K.}, \bibinfo{author}{{van Ardenne}, A.}, \bibinfo{author}{{Arts}, M.}, \bibinfo{author}{{Asgekar},
  A.}, \bibinfo{author}{{Avruch}, I.M.}, \bibinfo{author}{{Batejat}, F.}, \bibinfo{author}{{B{\"a}hren}, L.}, \bibinfo{author}{{Bell}, M.E.}, \bibinfo{author}{{Bell}, M.R.}, \bibinfo{author}{{van Bemmel}, I.}, \bibinfo{author}{{Bennema}, P.}, \bibinfo{author}{{Bentum}, M.J.}, \bibinfo{author}{{Bernardi}, G.}, \bibinfo{author}{{Best}, P.}, \bibinfo{author}{{B{\^i}rzan}, L.}, \bibinfo{author}{{Bonafede}, A.}, \bibinfo{author}{{Boonstra}, A.J.}, \bibinfo{author}{{Braun}, R.}, \bibinfo{author}{{Bregman}, J.}, \bibinfo{author}{{Breitling}, F.}, \bibinfo{author}{{van de Brink}, R.H.}, \bibinfo{author}{{Broderick}, J.}, \bibinfo{author}{{Broekema}, P.C.}, \bibinfo{author}{{Brouw}, W.N.}, \bibinfo{author}{{Br{\"u}ggen}, M.}, \bibinfo{author}{{Butcher}, H.R.}, \bibinfo{author}{{van Cappellen}, W.}, \bibinfo{author}{{Ciardi}, B.}, \bibinfo{author}{{Coenen}, T.}, \bibinfo{author}{{Conway}, J.}, \bibinfo{author}{{Coolen}, A.}, \bibinfo{author}{{Corstanje}, A.}, \bibinfo{author}{{Damstra}, S.}, \bibinfo{author}{{Davies},
  O.}, \bibinfo{author}{{Deller}, A.T.}, \bibinfo{author}{{Dettmar}, R.J.}, \bibinfo{author}{{van Diepen}, G.}, \bibinfo{author}{{Dijkstra}, K.}, \bibinfo{author}{{Donker}, P.}, \bibinfo{author}{{Doorduin}, A.}, \bibinfo{author}{{Dromer}, J.}, \bibinfo{author}{{Drost}, M.}, \bibinfo{author}{{van Duin}, A.}, \bibinfo{author}{{Eisl{\"o}ffel}, J.}, \bibinfo{author}{{van Enst}, J.}, \bibinfo{author}{{Ferrari}, C.}, \bibinfo{author}{{Frieswijk}, W.}, \bibinfo{author}{{Gankema}, H.}, \bibinfo{author}{{Garrett}, M.A.}, \bibinfo{author}{{de Gasperin}, F.}, \bibinfo{author}{{Gerbers}, M.}, \bibinfo{author}{{de Geus}, E.}, \bibinfo{author}{{Grie{\ss}meier}, J.M.}, \bibinfo{author}{{Grit}, T.}, \bibinfo{author}{{Gruppen}, P.}, \bibinfo{author}{{Hamaker}, J.P.}, \bibinfo{author}{{Hassall}, T.}, \bibinfo{author}{{Hoeft}, M.}, \bibinfo{author}{{Holties}, H.A.}, \bibinfo{author}{{Horneffer}, A.}, \bibinfo{author}{{van der Horst}, A.}, \bibinfo{author}{{van Houwelingen}, A.}, \bibinfo{author}{{Huijgen}, A.},
  \bibinfo{author}{{Iacobelli}, M.}, \bibinfo{author}{{Intema}, H.}, \bibinfo{author}{{Jackson}, N.}, \bibinfo{author}{{Jelic}, V.}, \bibinfo{author}{{de Jong}, A.}, \bibinfo{author}{{Juette}, E.}, \bibinfo{author}{{Kant}, D.}, \bibinfo{author}{{Karastergiou}, A.}, \bibinfo{author}{{Koers}, A.}, \bibinfo{author}{{Kollen}, H.}, \bibinfo{author}{{Kondratiev}, V.I.}, \bibinfo{author}{{Kooistra}, E.}, \bibinfo{author}{{Koopman}, Y.}, \bibinfo{author}{{Koster}, A.}, \bibinfo{author}{{Kuniyoshi}, M.}, \bibinfo{author}{{Kramer}, M.}, \bibinfo{author}{{Kuper}, G.}, \bibinfo{author}{{Lambropoulos}, P.}, \bibinfo{author}{{Law}, C.}, \bibinfo{author}{{van Leeuwen}, J.}, \bibinfo{author}{{Lemaitre}, J.}, \bibinfo{author}{{Loose}, M.}, \bibinfo{author}{{Maat}, P.}, \bibinfo{author}{{Macario}, G.}, \bibinfo{author}{{Markoff}, S.}, \bibinfo{author}{{Masters}, J.}, \bibinfo{author}{{McFadden}, R.A.}, \bibinfo{author}{{McKay-Bukowski}, D.}, \bibinfo{author}{{Meijering}, H.}, \bibinfo{author}{{Meulman}, H.},
  \bibinfo{author}{{Mevius}, M.}, \bibinfo{author}{{Middelberg}, E.}, \bibinfo{author}{{Millenaar}, R.}, \bibinfo{author}{{Miller-Jones}, J.C.A.}, \bibinfo{author}{{Mohan}, R.N.}, \bibinfo{author}{{Mol}, J.D.}, \bibinfo{author}{{Morawietz}, J.}, \bibinfo{author}{{Morganti}, R.}, \bibinfo{author}{{Mulcahy}, D.D.}, \bibinfo{author}{{Mulder}, E.}, \bibinfo{author}{{Munk}, H.}, \bibinfo{author}{{Nieuwenhuis}, L.}, \bibinfo{author}{{van Nieuwpoort}, R.}, \bibinfo{author}{{Noordam}, J.E.}, \bibinfo{author}{{Norden}, M.}, \bibinfo{author}{{Noutsos}, A.}, \bibinfo{author}{{Offringa}, A.R.}, \bibinfo{author}{{Olofsson}, H.}, \bibinfo{author}{{Omar}, A.}, \bibinfo{author}{{Orr{\'u}}, E.}, \bibinfo{author}{{Overeem}, R.}, \bibinfo{author}{{Paas}, H.}, \bibinfo{author}{{Pandey-Pommier}, M.}, \bibinfo{author}{{Pandey}, V.N.}, \bibinfo{author}{{Pizzo}, R.}, \bibinfo{author}{{Polatidis}, A.}, \bibinfo{author}{{Rafferty}, D.}, \bibinfo{author}{{Rawlings}, S.}, \bibinfo{author}{{Reich}, W.}, \bibinfo{author}{{de Reijer},
  J.P.}, \bibinfo{author}{{Reitsma}, J.}, \bibinfo{author}{{Renting}, G.A.}, \bibinfo{author}{{Riemers}, P.}, \bibinfo{author}{{Rol}, E.}, \bibinfo{author}{{Romein}, J.W.}, \bibinfo{author}{{Roosjen}, J.}, \bibinfo{author}{{Ruiter}, M.}, \bibinfo{author}{{Scaife}, A.}, \bibinfo{author}{{van der Schaaf}, K.}, \bibinfo{author}{{Scheers}, B.}, \bibinfo{author}{{Schellart}, P.}, \bibinfo{author}{{Schoenmakers}, A.}, \bibinfo{author}{{Schoonderbeek}, G.}, \bibinfo{author}{{Serylak}, M.}, \bibinfo{author}{{Shulevski}, A.}, \bibinfo{author}{{Sluman}, J.}, \bibinfo{author}{{Smirnov}, O.}, \bibinfo{author}{{Sobey}, C.}, \bibinfo{author}{{Spreeuw}, H.}, \bibinfo{author}{{Steinmetz}, M.}, \bibinfo{author}{{Sterks}, C.G.M.}, \bibinfo{author}{{Stiepel}, H.J.}, \bibinfo{author}{{Stuurwold}, K.}, \bibinfo{author}{{Tagger}, M.}, \bibinfo{author}{{Tang}, Y.}, \bibinfo{author}{{Tasse}, C.}, \bibinfo{author}{{Thomas}, I.}, \bibinfo{author}{{Thoudam}, S.}, \bibinfo{author}{{Toribio}, M.C.}, \bibinfo{author}{{van der Tol}, B.},
  \bibinfo{author}{{Usov}, O.}, \bibinfo{author}{{van Veelen}, M.}, \bibinfo{author}{{van der Veen}, A.J.}, \bibinfo{author}{{ter Veen}, S.}, \bibinfo{author}{{Verbiest}, J.P.W.}, \bibinfo{author}{{Vermeulen}, R.}, \bibinfo{author}{{Vermaas}, N.}, \bibinfo{author}{{Vocks}, C.}, \bibinfo{author}{{Vogt}, C.}, \bibinfo{author}{{de Vos}, M.}, \bibinfo{author}{{van der Wal}, E.}, \bibinfo{author}{{van Weeren}, R.}, \bibinfo{author}{{Weggemans}, H.}, \bibinfo{author}{{Weltevrede}, P.}, \bibinfo{author}{{White}, S.}, \bibinfo{author}{{Wijnholds}, S.J.}, \bibinfo{author}{{Wilhelmsson}, T.}, \bibinfo{author}{{Wucknitz}, O.}, \bibinfo{author}{{Yatawatta}, S.}, \bibinfo{author}{{Zarka}, P.}, \bibinfo{author}{{Zensus}, A.}, \bibinfo{author}{{van Zwieten}, J.}, \bibinfo{year}{2013}.
\newblock \bibinfo{title}{{LOFAR: The LOw-Frequency ARray}}.
\newblock \bibinfo{journal}{AAP} \bibinfo{volume}{556}, \bibinfo{pages}{A2}.
\newblock \DOIprefix\doi{10.1051/0004-6361/201220873}, \href{http://arxiv.org/abs/1305.3550}{{\tt arXiv:1305.3550}}.
\bibitem[{{van Straten}(2013)}]{straten2013}
\bibinfo{author}{{van Straten}, W.}, \bibinfo{year}{2013}.
\newblock \bibinfo{title}{{High-fidelity Radio Astronomical Polarimetry Using a Millisecond Pulsar as a Polarized Reference Source}}.
\newblock \bibinfo{journal}{APJs} \bibinfo{volume}{204}, \bibinfo{pages}{13}.
\newblock \DOIprefix\doi{10.1088/0067-0049/204/1/13}, \href{http://arxiv.org/abs/1212.3446}{{\tt arXiv:1212.3446}}.
\bibitem[{{Walker} et~al.(2013){Walker}, {Demorest} and {van Straten}}]{wdv2013}
\bibinfo{author}{{Walker}, M.A.}, \bibinfo{author}{{Demorest}, P.B.}, \bibinfo{author}{{van Straten}, W.}, \bibinfo{year}{2013}.
\newblock \bibinfo{title}{{Cyclic Spectroscopy of The Millisecond Pulsar, B1937+21}}.
\newblock \bibinfo{journal}{APJ} \bibinfo{volume}{779}, \bibinfo{pages}{99}.
\newblock \DOIprefix\doi{10.1088/0004-637X/779/2/99}, \href{http://arxiv.org/abs/1310.3535}{{\tt arXiv:1310.3535}}.
\bibitem[{{Wang} et~al.(2022){Wang}, {Shaifullah}, {Verbiest}, {Tiburzi}, {Champion}, {Cognard}, {Gaikwad}, {Graikou}, {Guillemot}, {Hu}, {Karuppusamy}, {Keith}, {Kramer}, {Liu}, {Lyne}, {Mickaliger}, {Stappers} and {Theureau}}]{wsv+22}
\bibinfo{author}{{Wang}, J.}, \bibinfo{author}{{Shaifullah}, G.M.}, \bibinfo{author}{{Verbiest}, J.P.W.}, \bibinfo{author}{{Tiburzi}, C.}, \bibinfo{author}{{Champion}, D.J.}, \bibinfo{author}{{Cognard}, I.}, \bibinfo{author}{{Gaikwad}, M.}, \bibinfo{author}{{Graikou}, E.}, \bibinfo{author}{{Guillemot}, L.}, \bibinfo{author}{{Hu}, H.}, \bibinfo{author}{{Karuppusamy}, R.}, \bibinfo{author}{{Keith}, M.J.}, \bibinfo{author}{{Kramer}, M.}, \bibinfo{author}{{Liu}, Y.}, \bibinfo{author}{{Lyne}, A.G.}, \bibinfo{author}{{Mickaliger}, M.B.}, \bibinfo{author}{{Stappers}, B.W.}, \bibinfo{author}{{Theureau}, G.}, \bibinfo{year}{2022}.
\newblock \bibinfo{title}{{A comparative analysis of pulse time-of-arrival creation methods}}.
\newblock \bibinfo{journal}{AAP} \bibinfo{volume}{658}, \bibinfo{pages}{A181}.
\newblock \DOIprefix\doi{10.1051/0004-6361/202141121}, \href{http://arxiv.org/abs/2111.13482}{{\tt arXiv:2111.13482}}.
\bibitem[{{Wang} et~al.(2017){Wang}, {Coles}, {Hobbs}, {Shannon}, {Manchester}, {Kerr}, {Yuan}, {Wang}, {Bailes}, {Bhat}, {Dai}, {Dempsey}, {Keith}, {Lasky}, {Levin}, {Os{\l}owski}, {Ravi}, {Reardon}, {Rosado}, {Russell}, {Spiewak}, {van Straten}, {Toomey}, {Wen}, {You} and {Zhu}}]{wch+2017}
\bibinfo{author}{{Wang}, J.B.}, \bibinfo{author}{{Coles}, W.A.}, \bibinfo{author}{{Hobbs}, G.}, \bibinfo{author}{{Shannon}, R.M.}, \bibinfo{author}{{Manchester}, R.N.}, \bibinfo{author}{{Kerr}, M.}, \bibinfo{author}{{Yuan}, J.P.}, \bibinfo{author}{{Wang}, N.}, \bibinfo{author}{{Bailes}, M.}, \bibinfo{author}{{Bhat}, N.D.R.}, \bibinfo{author}{{Dai}, S.}, \bibinfo{author}{{Dempsey}, J.}, \bibinfo{author}{{Keith}, M.J.}, \bibinfo{author}{{Lasky}, P.D.}, \bibinfo{author}{{Levin}, Y.}, \bibinfo{author}{{Os{\l}owski}, S.}, \bibinfo{author}{{Ravi}, V.}, \bibinfo{author}{{Reardon}, D.J.}, \bibinfo{author}{{Rosado}, P.A.}, \bibinfo{author}{{Russell}, C.J.}, \bibinfo{author}{{Spiewak}, R.}, \bibinfo{author}{{van Straten}, W.}, \bibinfo{author}{{Toomey}, L.}, \bibinfo{author}{{Wen}, L.}, \bibinfo{author}{{You}, X.P.}, \bibinfo{author}{{Zhu}, X.J.}, \bibinfo{year}{2017}.
\newblock \bibinfo{title}{{Comparison of pulsar positions from timing and very long baseline astrometry}}.
\newblock \bibinfo{journal}{MNRAS} \bibinfo{volume}{469}, \bibinfo{pages}{425--434}.
\newblock \DOIprefix\doi{10.1093/mnras/stx837}, \href{http://arxiv.org/abs/1704.01011}{{\tt arXiv:1704.01011}}.
\bibitem[{{Wang, Pingli} et~al.(2019){Wang, Pingli}, {Wang, Guangli}, {Gao, Yuping}, {Cai, Hongbing} and {Liu, Na}}]{refId0}
\bibinfo{author}{{Wang, Pingli}}, \bibinfo{author}{{Wang, Guangli}}, \bibinfo{author}{{Gao, Yuping}}, \bibinfo{author}{{Cai, Hongbing}}, \bibinfo{author}{{Liu, Na}}, \bibinfo{year}{2019}.
\newblock \bibinfo{title}{Comparison of vlbi and gnss common view for time transfer}.
\newblock \bibinfo{journal}{Int. J. Metrol. Qual. Eng.} \bibinfo{volume}{10}, \bibinfo{pages}{15}.
\newblock \URLprefix \url{https://doi.org/10.1051/ijmqe/2019014}, \DOIprefix\doi{10.1051/ijmqe/2019014}.
\bibitem[{{Wayth} et~al.(2018){Wayth}, {Tingay}, {Trott}, {Emrich}, {Johnston-Hollitt}, {McKinley}, {Gaensler}, {Beardsley}, {Booler}, {Crosse}, {Franzen}, {Horsley}, {Kaplan}, {Kenney}, {Morales}, {Pallot}, {Sleap}, {Steele}, {Walker}, {Williams}, {Wu}, {Cairns}, {Filipovic}, {Johnston}, {Murphy}, {Quinn}, {Staveley-Smith}, {Webster} and {Wyithe}}]{Wayth2018}
\bibinfo{author}{{Wayth}, R.B.}, \bibinfo{author}{{Tingay}, S.J.}, \bibinfo{author}{{Trott}, C.M.}, \bibinfo{author}{{Emrich}, D.}, \bibinfo{author}{{Johnston-Hollitt}, M.}, \bibinfo{author}{{McKinley}, B.}, \bibinfo{author}{{Gaensler}, B.M.}, \bibinfo{author}{{Beardsley}, A.P.}, \bibinfo{author}{{Booler}, T.}, \bibinfo{author}{{Crosse}, B.}, \bibinfo{author}{{Franzen}, T.M.O.}, \bibinfo{author}{{Horsley}, L.}, \bibinfo{author}{{Kaplan}, D.L.}, \bibinfo{author}{{Kenney}, D.}, \bibinfo{author}{{Morales}, M.F.}, \bibinfo{author}{{Pallot}, D.}, \bibinfo{author}{{Sleap}, G.}, \bibinfo{author}{{Steele}, K.}, \bibinfo{author}{{Walker}, M.}, \bibinfo{author}{{Williams}, A.}, \bibinfo{author}{{Wu}, C.}, \bibinfo{author}{{Cairns}, I.H.}, \bibinfo{author}{{Filipovic}, M.D.}, \bibinfo{author}{{Johnston}, S.}, \bibinfo{author}{{Murphy}, T.}, \bibinfo{author}{{Quinn}, P.}, \bibinfo{author}{{Staveley-Smith}, L.}, \bibinfo{author}{{Webster}, R.}, \bibinfo{author}{{Wyithe}, J.S.B.}, \bibinfo{year}{2018}.
\newblock \bibinfo{title}{{The Phase II Murchison Widefield Array: Design overview}}.
\newblock \bibinfo{journal}{PASA} \bibinfo{volume}{35}, \bibinfo{pages}{e033}.
\newblock \DOIprefix\doi{10.1017/pasa.2018.37}, \href{http://arxiv.org/abs/1809.06466}{{\tt arXiv:1809.06466}}.
\bibitem[{Xu et~al.(2023)Xu, Chen, Guo, Jiang, Wang, Xu, Xue, Caballero, Yuan, Xu, Wang, Hao, Luo, Lee, Han, Jiang, Shen, Wang, Wang, Xu, Wu, Manchester, Qian, Guan, Huang, Sun and Zhu}]{Xu2023}
\bibinfo{author}{Xu, H.}, \bibinfo{author}{Chen, S.}, \bibinfo{author}{Guo, Y.}, \bibinfo{author}{Jiang, J.}, \bibinfo{author}{Wang, B.}, \bibinfo{author}{Xu, J.}, \bibinfo{author}{Xue, Z.}, \bibinfo{author}{Caballero, R.N.}, \bibinfo{author}{Yuan, J.}, \bibinfo{author}{Xu, Y.}, \bibinfo{author}{Wang, J.}, \bibinfo{author}{Hao, L.}, \bibinfo{author}{Luo, J.}, \bibinfo{author}{Lee, K.}, \bibinfo{author}{Han, J.}, \bibinfo{author}{Jiang, P.}, \bibinfo{author}{Shen, Z.}, \bibinfo{author}{Wang, M.}, \bibinfo{author}{Wang, N.}, \bibinfo{author}{Xu, R.}, \bibinfo{author}{Wu, X.}, \bibinfo{author}{Manchester, R.}, \bibinfo{author}{Qian, L.}, \bibinfo{author}{Guan, X.}, \bibinfo{author}{Huang, M.}, \bibinfo{author}{Sun, C.}, \bibinfo{author}{Zhu, Y.}, \bibinfo{year}{2023}.
\newblock \bibinfo{title}{Searching for the nano-hertz stochastic gravitational wave background with the chinese pulsar timing array data release i}.
\newblock \bibinfo{journal}{Research in Astronomy and Astrophysics} \bibinfo{volume}{23}, \bibinfo{pages}{075024}.
\newblock \URLprefix \url{https://dx.doi.org/10.1088/1674-4527/acdfa5}, \DOIprefix\doi{10.1088/1674-4527/acdfa5}.
\bibitem[{{Xue} et~al.(2019){Xue}, {Shi}, {Guo}, {Huang}, {Peng}, {Luo}, {Shentu} and {Chen}}]{Mengfan2019}
\bibinfo{author}{{Xue}, M.}, \bibinfo{author}{{Shi}, Y.}, \bibinfo{author}{{Guo}, Y.}, \bibinfo{author}{{Huang}, N.}, \bibinfo{author}{{Peng}, D.}, \bibinfo{author}{{Luo}, J.}, \bibinfo{author}{{Shentu}, H.}, \bibinfo{author}{{Chen}, Z.}, \bibinfo{year}{2019}.
\newblock \bibinfo{title}{{X-Ray Pulsar-Based Navigation Considering Spacecraft Orbital Motion and Systematic Biases}}.
\newblock \bibinfo{journal}{Sensors} \bibinfo{volume}{19}, \bibinfo{pages}{1877}.
\newblock \DOIprefix\doi{10.3390/s19081877}.
\bibitem[{{You} et~al.(2012){You}, {Coles}, {Hobbs} and {Manchester}}]{You2012}
\bibinfo{author}{{You}, X.P.}, \bibinfo{author}{{Coles}, W.A.}, \bibinfo{author}{{Hobbs}, G.B.}, \bibinfo{author}{{Manchester}, R.N.}, \bibinfo{year}{2012}.
\newblock \bibinfo{title}{{Measurement of the electron density and magnetic field of the solar wind using millisecond pulsars}}.
\newblock \bibinfo{journal}{MNRAS} \bibinfo{volume}{422}, \bibinfo{pages}{1160--1165}.
\newblock \DOIprefix\doi{10.1111/j.1365-2966.2012.20688.x}, \href{http://arxiv.org/abs/1202.2263}{{\tt arXiv:1202.2263}}.
\bibitem[{You et~al.(2007)You, Hobbs, Coles, Manchester and Han}]{You2007}
\bibinfo{author}{You, X.P.}, \bibinfo{author}{Hobbs, G.B.}, \bibinfo{author}{Coles, W.A.}, \bibinfo{author}{Manchester, R.N.}, \bibinfo{author}{Han, J.L.}, \bibinfo{year}{2007}.
\newblock \bibinfo{title}{An improved solar wind electron density model for pulsar timing}.
\newblock \bibinfo{journal}{The Astrophysical Journal} \bibinfo{volume}{671}, \bibinfo{pages}{907}.
\newblock \URLprefix \url{https://dx.doi.org/10.1086/522227}, \DOIprefix\doi{10.1086/522227}.
\bibitem[{{Zheng} et~al.(2019){Zheng}, {Zhang}, {Lu}, {Wang}, {Gao}, {Li}, {Song}, {Ge}, {Han}, {Chen}, {Xu}, {Cao}, {Liu}, {Zhang}, {Qu}, {Chang}, {Chen}, {Chen}, {Chen}, {Chen}, {Chen}, {Cui}, {Cui}, {Deng}, {Dong}, {Du}, {Fu}, {Gao}, {Gao}, {Gao}, {Gu}, {Guan}, {Gungor}, {Guo}, {Han}, {Hu}, {Huang}, {Huo}, {Ji}, {Jia}, {Jiang}, {Jiang}, {Jin}, {Jin}, {Li}, {Li}, {Li}, {Li}, {Li}, {Li}, {Li}, {Li}, {Li}, {Li}, {Li}, {Liang}, {Liao}, {Liu}, {Liu}, {Liu}, {Liu}, {Liu}, {Liu}, {Lu}, {Lu}, {Luo}, {Ma}, {Meng}, {Nang}, {Nie}, {Ou}, {Sai}, {Shang}, {Sun}, {Tan}, {Tao}, {Tao}, {Tuo}, {Wang}, {Wang}, {Wang}, {Wang}, {Wen}, {Wu}, {Wu}, {Xiao}, {Xiong}, {Xu}, {Yan}, {Yang}, {Yang}, {Yang}, {Zhang}, {Zhang}, {Zhang}, {Zhang}, {Zhang}, {Zhang}, {Zhang}, {Zhang}, {Zhang}, {Zhang}, {Zhang}, {Zhang}, {Zhang}, {Zhang}, {Zhang}, {Zhang}, {Zhang}, {Zhang}, {Zhao}, {Zhao}, {Zhao}, {Zhu}, {Zhu} and {Zou}}]{zzl+2019}
\bibinfo{author}{{Zheng}, S.J.}, \bibinfo{author}{{Zhang}, S.N.}, \bibinfo{author}{{Lu}, F.J.}, \bibinfo{author}{{Wang}, W.B.}, \bibinfo{author}{{Gao}, Y.}, \bibinfo{author}{{Li}, T.P.}, \bibinfo{author}{{Song}, L.M.}, \bibinfo{author}{{Ge}, M.Y.}, \bibinfo{author}{{Han}, D.W.}, \bibinfo{author}{{Chen}, Y.}, \bibinfo{author}{{Xu}, Y.P.}, \bibinfo{author}{{Cao}, X.L.}, \bibinfo{author}{{Liu}, C.Z.}, \bibinfo{author}{{Zhang}, S.}, \bibinfo{author}{{Qu}, J.L.}, \bibinfo{author}{{Chang}, Z.}, \bibinfo{author}{{Chen}, G.}, \bibinfo{author}{{Chen}, L.}, \bibinfo{author}{{Chen}, T.X.}, \bibinfo{author}{{Chen}, Y.B.}, \bibinfo{author}{{Chen}, Y.P.}, \bibinfo{author}{{Cui}, W.}, \bibinfo{author}{{Cui}, W.W.}, \bibinfo{author}{{Deng}, J.K.}, \bibinfo{author}{{Dong}, Y.W.}, \bibinfo{author}{{Du}, Y.Y.}, \bibinfo{author}{{Fu}, M.X.}, \bibinfo{author}{{Gao}, G.H.}, \bibinfo{author}{{Gao}, H.}, \bibinfo{author}{{Gao}, M.}, \bibinfo{author}{{Gu}, Y.D.}, \bibinfo{author}{{Guan}, J.}, \bibinfo{author}{{Gungor}, C.},
  \bibinfo{author}{{Guo}, C.C.}, \bibinfo{author}{{Han}, D.W.}, \bibinfo{author}{{Hu}, W.}, \bibinfo{author}{{Huang}, Y.}, \bibinfo{author}{{Huo}, J.}, \bibinfo{author}{{Ji}, J.F.}, \bibinfo{author}{{Jia}, S.M.}, \bibinfo{author}{{Jiang}, L.H.}, \bibinfo{author}{{Jiang}, W.C.}, \bibinfo{author}{{Jin}, J.}, \bibinfo{author}{{Jin}, Y.J.}, \bibinfo{author}{{Li}, B.}, \bibinfo{author}{{Li}, C.K.}, \bibinfo{author}{{Li}, G.}, \bibinfo{author}{{Li}, M.S.}, \bibinfo{author}{{Li}, W.}, \bibinfo{author}{{Li}, X.}, \bibinfo{author}{{Li}, X.B.}, \bibinfo{author}{{Li}, X.F.}, \bibinfo{author}{{Li}, Y.G.}, \bibinfo{author}{{Li}, Z.J.}, \bibinfo{author}{{Li}, Z.W.}, \bibinfo{author}{{Liang}, X.H.}, \bibinfo{author}{{Liao}, J.Y.}, \bibinfo{author}{{Liu}, G.Q.}, \bibinfo{author}{{Liu}, H.W.}, \bibinfo{author}{{Liu}, S.Z.}, \bibinfo{author}{{Liu}, X.J.}, \bibinfo{author}{{Liu}, Y.}, \bibinfo{author}{{Liu}, Y.N.}, \bibinfo{author}{{Lu}, B.}, \bibinfo{author}{{Lu}, X.F.}, \bibinfo{author}{{Luo}, T.}, \bibinfo{author}{{Ma}, X.},
  \bibinfo{author}{{Meng}, B.}, \bibinfo{author}{{Nang}, Y.}, \bibinfo{author}{{Nie}, J.Y.}, \bibinfo{author}{{Ou}, G.}, \bibinfo{author}{{Sai}, N.}, \bibinfo{author}{{Shang}, R.C.}, \bibinfo{author}{{Sun}, L.}, \bibinfo{author}{{Tan}, Y.}, \bibinfo{author}{{Tao}, L.}, \bibinfo{author}{{Tao}, W.}, \bibinfo{author}{{Tuo}, Y.L.}, \bibinfo{author}{{Wang}, G.F.}, \bibinfo{author}{{Wang}, J.}, \bibinfo{author}{{Wang}, W.S.}, \bibinfo{author}{{Wang}, Y.S.}, \bibinfo{author}{{Wen}, X.Y.}, \bibinfo{author}{{Wu}, B.B.}, \bibinfo{author}{{Wu}, M.}, \bibinfo{author}{{Xiao}, G.C.}, \bibinfo{author}{{Xiong}, S.L.}, \bibinfo{author}{{Xu}, H.}, \bibinfo{author}{{Yan}, L.L.}, \bibinfo{author}{{Yang}, J.W.}, \bibinfo{author}{{Yang}, S.}, \bibinfo{author}{{Yang}, Y.J.}, \bibinfo{author}{{Zhang}, A.M.}, \bibinfo{author}{{Zhang}, C.L.}, \bibinfo{author}{{Zhang}, C.M.}, \bibinfo{author}{{Zhang}, F.}, \bibinfo{author}{{Zhang}, H.M.}, \bibinfo{author}{{Zhang}, J.}, \bibinfo{author}{{Zhang}, Q.}, \bibinfo{author}{{Zhang}, T.},
  \bibinfo{author}{{Zhang}, W.}, \bibinfo{author}{{Zhang}, W.C.}, \bibinfo{author}{{Zhang}, W.Z.}, \bibinfo{author}{{Zhang}, Y.}, \bibinfo{author}{{Zhang}, Y.}, \bibinfo{author}{{Zhang}, Y.F.}, \bibinfo{author}{{Zhang}, Y.J.}, \bibinfo{author}{{Zhang}, Z.}, \bibinfo{author}{{Zhang}, Z.}, \bibinfo{author}{{Zhang}, Z.L.}, \bibinfo{author}{{Zhao}, H.S.}, \bibinfo{author}{{Zhao}, J.L.}, \bibinfo{author}{{Zhao}, X.F.}, \bibinfo{author}{{Zhu}, Y.}, \bibinfo{author}{{Zhu}, Y.X.}, \bibinfo{author}{{Zou}, C.L.}, \bibinfo{year}{2019}.
\newblock \bibinfo{title}{{In-orbit Demonstration of X-Ray Pulsar Navigation with the Insight-HXMT Satellite}}.
\newblock \bibinfo{journal}{APJs} \bibinfo{volume}{244}, \bibinfo{pages}{1}.
\newblock \DOIprefix\doi{10.3847/1538-4365/ab3718}, \href{http://arxiv.org/abs/1908.01922}{{\tt arXiv:1908.01922}}.
\bibitem[{{Zhuravlev} et~al.(2020){Zhuravlev}, {Yermolaev} and {Andrianov}}]{Zhuravlev2020}
\bibinfo{author}{{Zhuravlev}, V.I.}, \bibinfo{author}{{Yermolaev}, Y.I.}, \bibinfo{author}{{Andrianov}, A.S.}, \bibinfo{year}{2020}.
\newblock \bibinfo{title}{{Probing the ionosphere by the pulsar B0950+08 with help of RadioAstron ground-space baselines}}.
\newblock \bibinfo{journal}{MNRAS} \bibinfo{volume}{491}, \bibinfo{pages}{5843--5851}.
\newblock \DOIprefix\doi{10.1093/mnras/stz3370}, \href{http://arxiv.org/abs/1906.10435}{{\tt arXiv:1906.10435}}.

\end{thebibliography}







\end{document}